\documentclass[11pt,letter]{article} 
 
\usepackage{jheppub}  
\usepackage{colortbl} 
\usepackage[section] {placeins}
 \usepackage{arydshln}

 \def\be{\begin{equation}}
\def\ee{\end{equation}}
\def\bea{\begin{eqnarray}}
\def\eea{\end{eqnarray}}
 \definecolor{darkshade}{RGB}{153,153,153}
 \definecolor{lightshade}{RGB}{204,204,204}
 
\title{Exactly stable non-BPS spinors in heterotic string theory on tori}
\author{Jihye Seo} 
\affiliation{Ernest Rutherford Physics Department, McGill University \\
3600 rue University, Montr\'{e}al, Qu\'{e}bec, Canada
H3A 2T8}
\affiliation{Centre de recherches math\'{e}matiques,
Universit\'{e} de Montr\'{e}al \\
C.P. 6128, succ. centre-ville,
Montr\'{e}al, Qu\'{e}bec, Canada
H3C 3J7}
\emailAdd{jihyeseo@hep.physics.mcgill.ca} 
\emailAdd{seo@crm.umontreal.ca}
\abstract{Considering $SO(32)$ heterotic string theory compactified on $T^d$ with $ d \le 4$, stability of non-supersymmetric states is studied. A non-supersymmetric state with robust stability is constructed, and its exact stability is proven in a large region of moduli space against all the possible decay mechanisms allowed by charge conservation. Using various $T$-duality transform matrices of \cite{SeoPhD}, we translate various selection rules about conserved charges into simpler problems resembling partition and parity of integers. For heterotic string on $T^4$, we give a complete list of BPS atoms with elementary excitations, and we study BPS and non-BPS molecules with various binding energies. Using string-string duality, the results are interpreted in terms of Dirichlet-branes in type IIA string theory compactified on an orbifold limit of a K3 surface.}
\begin{document}
\maketitle
\flushbottom
 \toccontinuoustrue


\section{Introduction}

Superstring theories come equipped with supersymmetry and extra dimensions, but to accommodate the nature we have to find a nonsupersymmetric vacuum with long lifetime in four dimensions. 
There have been lots of effort to search for metastable or exactly stable nonsupersymmetric configurations in string theories, as reviewed in \cite{SenNonBPSReview,SchwarzTASInonBPS,GaberdielLecture}. 
 However usually supersymmetric vacua are ground states of the system, and it is often difficult to conceive a nonsupersymmetric vacuum which is also energetically favored (or allowed for some time).

In this paper, we continue on this effort, in heterotic string theory on tori, where it is straightforward to compute masses and check for supersymmetry. We construct a non-BPS state which is exactly stable in a huge region, which connects every other corner of the moduli space, containing the point of self-dual radii. Therefore it gives hope that supersymmetry breaking in string theory does not necessarily require a fine tuning of parameters. To build an exactly stable non-BPS state, there is going to be a competition between BPS states and non-BPS states to have less mass, with the rule of game to be charge conservation. To simplify the analysis we will consider the competition between the best players, namely we will consider the lightest non-BPS states (with constant mass) against all possible BPS states (whose mass depend on moduli). 
 A lightest non-BPS state may decay into a collection of $n$ BPS states
\begin{equation}
{\rm ( state)}_{\mathrm{lightest \ non-BPS}} \rightarrow \sum_{i=1}^n {\rm ( state)}_{{\rm BPS},i}
\end{equation}
subject to charge conservation and non-creation of mass
\begin{eqnarray}
\mathbf{P}_{\rm non-BPS}   =   \sum_{i=1}^n \mathbf{P}_{{\rm BPS},i}, \qquad m_{\rm non-BPS} \ge  \sum_{i=1}^n m_{{\rm BPS},i} \label{masscomp}.
\end{eqnarray}
If any of \eqref{masscomp} fails to hold in a moduli region, then the non-BPS state is exactly stable there. We take the following strategy to test exact stability region of non-BPS states:
\begin{enumerate}
 \item Start with a lightest possible non-BPS state.
 \item Classify a collection of BPS states that holds charge conservation of \eqref{masscomp} and identify the lightest possible collection of BPS states in each class.
 \item Compare masses and find conditions on moduli which makes non-BPS state to be lighter than {\emph {every}} class of BPS states. (If any set of BPS states is lighter, than non-BPS state will decay into that set which carries the same charge and less mass.)
 \end{enumerate} 

In this paper, by ``selection rules'' we will mean partitioning a given charge vector into a set of other charge vectors satisfying the charge conservation and various rules from heterotic string on a torus. 
Each state in heterotic string carries a charge on a multidimensional lattice, subject to various restrictions from Wilson lines etc. Since charge is quantized, partitioning the charge somewhat resembles partitioning an integer. For integer partition, one considers a positive integer and a way of writing it as a sum of other positive integers. (For heterotic charge partition, each entry of charge vector can be an integer or a half-integer with any sign.) For integer partition, a restricted partition is a partition with rules on the parts: for example, odd partition is where every part must be odd, and 
distinct partition is where all parts are distinct from one another. In fact these two rules are equivalent as shown by Euler. Perhaps two facts about integer partition are particularly relevant for us:
\begin{description}
\item[Integer partition and parity] A partition of an odd integer must contain at least one odd integer. (By summing up even numbers only, one never obtains an odd number.)
\item[Transposition of restricted partition] A restricted partition can be translated into another restricted partition. (e.g. odd partition and distinct partition mentioned above)
\end{description}
From those,
we are taking these two lessons for charge conservation of heterotic string theory. 
\begin{description}
\item[Parity of charge in the internal lattice, whose elements are half-integer or integer] If a state has half-integer charge along certain direction, the decay products must contain one with half-integer charge.  
\item[Transformation of internal lattice] We will perform various transformations (which come from $T$-dualities) on the internal lattice, so that obscure selection rule will be translated into a simpler selection of the first type above. That way we will have more straightforward way of protecting a non-BPS state with many selection rules. 
\end{description}
Instead of adding up integers, we will be adding up vectors on a lattice, which are conserved charges. Extra restriction is that when we partition a given charge vector into a set of charge vectors, the sum of mass cannot increase.

A heterotic string has a left-moving bosonic mover and a right-moving fermionic mover, and the supersymmetry (or BPS) is equivalent to having the right-mover to be in the ground state. 
For a given set of charges, if a BPS state exists it is the lightest possible, therefore energetically preferred. However with certain choice of charges, the choice itself already prevents BPS condition, and the state is nonsupersymmetric (i.e. non-BPS). It may decay into a set of other BPS or non-BPS states, if it is allowed by energy (non-increasing) and charge conservation. We will study non-BPS states, whose charge is protected by various selection rules. 
Since the left-mover has 16 extra compact directions to store conserved charges, one can concoct a stubborn non-BPS state with limited possibilities for decay into a set of lighter states. 

Continuing along \cite{SeoPhD}, we study BPS and non-BPS states in $SO(32)$ heterotic string theory compactified on $T^4$ and map them to the dual type IIA theory on K3 surface as in \cite{BG,GaberdielLecture}. We consider BPS atoms with elementary excitations, by adding them together (charge-wise) we build BPS and non-BPS molecules with various binding energy. 
We study lightest non-BPS states and systematically exhaust all the possible decay channels allowed by charge conservation, and then we find that a certain spinor representation has a large stability region, which contains those of other less stable non-BPS states as well.
We interpret these non-BPS and BPS heterotic string states in terms of D-branes wrapped over orbifold limit of K3 surface in type IIA string theory.

We introduce a set of transformation matrices in heterotic string side, which is equivalent to taking even number of $T$-dualities on $T^4$.  These $16 \times 16$ matrices form a subgroup of isometry group of compactified 16 dimensional momentum vector of the left-mover of a heterotic string state. The momentum is a conserved charge and limits possible decay modes. With
these new tools, one can study possible decay channels of given non-BPS states in a
systematic way. We also consider heterotic string compactified on smaller dimensional tori, and build a non-BPS state which is exactly stable in a large region of moduli space. By observing the action of $T$-duality transformation matrices originally introduced in \cite{SeoPhD}, we also nail down the allowed alternative possible forms of the matrices. 

Stability region of non-BPS states in heterotic string theory turns out to be large. Every other corner of moduli space allows stable non-BPS
states, and these corners are connected into one huge region of the moduli
space where a non-BPS is stable against all the possible decays allowed by charge conservation. This may be a fertile path for studying
non-supersymmetric field theory and supersymmetry-breaking hidden sectors for
realistic model building in string theory.
 
The robust non-BPS state we build on tori is originated from the spinor representation of $SO(32)$ heterotic string theory in 10 dimensions, which does not decay due to conserved charge \cite{SpinorNotDecay}, which we will review now.


\subsection{Exactly stable non-BPS state in heterotic string theory \label{unComp}}
Heterotic string theory has a fermionic 10-dimensional right-mover and a bosonic 26-dimensional left-mover. When compactified on a circle, $SO(32)$ heterotic string is $T$-dual to $E_8 \times E_8$ heterotic string. Since our main focus is on heterotic string theory compactified on a torus, without loss of generality we will consider $SO(32)$ heterotic string theory. The left-mover has extra 16 dimensions, whose momentum is quantized as a 16-dimensional vector $P_L$, which is given as 
\be {P}_{L} =V , \qquad {\rm for \ an uncompactified \ string} \label{Vnoncomp} \ee
where $V$ is an element of a 16-dimensional even self-dual lattice $\Gamma^{16}$ 
\be V \in \Gamma ^{16}=\left \{  (\gamma_1, \ldots , \gamma_{16})  \bigg| \sum_i \gamma_i \in 2 \mathbb{Z},  \gamma_i \in   \mathbb{Z} \cup  \left( \mathbb{Z} + \frac{1}{2} \right) \right \} . \label{Gamma16}
\end{equation}
 Notice that each entry allows for half-integer or integer. Later on, we will see a no-go theorem for decay of certain states from a simple fact that sum of integers cannot give a half-integer.

 Along a closed string, there is no preferred location for the origin of a coordinate, which translates into the level matching condition between left and right movers, which has to be satisfied by any physical state \cite{SpinorNotDecay}:
\begin{equation}
{\frac{1}{2}} {P}_{L}^{2}+N_{L}-1=  N_{R}-C_{R}, \label{HetMatch}
\end{equation}%
where $-1$ on the left hand side and 
\begin{equation}
C_{R}=\frac{1}{2}, \ \ {\rm (NS)};  \qquad C_{R}=  0,  \ \  {\rm(R)} \label{RNS}
\end{equation}
on the right hand side are added to cancel the zero-point energy of the left-mover and the right-mover. 
In \eqref{RNS}, Ramond and Neveu-Schwarz sectors are denoted by R and NS, with the periodic and anti-periodic boundary conditions for right-moving fermions, respectively.  Non-negative integers $N_{L}$ and $N_R-C_R$ in \eqref{HetMatch} denote oscillation numbers on the
bosonic left-mover and the fermionic right-mover, respectively.  

The heterotic string state has mass given by:%
\begin{equation}
\frac{1}{8}m_{h}^{2}={\frac{1}{2}} {P}_{L}^{2}+N_{L}-1= N_{R}-C_{R} .  \label{HetMass}
\end{equation}
It is manifest that condition for BPS is straightforward for heterotic string theory: it is equivalent to having the right-mover in the ground state
\begin{equation}
N_{R}=C_{R}.  \label{BPScondition}
\end{equation}
For an uncompactified heterotic string, the states which saturates BPS condition have zero-mass according to \eqref{HetMass} and preserve supersymmetry. The amount of supersymmetry can be different depending on $N_L$:
BPS states with $N_{L}=1$ form a short multiplet and are called half-BPS states (preserving half of supersymmetry), and BPS states with $N_{L}=0$ form a ultrashort multiplet and are called quarter-BPS states (preserving quarter of the supersymmetry). 

All BPS states have $P_L^2 \le 2$, and the components of $P_L$ have to be all integers: since living on $\Gamma^{16}$, if $P_L$ contains a half-integer, all components have to be half-integers, and the norm will become too large $P_L^2 \ge 4$.
There is only one half-BPS state, given by $P_L=0$.
 One obtains quarter-BPS states from \be P_L^2=(\pm 1, \pm 1, 0^{14}) \label{masslessBPS} \ee or by swapping the elements around: there are total 
 $\frac{16 \times 15}{2} \times 2^2$ such states. Superscripts on the elements of the charge vector denote repetition of components. (As we will see later, some of these quarter-BPS states remain as massless BPS states compactified on few-tori, but for a 4-torus, all of them become massive BPS states which have momenta excitations inside the torus.)  

As explained in \cite{SpinorNotDecay}, a heterotic state (spinor) with $P_L =\left( \left(\frac{1}{2}\right)^{16} \right)$, or flipping any signs even times, is a lightest non-BPS state, which cannot decay into lighter BPS states, due to charge conservation. In other words, BPS states (with integer charge) cannot add up to give the charge of this non-BPS state (with half-integer charge). Therefore this non-BPS state is always protected by the topology of conserved charges.

 Now we consider compactification on tori, which will allow more radial moduli, we will see that the stability of the spinor state survives, though not everywhere in the moduli space.
  
\subsection{Organization of the paper}
The organization of the paper is as follows.  
Section \ref{HetT4} studies heterotic string theory compactified on $T^4$. 
First we look at non-BPS states in subsection \ref{nonBPS} with focus on the lightest non-BPS states in
     subsection \ref{lightnonBPS}. BPS states are treated in
 subsection \ref{BPSt4}. First in   
 subsection \ref{atomsub}, we look at the BPS atoms, or the simplest and most elementary BPS excitations, from which one can build any BPS states.
  Subsection \ref{Tmat} introduces various $T$-duality transformation matrices, which manifests symmetries among BPS atoms. 
   Subsection \ref{BPSmole} talks about BPS and non-BPS molecules, which bound states made from the BPS atoms.  
  In subsection \ref{stbnon}, stability of non-BPS states is discussed. 
  In section \ref{sec:HeteroticOrganizer}, we will see that $T$-duality transformation matrices translate various selection rules into simple parity problems, and we can keep track of certain properties of conserved charges and analyzes various non-BPS and BPS states.
   Finally, the stability region of non-BPS states is discussed in subsection \ref{HetRegion}.  
    In subsection \ref{smaller}, we consider heterotic string theory compactified on smaller dimensional tori.
 
        Section \ref{HetIIAduality} studies type IIA string theory on K3 from the heterotic string perspective.
     Subsection \ref{dualitychain} reviews the duality chain
between heterotic string theory and type IIA string theory.  
        Subsection \ref{BPSmole2a} interprets BPS bound states in type IIA using that in heterotic string.
          Subsection \ref{stab2anon} translates stable non-BPS bound states of heterotic string theory into that of type IIA string theory.
  
    Conclusion and open problems are discussed in 
section\ref{Conclusion}.

Appendix \ref{0eigen} deals with further properties of various matrices used in the paper: subsection \ref{alter} discusses alternative choices for $T$-duality transformation matrices, and eigenvectors and eigenvalues are listed in subsection \ref{eigen}.

\section{Stability of non-BPS states in Heterotic String theory on $T^4$ \label{HetT4}}
Here we will consider similar questions of stability of non-BPS states, but now with compactification on torus. We start with $T^4$ in this section, and in later subsections we will go down to smaller dimensions. We will again see that the non-BPS states in spinor representation play an important role.

Compactifying heterotic string theory on torus allows excitations along the circles and freedom to choose Wilson lines. A heterotic string compactified on a 4-torus \cite{HetTorus,NarainMassLevel} has Kaluza-Klein and winding excitations $n^i , w_i \in \mathbb{Z}$ in each direction $x_i$ ($i \in \{1,2,3,4 \}$) of a 4-torus $T^4$. 
Because there are 4 circles in the torus, we also choose four Wilson lines $A^i$: since the fundamental group of the torus is Abelian, these four Wilson lines commute \cite{HetTorus} and can be written as four diagonal matrices as:
\begin{eqnarray}
A^{1} &=&\left( \left( \frac{1}{2}\right) ^{8},0^{8}\right) , \qquad
A^{2}=\left( \left( \left( \frac{1}{2}\right) ^{4},0^{4} \right)^2   \right) \nonumber \\
A^{3} &=&\left( \left( \left( \frac{1}{2}\right) ^{2},0^{2} \right)^4   \right), \qquad
A^{4} =\left( \left( \frac{1}{2},0\right) ^{8}\right), \label{FixWilson}
\end{eqnarray}
so that $SO(32)$ breaks down to $SO(2^{5-4})^{2^4}=U(1)^{16}$ 
\cite{PolchinskiWittenHetIduality}, 
and the theory is dual to type IIA string theory compactified on an orbifold limit of a K3 surface \cite{BG} (more details to come later section \ref{HetIIAduality}, where we discuss IIA on K3).

The choice of Wilson lines in \eqref{FixWilson} motivates this following notation for 16 directions in $\Gamma^{16}$ of \eqref{Gamma16}: 
\bea
 (\gamma_1, \ldots, \gamma_{16} ) &=& \sum_{i=1}^{16} \gamma_i E_{16-i} = \gamma_1 E_{15} +\gamma_2 E_{14} + \ldots + \gamma_{16} E_0 \nonumber \\
 &=& \gamma_1 E_{1111_{(2)}} +\gamma_2 E_{1110_{(2)}} + \ldots + \gamma_{16} E_{0000_{(2)}} \nonumber \\ &=&\sum_{a,b,c,d=0}^{1} \gamma_{16-8 a - 4 b -2 c -d} E_{abcd_{(2)}}. \label{Enotation}
\eea
For further convenience, we denote the subscripts of $E$'s using the base-2 number system in the second and the third lines of \eqref{Enotation}. From now on, we will even drop the subscript (2) from the subscripts of $E$'s.
Wilson lines of \eqref{FixWilson} can be written in a much simpler form now
\be 
A^{1} =  \frac{1}{2} \sum_{b,c,d=0}^{1} E_{1bcd}, \quad
A^{2}= \frac{1}{2} \sum_{a,c,d=0}^{1}E_{a1cd} , \quad
A^{3} = \frac{1}{2} \sum_{a,b,d=0}^{1}E_{ab1d}, \quad
A^{4} = \frac{1}{2} \sum_{a,b,c=0}^{1}E_{abc1} \label{FixWilsonBi}.
\ee

Let us denote the left- and right-moving momenta in $(16+4, 4)$ internal directions by
\begin{equation}
\mathbf{P}_{L}=(P_{L},p_{L})    , \qquad \mathbf{P}_{R} = p_{R}.
\end{equation}%
The momentum of the left-mover on 16-dimensional lattice 
\begin{equation} P_L = V+ A^{i}w_{i}  \label{PLAw} \end{equation}
 is shifted by Wilson lines and winding, away from the uncompactified case of \eqref{Vnoncomp}, where summation over $i \in \{1,2,3,4 \}$ is implied. Note that $P_L$ can have 8 integers and 8 half-integers (although their locations are subject to various constraint) as its elements if any of $w_i$ is odd.
Components for left- and right-moving momenta in $T^4$, $p_{L,R}$ are given as:
\begin{equation}
p_L^i =   \frac{p^{i}}{%
2R_{i}}+w_{i}R_{i}  , \qquad
p_R^i  =  \frac{p^{i}}{2R_{i}}%
-w_{i}R_{i} , \label{t4mom}
\end{equation} where the index $i \in \{1,2,3,4 \}$ is {\emph{not}} contracted.
  The physical momentum $p^{i}$ in the $T^{4}$ is also shifted by winding excitations and choice of Wilson lines as
\begin{equation}
p^i = n^i + B^{ij} w_j - V A^i -\frac{1}{2} A^i A^j w_j , \label{pVAw}
\end{equation}  with $w_i, n_i \in {\mathbb{Z} }$ and contraction over the index $j \in \{1,2,3,4 \}$.
For simplicity, we will assume $B^{ij}=0$ throughout this paper. 

Similarly to the uncompactified case of \eqref{HetMatch} and \eqref{HetMass}, the level matching condition has to be satisfied  
\begin{equation}
{\frac{1}{2}}\mathbf{P}_{L}^{2}+N_{L}-1={\frac{1}{2}}\mathbf{P}_{R}^{2}+N_{R}-C_{R}, \label{HetMatchT4}
\end{equation}%
with
\begin{equation}
C_{R}=\frac{1}{2}, \ \ {\rm (NS)};  \qquad C_{R}=  0, \ \  {\rm(R)}
\end{equation} 
which can be written more explicitly as   
\bea
{\frac{1}{2}} {P}_{L}^{2}+N_{L}-1&=&{\frac{1}{2}} \left( {p}_{R}^{2}-p_L^2\right)+N_{R}-C_{R} \nonumber \\ &=& {\frac{1}{2}} \left( \left(  \frac{p^{i}}{2R_{i}}%
-w_{i}R_{i}  \right)^{2}-\left(   \frac{p^{i}}{%
2R_{i}}+w_{i}R_{i} \right)^2\right)+N_{R}-C_{R}   \nonumber \\ &=&  \left( -  {p^{i}} 
 w_{i}   \right)+N_{R}-C_{R} .
  \label{HetMatchT4mod}
\eea

The mass is given by \begin{equation}
\frac{1}{8}m_{h}^{2}={\frac{1}{2}}\mathbf{P}_{L}^{2}+N_{L}-1={\frac{1}{2}}%
\mathbf{P}_{R}^{2}+N_{R}-C_{R} .  \label{HetMassT4}
\end{equation}  
And again, half-BPS (resp. quarter-BPS) states form a short (resp. ultrashort) multiplet and satisfy $N_{L}=1$ (resp. $N_{L}=0$), and all BPS states have $N_{R}=C_{R}$. Note that having an extra excitation $N_L=1$ can cost a bit more mass energy, as we will see in BPS bound states in \eqref{twoways}, \eqref{m12h}, and \eqref{m12q}.
%
%


For our choice of Wilson lines as in \eqref{FixWilsonBi}, note here that $p^i$ can receive contribution of half-integer from last two terms in \eqref{pVAw}. From \eqref{FixWilsonBi}, we have 
\be A^i A^j = 1 + \delta_{ij} \ee where $\delta$ is a Kronecker delta symbol. 
By multiplying \eqref{pVAw} by 2, we have
\begin{equation}
2p^i = 2n^i - 2 V A^i -  A^i A^j w_j  \label{pVAw1}
\equiv   2 V A^i  - w_i + \sum_{j=1}^4 w_j   \quad ({\rm mod}\ 2),
\end{equation} where `$\equiv$' with (mod 2) denotes congruence modulo 2: i.e. integers have same remainder under division by 2.
By multiplying \eqref{PLAw} with $2 A^i$, we have 
\begin{equation} 2 P_L  A^i=  2V  A^i+ 2  A^i A^{j}w_{j}  \equiv 2V  A^i \quad  ({\rm mod}\ 2). \label{2PLAw} \end{equation}
By combining \eqref{pVAw1} and \eqref{2PLAw}, we obtain the following: 
\bea 2p^i\equiv 2 P_L A^i - w_i + \sum_{j }  w_j  \equiv 2 V A^i - w_i + \sum_{j }  w_j   \quad  ({\rm mod}\ 2).  \label{pQuanta}\eea  
In \eqref{pQuanta}, we dropped the range of $j$.  In later subsections we will use this formula for smaller dimensional tori. In all cases, $j$ is supposed to be added for all circle directions.
 If one satisfies \eqref{pQuanta}, then by proper choice of $n$, one can hold \eqref{PLAw}.
As we can see from the momenta formulas which is linear in $p, w, V$, charge is completely additive (one can superpose), but BPS condition and mass are not.
Later in subsection \ref{BPSmole}, we will consider adding multiple BPS states to form bound states with various binding energy.

\subsection{Non-BPS states \label{nonBPS}}
Non-BPS heterotic string states still satisfy \eqref{HetMassT4}, but their right-movers are excited $N_R> C_R$. There are two classes of non-BPS states:
\begin{itemize}
\item $N_L=0$: left-mover is in the ground state. This heterotic string carries charge which cannot satisfy BPS condition. 
\item $N_L=1$: left-mover and right-mover are both excited, and the non-BPS state can release energy and go down to lighter BPS or non-BPS state with smaller $N_{L,R}$.
\end{itemize}
The first type is more interesting for us, since we are competing for the lighter mass. 

The second type reminds us that we can build an infinite tower of non-BPS states on top of any heterotic string state, by increasing $N_{L,R}$.
From any BPS state, we can add $N_L$ and $N_R$ (keeping the same charge), and it can become a non-BPS state, carrying same charge but more mass. Therefore generically non-BPS states are thought to be heavier objects than BPS states. Perhaps this fact gave non-BPS fame of being massive and unstable. 

However, we will focus our attention on the lightest possible non-BPS state, and see that they {\emph {can}} be exactly stable in a huge region of moduli space. 
\subsubsection{Lightest non-BPS states \label{lightnonBPS}}
From \eqref{HetMassT4}, we learn that the lightest possible heterotic string state is given when $N_R = C_R+1$ and $p_L=p_R=0$, with $m_{\emptyset}=\sqrt{8}$ ($\emptyset$ denoting zero supersymmetry). The $P_L$ and $N_L$ need to satisfy one of the following:
\begin{itemize}
\item  $P_L^2=0$ and $N_L=2$,
\item $P_L^2=2$ and $N_L=1$,
\item $P_L^2=4$ and $N_L=0$.
\end{itemize}
But first two can decay immediately to a BPS state with smaller $N_{L,R}$ and carrying the same charges\footnote{The corresponding BPS states are massless $N_L=1, P_L=0$ in gravity multiplet and   massive $N_L=0, P_L^2=2$ discussed in \cite{BG}.}. Therefore we will focus on the third option. 
 There are three possible forms of $P_L$ as below (with proper constraint, which is explained shortly):
 \begin{itemize}
\item one $\pm 2$ and 15 $0$'s. For example, $P_L=\left( 0, -2, 0^{14} \right)$,
\item four $\pm 1$ and 12 $0$'s. For example, $P_L=\left( 1,-1,1,-1, 0^{14} \right)$,
\item sixteen $\pm \frac{1}{2}$'s. For example, $P_L =\left( \left(  \frac{1}{2} \right)^{16} \right)$.
\end{itemize} 

In order not to excite $p$ and $w$ modes, $P_L$ needs to be chosen certain ways keeping in mind the restrictions from \eqref{pQuanta}. The first case, any choice works: there are 32 such states from a choice of 16 possible locations for the nonzero element in the vector and a choice of 2 possible signs.
The 
second case, signs do not matter at all, and only locations matter: again borrowing the notation of \eqref{Enotation}, $P_L$ charges needs to take a form of 
\be e_a E_{a_1 a_2 a_3 a_4}+e_b E_{b_1 b_2 b_3 b_4}+e_c E_{c_1 c_2 c_3 c_4}+e_d E_{d_1 d_2 d_3 d_4} \ee where $\vec{a}=(a_1,a_2,a_3,a_4), \vec{b}, \vec{c}, \vec{d}$ are all different, and $a_i+b_i+c_i+d_i$ is even for all $i$.
The third case, we can only tamper with signs, with $2^{16-4-1}=2^{11}$ allowed combinations: this counting comes from the fact that $2P_L A^i $ needs to be an even number rather than odd, and that negative signs can appear even times among 16 places. 

\subsection{BPS states \label{BPSt4}}
 
Let us see if it is possible to have $P_L\ne 0, w=p=0$ for BPS states. One could have $P_L=0$ and $N_L=1$, which gives a massless state in gravity multiplet as given in \cite{BG}. However since this carries no charge and has no mass, it is not meaningful to consider this for the purpose of the paper. (Recall that our aim is to find a non-BPS state whose {\emph{charge}} conservation protects it from decaying into lighter states.) The remaining option is $P_L^2=2, N_L=0$, and $m_h=2$. Since there is no winding excitation, $P_L = V \in \Gamma^{16}$, and the entries can be all integers or all half-integers. If they are all half-integers, then the norm is too large. Therefore it has to be all integers. To have the correct norm, the possible 16 elements are two $\pm 1$'s and fourteen $0$'s, similarly to \eqref{masslessBPS}, subject to an extra constraint. Using the notation of \eqref{Enotation}, it takes the following form: \be P_L=e_a E_{a_1 a_2 a_3 a_4} + e_b E_{b_1 b_2 b_3 b_4}  \ee
with $\vec{a} \ne \vec{b}$ (so that the two non-vanishing elements sit at two {\emph{different}} locations) and $e_a, e_b =\pm 1$.  
However, 
from \eqref{pQuanta} for $w=0$, 
\be  0=2p^i\equiv 2 V A^i   = 2 P_L A^i   \qquad  ({\rm mod}\ 2) , \label{2pviol} \ee 
  \be 2 P_L  A^i =  \left( e_a \delta_{a_i 1}  + e_b \delta_{b_i 1} \right)   \equiv     \delta_{a_i 1}  + \delta_{b_i 1}  \qquad  ({\rm mod}\ 2) \label{nowp} \ee needs to hold for any $i \in \{1,2,3,4 \}$.

We will show that no matter where we place $\pm 1$ in $P_L$, we are meant to violate \eqref{nowp}. 
In order to have $\vec{a} \ne \vec{b}$, at least one of the component needs to be different. If $a_j \ne b_j$ without loss of generality, then $(a_j , b_j)=(0,1)$ or $(a_j , b_j)=(1,0)$. In other words, one of them has to equal to $1$, and we have $  \delta_{a_j 1}  + \delta_{b_j 1} =1$. 

 Therefore we conclude that 
it is not possible to have BPS states with $P_L\ne 0$ if $w=p=0$. Therefore BPS states must have non-zero elements for $w$ or $p$. 
However later for smaller tori, we will learn that  it is possible to have these types of BPS states, since \eqref{nowp} needs to hold for only three or less $i$'s.  

Now we learned that BPS states of heterotic string theory on $T^4$ must carry charges in $w$ or $p$ directions. (in 4-torus). Having $w$ and $p$ on the same circle can complicate the BPS condition as in \eqref{HetMatchT4mod} and it makes one worry about signs of $p,w$ when computing masses. Therefore we will first consider BPS states which does not carry both $w$ and $p$ on the same circle. And then later in next subsection \ref{BPSmole} we will consider the effect of combining them.

\subsubsection{BPS atoms \label{atomsub}} 
For each circle we can have $w$ or $p$ excitation, and there are 4 circles; we have total 8 directions to have excitations. Both $w$ and $p$ are quantized, each in units of 1 and $\frac{1}{2}$. We will consider BPS atoms which has the smallest excitation among one of these eight, and they all are quarter-BPS states. And we will call them BPS atoms, and we will study these 8 sets of atoms here. By adding these atoms, we can make any BPS states, therefore these atoms are BPS building blocks. From these building blocks we can also make any non-BPS states which are mentioned in this paper, including the most stable one. Afterwards in subsection \ref{BPSmole}, we will consider combining various atoms together with various binding energy. 

\begin{table}[!h]
\begin{center}
\renewcommand{\arraystretch}{1.5}
\begin{tabular}{|c|c||c|} 
\hline
$p_{L}$ & $P_{L}$ & symbol  \\ \hline
$ \left( R_{h1},0,0,0 \right)  $ &  $\frac{e_1}{2}  \left( \sum_{c,d=0}^1 E_{A1cd}-2E_{A1CD}\right)+\frac{e_0}{2}  \left( \sum_{c,d=0}^1 E_{A0cd}-2E_{A0\bar{C}\bar{D}}\right)$ & $W_{1}$     \\ \hdashline
$\left(0,R_{h2},0,0 \right)$ & $\frac{e_1}{2}  \left( \sum_{c,d=0}^1 E_{1Bcd}-2E_{1BCD}\right)+\frac{e_0}{2}  \left( \sum_{c,d=0}^1 E_{0Bcd}-2E_{0B\bar{C}\bar{D}}\right)$  & $W_{2}$ \\ \hdashline
$\left(0,0,R_{h3},0 \right)$ & $\frac{e_1}{2}  \left( \sum_{a,b=0}^1 E_{abC1}-2E_{ABC1}\right)+\frac{e_0}{2}  \left( \sum_{a,b=0}^1 E_{abC0}-2E_{\bar{A}\bar{B}C0}\right)$ & $W_{3}$  \\ \hdashline
$\left(0,0,0,R_{h4} \right)$ & $\frac{e_1}{2}  \left( \sum_{a,b=0}^1 E_{ab1D}-2E_{AB1D}\right)+\frac{e_0}{2}  \left( \sum_{a,b=0}^1 E_{ab0D}-2E_{\bar{A}\bar{B}0D}\right)$  & $W_{4}$    \\ \hline
$\left(\frac{1}{4R_{h1}},0,0,0 \right)$ & $e_1 E_{1BCD} +e_0 E_{0BCD} $& $M_{1}$   \\ \hdashline
$\left(0,\frac{1}{4R_{h2}},0,0 \right)$ & $e_1 E_{A1CD}+e_0  E_{A0CD}  $ &  $M_{2}$  \\ \hdashline
$\left(0,0,\frac{1}{4R_{h3}},0 \right)$ & $ e_1 E_{AB1D} +e_0 E_{AB0D} $  & $M_{3}$   \\ \hdashline
$\left(0,0,0,\frac{1}{4R_{h4}} \right)$ & $ e_1 E_{ABC1} +e_0 E_{ABC0} $ & $M_{4}$   \\ \hline   \end{tabular} 
\renewcommand{\arraystretch}{1}
\caption{ BPS atoms in heterotic string theory compactified on $T^4$} \label{table:hetBPSt4} 
\end{center} 
\end{table}
Here we list 8 classes of BPS atoms with an elementary excitation inside the 4-torus in {\bf table} \ref{table:hetBPSt4}.  
First two columns correspond to $p_L$ and $P_L$ of heterotic string states.  A symbol $W_i$ denotes a set of these BPS objects with $w_i=1$ and $w_j=p=0$. Similarly, a set $M_i$ consists of the BPS excitation modes with minimal physical momentum $p^i =\frac{1}{2}$ in one of $T^4$ directions, with no other excitations $p^j = w =0$. A bar $\bar{\phantom{C}}$ means $\bar{{A}}=1-A$.
Also not shown in the table, masses of each atom are given from $m_h = 2|p_R|=2|p_L|$ as $m_{W_i,\frac{1}{4}}=2R_{hi}, m_{W_i,\frac{1}{4}}=\frac{1}{2R_{hi}}$.
    
We now show that the choice of $P_L$ for $M_1$ and $W_1$ indeed satisfies \eqref{pQuanta},  \bea 2p^i\equiv 2 P_L A^i - w_i + \sum_{j }  w_j  \equiv 2 V A^i - w_i + \sum_{j }  w_j   \quad  ({\rm mod}\ 2).   \eea   

For $M_1$ case, 
 \bea
2 P_{L,M_1} A^1   =  e_1 \equiv 1, \qquad
2 P_{L,M_1} A^2   =  \delta_{B1} (e_1  +e_0 ) \equiv 0  \eea
and by symmetry $ 2 P_{L,M_1} A^2 \equiv 2 P_{L,M_1} A^3  \equiv 2 P_{L,M_1} A^4  \equiv 0 $ as desired by \eqref{pQuanta}.

For $W_1$, $- w_i + \sum_{j }  w_j $ is even for $i=1$ and odd for $i\ne 1$.
Multiplying $P_L$ of $W_1$ in {\bf table} \ref{table:hetBPSt4} (now denoted $P_{L, W_1} $) with Wilson lines we obtain  
\bea
2 P_{L, W_1}  A^1 &=& \delta_{A 1} \left( \frac{e_1}{2}  \left( \sum_{c,d=0}^1 1-2 \right)+\frac{e_0}{2}  \left( \sum_{c,d=0}^1 1-2 \right) \right) =\delta_{A 1} \left(  {e_1} + {e_0}  \right) \equiv 0 \nonumber , \\
2P_{L, W_1}  A^2 &=& \frac{e_1}{2}  \left( 4-2 \right) = e_1 \equiv 1 \nonumber , \\
2P_{L, W_1}  A^3 &=& \frac{e_1}{2}  \left( 2-2 \delta_{C 1}\right)+\frac{e_0}{2}  \left( 2 -2 \delta_{\bar{C} 1}\right) = {e_1}{ }  \left( 1-   \delta_{C 1}\right)+ {e_0}{ }  \left( 1 -  \delta_{\bar{C} 1}\right)\equiv 1 . \eea 

The expression for $P_L$ is manifestly symmetric between change of third and fourth digits ($c \leftrightarrow d, C \leftrightarrow D$) of index of $E$. Therefore $ 2 P_L A^4 $ is also odd. In fact, $P_L$ is symmetric under exchange of second, third, and fourth digits (among $b,c,d$ and $B,C,D$) and $P_L$ for $W_1$ could have been written in equivalent forms such as 
\bea 
 P_{L, W_1} &=&  \frac{e_1}{2}  \left( \sum_{b,d=0}^1 E_{Ab1d}-2E_{AB1D}\right)+\frac{e_0}{2}  \left( \sum_{b,d=0}^1 E_{Ab0d}-2E_{A\bar{B}0\bar{D}}\right) , \nonumber \\
{\rm or}  \quad
 P_{L, W_1} &=& \frac{e_1}{2}  \left( \sum_{b,c=0}^1 E_{Abc1}-2E_{ABC1}\right)+\frac{e_0}{2}  \left( \sum_{b,c=0}^1 E_{Abc0}-2E_{A\bar{B}\bar{C}D}\right)\label{actusym}. \eea
In each expression a symmetry between different two circles is manifest.

We have freedom to choose from $0,1$ for each capitalized index ($A,B,C,D$). Indices in lower cases ($a,b,c,d$) are dummy variables. We can also choose signs for each $\pm$ symbol and $e_i=\pm 1$. Therefore for each row we have 32 choices. This is expected because we have $2^{7-3}$ choices of sign combination: negative signs can appear even times, and that guarantees $2P_L A^1$ to be even ($  2^{8-1}$). To ensure $2P_L A^{2,3,4}$ is odd, it takes $2^3$ less options ($\times 2^{-3}$).  With a choice of sign for a nonzero component of $p$ or $ w$, we obtain 64 choices, which is equal to the counting of \cite{BG}.  Since the expression of $P_L$ in {\bf table} \ref{table:hetBPSt4} satisfies the correct rules and also they give correct number of solutions, therefore it is a {\bf complete} list of correct solutions.

As seen in \eqref{actusym}, there is a symmetry of exchanging 4 circles. Also 
from the shape of \eqref{FixWilson} it is clear that the Wilson lines can get exchanged by permuting components in $P_L$, $\Gamma^{16}$. 

We introduce matrices which swap internal directions of 4-torus: action of the unitary matrix $U_{ij}$ will swap $x^{i,j}$ directions and also exchange the form of Wilson lines $A^{i,j}$, without affecting other two directions of the torus: \cite{SeoPhD}
\begin{eqnarray}
U_{12}  & = &  \left(
\begin{array}{cccc}
{{\mathsf{1\!\!1}}}_4 & 0 & 0 & 0 \\
0 & 0 & {\mathsf{1\!\!1}}_4 & 0 \\
0 & {\mathsf{1\!\!1}}_4 & 0 & 0 \\
0 & 0 & 0 & {\mathsf{1\!\!1}}_4 %
\end{array} \right), \quad
U_{23}    =    \left(
\begin{array}{cc}
u_{23} & 0   \\
0 & u_{23}
\end{array} \right), \quad U_{34}   =   \left(
\begin{array}{cccc}
u_{34}  & 0 & 0 & 0 \\
0 & u_{34} & 0  & 0 \\
0 & 0 & u_{34} & 0 \\
0 & 0 & 0 & u_{34}  %
\end{array} \right), \label{Udef}
\end{eqnarray}
with following $8\times 8$ and $4\times 4$ submatrices:
\begin{equation}
u_{23}   =   \left(
\begin{array}{cccc}
{\mathsf{1\!\!1}}_2 & 0 & 0 & 0 \\
0 & 0 & {\mathsf{1\!\!1}}_2 & 0 \\
0 & {\mathsf{1\!\!1}}_2 & 0 & 0 \\
0 & 0 & 0 & {\mathsf{1\!\!1}}_2 %
\end{array} \right), \qquad u_{34}   =   \left(
\begin{array}{cccc}
1  & 0 & 0 & 0 \\
0 & 0 & 1  & 0 \\
0 & 1  & 0 & 0 \\
0 & 0 & 0 & 1  %
\end{array} \right), \label{sigma4def}\end{equation}
where ${\mathsf{1\!\!1}}_k$ is an identity matrix of size $k$. Although we are defining matrices here, we are not using $\equiv$ symbols, because we reserve it only when we mean congruence moduo 2, as in \eqref{pQuanta}. It is straightforward to see
that $U_{12}$ exchanges $E_{abcd}$ with $E_{bacd}$, $U_{13}$ exchanges $E_{abcd}$ with $E_{cbad}$, etc 
in terms of notation of basis vector for $\Gamma^{16}$ given in \eqref{Enotation}.

We will now analyze $W_1$ and $M_4$ cases of {\bf table} \ref{table:hetBPSt4} in detail below. In \eqref{m1v2}, \eqref{2w1v2}, \eqref{1w1v2} to come, signs of $p,w,n,V$ are all correlated.
Since all circle directions are interchangeable, the results can be easily generalized into $W_i$ and $M_i$ cases, in other words the full list of BPS atoms.
 
\paragraph{$M_4$: where $p^{4}=\pm \frac{1}{2}$ is the only excitation in 4-torus}
   
   \be  \renewcommand{\arraystretch}{1.2}
\begin{tabular}{|c:c:c:c:c|} \hline
$\pm V$ & $\pm n_{1}$ & $\pm n_{2}$ & $\pm n_{3}$ & $\pm n_{4}$ \\ \hline
$\left( 0^{2a},1,-1,0^{14-2a}\right) $ & $0$ & $0$ & $0$ & $1$ \\ \hdashline
$\left( 0^{2a},-1,+1,0^{14-2a}\right) $ & $0$ & $0$ & $0$ & $0$  \\ \hdashline
$\left( 1,1,0^{14}\right) $ & $1$ & $1$ & $1$ & $1$ \\ \hdashline
$\left( 0^{2},1,1,0^{12}\right) $ & $1$ & $1$ & $0$ & $1$ \\ \hdashline
$\left( 0^{4},1,1,0^{10}\right) $ & $1$ & $0$ & $1$ & $1$ \\ \hdashline
$\left( 0^{6},1,1,0^{8}\right) $ & $1$ & $0$ & $0$ & $1$ \\ \hdashline
$\left( 0^{8},1,1,0^{6}\right) $ & $0$ & $1$ & $1$ & $1$ \\ \hdashline
$\left( 0^{10},1,1,0^{4}\right) $ & $0$ & $1$ & $0$ & $1$ \\ \hdashline
$\left( 0^{12},1,1,0^{2}\right) $ & $0$ & $0$ & $1$ & $1$ \\ \hdashline
$\left( 0^{14},1,1\right) $ & $0$ & $0$ & $0$ & $1$  \\ \hdashline
$\left( -1,-1,0^{14}\right) $ & $-1$ & $-1$ & $-1$ & $0$ \\ \hdashline
$\left( 0^{2},-1,-1,0^{12}\right) $ & $-1$ & $-1$ & $0$ & $0$ \\ \hdashline
$\left( 0^{4},-1,-1,0^{10}\right) $ & $-1$ & $0$ & $-1$ & $0$ \\ \hdashline
$\left( 0^{6},-1,-1,0^{8}\right) $ & $-1$ & $0$ & $0$ & $0$ \\ \hdashline
$\left( 0^{8},-1,-1,0^{6}\right) $ & $0$ & $-1$ & $-1$ & $0$ \\ \hdashline
$\left( 0^{10},-1,-1,0^{4}\right) $ & $0$ & $-1$ & $0$ & $0$ \\ \hdashline
$\left( 0^{12},-1,-1,0^{2}\right) $ & $0$ & $0$ & $-1$ & $0$ \\ \hdashline
$\left( 0^{14},-1,-1\right) $ & $0$ & $0$ & $0$ & $0$ \\ \hline
\end{tabular} \renewcommand{\arraystretch}{1}
\label{m1v2} \ee

From this one can similarly write down information about all other $M_i$'s, thanks to $U_{ij}$ matrices. In fact, we will soon introduce matrices which makes dualities between $M$ and $W$ modes. So almost we do not need to write down detailed information about $W_i$'s at all. They act nicely on $P_L$ but it is not clear what it does for $n$'s, so we will list the information about $W_1$ nonetheless, for the sake of $n$'s and completeness. 

\paragraph{$W_1$: where $w_{1}=\pm 1$ is the only excitation in 4-torus}
Note that thanks to Wilson lines and winding numbers, we can have various form of $P_L$, such as with 8 integer and 8 half-integer elements, which does not necessarily live inside $\Gamma^{16}$. As we consider smaller dimensional tori later, we lose some Wilson lines, and the form of $P_L$ will be more restricted. The form of $P_L$ was given in a compact form in {\bf table} \ref{table:hetBPSt4} already, but here we will explicitly write down every single case of $V$. 
Below we list the information of $V$ and $n$'s from the $W_1$ of {\bf table} \ref{table:hetBPSt4} with $A=0$ ($A$ appears in the subscript of $E$, this is not to be confused with Wilson lines $A^i$.), whose $P_L$ has 0 for the first 8 components and half-integers for the last 8 components:
\be 
\renewcommand{\arraystretch}{1.5}
\begin{tabular}{|c:c:c:c:c|} \hline 
$\pm V$ & $\pm n_{1}$ & $\pm n_{2}$ & $\pm n_{3}$ & $\pm n_{4}$ \\ \hline
$\left( \left( -\frac{1}{2}\right) ^{8},+\frac{1}{2},+\frac{1}{2},+\frac{1}{2},-\frac{1}{2};+\frac{1}{2},-\frac{1}{2},-\frac{1}{2},-\frac{1}{2}%
\right) $ & $-1$ & $0$ & $0$ & $0$ \\ \hdashline
$\left( \left( -\frac{1}{2}\right) ^{8},+\frac{1}{2},+\frac{1}{2},+%
\frac{1}{2},-\frac{1}{2};-\frac{1}{2},+\frac{1}{2},+\frac{1}{2},+\frac{1}{2}%
\right) $ & $-1$ & $0$ & $0$ & $0$ \\ \hdashline
$\left( \left( -\frac{1}{2}\right) ^{8},+\frac{1}{2},+\frac{1}{2},-%
\frac{1}{2},+\frac{1}{2};+\frac{1}{2},-\frac{1}{2},+\frac{1}{2},+\frac{1}{2}%
\right) $ & $-1$ & $0$ & $0$ & $0$ \\ \hdashline
$\left( \left( -\frac{1}{2}\right) ^{8},+\frac{1}{2},-\frac{1}{2},+%
\frac{1}{2},+\frac{1}{2};+\frac{1}{2},+\frac{1}{2},-\frac{1}{2},+\frac{1}{2}%
\right) $ & $-1$ & $0$ & $0$ & $0$ \\ \hdashline
$\left( \left( -\frac{1}{2}\right) ^{8},-\frac{1}{2},+\frac{1}{2},+%
\frac{1}{2},+\frac{1}{2};+\frac{1}{2},+\frac{1}{2},
+\frac{1}{2},-\frac{1}{2}\right) $ & $-1$ & $0$ & $0$ & $0$ \\ \hdashline
$\left( \left( -\frac{1}{2}\right) ^{8},+\frac{1}{2},-\frac{1}{2},-%
\frac{1}{2},-\frac{1}{2};+\frac{1}{2},+\frac{1}{2},+\frac{1}{2},-\frac{1}{2}%
\right) $ & $-1$ & $-1$ & $0$ & $0$ \\ \hdashline
$\left( \left( -\frac{1}{2}\right) ^{8},+\frac{1}{2},-\frac{1}{2},+%
\frac{1}{2},+\frac{1}{2};-\frac{1}{2},-\frac{1}{2},
+\frac{1}{2},-\frac{1}{2}\right) $ & $-1$ & $0$ & $-1$ & $0$ \\ \hdashline
$\left( \left( -\frac{1}{2}\right) ^{8},+\frac{1}{2},+\frac{1}{2},-%
\frac{1}{2},+\frac{1}{2};-\frac{1}{2},+\frac{1}{2},-\frac{1}{2},-\frac{1}{2}%
\right) $ & $-1$ & $0$ & $0$ & $-1$ \\ \hdashline
$\left( \left( -\frac{1}{2}\right) ^{8},-\frac{1}{2},-\frac{1}{2},
+\frac{1}{2},-\frac{1}{2};+\frac{1}{2},-\frac{1}{2},+\frac{1}{2},+\frac{1}{2}%
\right) $ & $-1$ & $-1$ & $-1$ & $0$ \\ \hdashline
$\left( \left( -\frac{1}{2}\right) ^{8},-\frac{1}{2},+\frac{1}{2},-%
\frac{1}{2},-\frac{1}{2};+\frac{1}{2},+\frac{1}{2},-\frac{1}{2},+\frac{1}{2}%
\right) $ & $-1$ & $-1$ & $0$ & $-1$ \\ \hdashline
$\left( \left( -\frac{1}{2}\right) ^{8},-\frac{1}{2},+\frac{1}{2},+%
\frac{1}{2},+\frac{1}{2};-\frac{1}{2},-\frac{1}{2},-\frac{1}{2},+\frac{1}{2}%
\right) $ & $-1$ & $0$ & $-1$ & $-1$ \\ \hdashline
$\left( \left( -\frac{1}{2}\right) ^{8},+\frac{1}{2},-\frac{1}{2},-%
\frac{1}{2},-\frac{1}{2};-\frac{1}{2},-\frac{1}{2},-\frac{1}{2},+\frac{1}{2}%
\right) $ & $-1$ & $-1$ & $-1$ & $-1$ \\ \hdashline
$\left( \left( -\frac{1}{2}\right) ^{8},-\frac{1}{2},+\frac{1}{2},-%
\frac{1}{2},-\frac{1}{2};-\frac{1}{2},-\frac{1}{2},
+\frac{1}{2},-\frac{1}{2}\right) $ & $-1$ & $-1$ & $-1$ & $-1$ \\ \hdashline
$\left( \left( -\frac{1}{2}\right) ^{8},-\frac{1}{2},-\frac{1}{2},%
+\frac{1}{2},-\frac{1}{2};-\frac{1}{2},+\frac{1}{2},-\frac{1}{2},-\frac{1}{2}%
\right) $ & $-1$ & $-1$ & $-1$ & $-1$ \\ \hdashline
$\left( \left( -\frac{1}{2}\right) ^{8},-\frac{1}{2},-\frac{1}{2},-%
\frac{1}{2},+\frac{1}{2};+\frac{1}{2},-\frac{1}{2},-\frac{1}{2},-\frac{1}{2}%
\right) $ & $-1$ & $-1$ & $-1$ & $-1$ \\ \hdashline
$\left( \left( -\frac{1}{2}\right) ^{8},-\frac{1}{2},-\frac{1}{2},-%
\frac{1}{2},+\frac{1}{2};-\frac{1}{2},+\frac{1}{2},+\frac{1}{2},+\frac{1}{2}%
\right) $ & $-1$ & $-1$ & $-1$ & $-1$ \\ \hline
\end{tabular}
\renewcommand{\arraystretch}{1} \label{1w1v2}
 \ee

The signs on the last eight components are given with the following rules:
\begin{itemize}
\item For $v_{9,\ldots, 12}$, negative sign appears odd times. (This rule is explicitly reflected in the first half of $P_L$ of $W_1$ of {\bf table} \ref{table:hetBPSt4}.) This relation ensures that $2P_L A^2$ has a correct value molulo 2. 
\item Between $v_{9,\ldots, 12}$ and $v_{13,\ldots, 16}$, 
\bea v_{12+i}&=&v_{13-i}, \qquad \forall i ,\nonumber \\ {\rm or} \qquad
 v_{12+i}&=&-v_{13-i},\qquad \forall  i. \label{forall} \eea
  (This is why the $\bar{\phantom{A}}$'s appear in the second half of $P_L$ of $W_1$ of {\bf table} \ref{table:hetBPSt4}.)  
This relation ensures that $2P_L A^{3,4}$ has a correct value molulo 2. 
\end{itemize}

Second set is from the $W_1$ of {\bf table} \ref{table:hetBPSt4} with $A=1$, whose $P_L$ has half-integers for the first 8 components and zeros for the last 8 components. The information of $V$ and $n$'s is given as below:
\be 
\renewcommand{\arraystretch}{1.2}
\begin{tabular}{|c:c:c:c:c|} 
\hline
$\pm V$ & $\pm n_{1}$ & $\pm n_{2}$ & $\pm n_{3}$ & $\pm n_{4}$ \\ \hline
$\left( 0,0,0,-1;-1,0,0,0;0^{8}\right) $ & $0$ & $0$ & $0$ & $0$ \\ \hdashline
$\left( 0,0,-1,0;0,-1,0,0;0^{8}\right) $ & $0$ & $0$ & $0$ & $0$ \\ \hdashline
$\left( 0,-1,0,0;0,0,-1,0;0^{8}\right) $ & $0$ & $0$ & $0$ & $0$ \\ \hdashline
$\left( -1,0,0,0;0,0,0,-1;0^{8}\right) $ & $0$ & $0$ & $0$ & $0$ \\ \hdashline
$\left( 0,0,0,-1;0,-1,-1,-1;0^{8}\right) $ & $-1$ & $0$ & $0$ & $0$ \\ \hdashline
$\left( 0,-1,-1,-1;0,0,0,-1;0^{8}\right) $ & $-1$ & $-1$ & $0$ & $0$ \\ \hdashline
$\left( 0,-1,0,0;-1,-1,0,-1;0^{8}\right) $ & $-1$ & $0$ & $-1$ & $0$ \\ \hdashline
$\left( 0,0,-1,0;-1,0,-1,-1;0^{8}\right) $ & $-1$ & $0$ & $0$ & $-1$ \\ \hdashline
$\left( -1,-1,0,-1;0,-1,0,0;0^{8}\right) $ & $-1$ & $-1$ & $-1$ & $0$ \\ \hdashline
$\left( -1,0,-1,-1;0,0,-1,0;0^{8}\right) $ & $-1$ & $-1$ & $0$ & $-1$\\ \hdashline
$\left( -1,0,0,0;-1,-1,-1,0;0^{8}\right) $ & $-1$ & $0$ & $-1$ & $-1$\\ \hdashline
$\left( -1,-1,-1,0;-1,0,0,0;0^{8}\right) $ & $-1$ & $-1$ & $-1$ & $-1$  \\ \hdashline
 $\left( 0,-1,-1,-1;-1,-1,-1,0;0^{8}\right) $ & $-2$ & $-1$ & $-1$ & $-1 $ \\ \hdashline
$\left( -1,0,-1,-1;-1,-1,0,-1;0^{8}\right) $ & $-2$ & $-1$ & $-1$ & $-1$ \\ \hdashline
$\left( -1,-1,0,-1;-1,0,-1,-1;0^{8}\right) $ & $-2$ & $-1$ & $-1$ & $-1$ \\ \hdashline
$\left( -1,-1,-1,0;0,-1,-1,-1;0^{8}\right) $ & $-2$ & $-1$ & $-1$ & $-1 $ \\ \hline
\end{tabular}
\renewcommand{\arraystretch}{1}
\label{2w1v2} \ee
After adding with $A^iw_i$ to obtain $P_L$, one immediately notices the similarities between \eqref{1w1v2} and \eqref{2w1v2} in the shape of $P_L$. However the forms of $n$'s does not manifest a nice pattern.

\subsubsection{$T$-duality transformation matrices \label{Tmat}}

The transformations for 16-dimensional momentum vector $P_L$ of the left-mover of heterotic string states are given by the $16\times 16$ matrices ${\mathsf{1\!\!1}}_{16}, T_{1234}$, and $ T_{ij}$ with $i < j \in \{1,2,3,4\}$, given by the following formulas \cite{SeoPhD}:
\bea
T_{12}  &  = &  \left(
\begin{array}{cccc}
L & 0 & 0 & 0 \\
0 & 0 & R & 0 \\
0 & R & 0 & 0 \\
0 & 0 & 0 & L%
\end{array} \right), \qquad T_{13}  =   U_{23} \cdot T_{12} \cdot U_{23}, \qquad
T_{23}  =   U_{12} \cdot T_{13} \cdot U_{12}, \nonumber \\ 
T_{14}  & = &  U_{34} \cdot T_{13} \cdot U_{34} , \qquad
T_{24}  =   U_{34} \cdot T_{23} \cdot U_{34}, \qquad
T_{34}  =   U_{23} \cdot T_{24} \cdot U_{23}, \nonumber \\
T_{1234} &=  & T_{12} \cdot T_{34} = T_{13} \cdot T_{24} =T_{14} \cdot T_{23} ,  \label{T12def}
\end{eqnarray}
where $L$ and $R$ are $4 \times 4$ matrices given below \cite{SeoPhD}:%
\begin{equation}
L  =  {\frac{1}{2}} \left(
\begin{array}{cccc}
+1 & +1 & +1 & -1 \\
+1 & +1 & -1 & +1 \\
+1 & -1 & +1 & +1 \\
-1 & +1 & +1 & +1%
\end{array}%
\right), \qquad R = {\frac{1}{2}} \left(
\begin{array}{cccc}
+1 & -1 & -1 & -1 \\
-1 & +1 & -1 & -1 \\
-1 & -1 & +1 & -1 \\
-1 & -1 & -1 & +1%
\end{array}%
\right). \label{LRdef}
\end{equation}
All of $T$'s commute with one another, have unit determinant, and square to an identity matrix,
\begin{equation}
T_{1234}^{2}  =  T_{ij}^{2}={\mathsf{1\!\!1}}_{16} , \qquad i < j \in \{1,2,3,4\}  .
\end{equation}%
Also \be T_{ij} T_{jk} =T_{ik} \ee %
 A careful look at $T_{12}, L, R$ in \eqref{T12def} tells us that the rules given near \eqref{forall} are satisfied and that from $M_1$ class one obtains $W_1$ class given in {\bf table} \ref{table:hetBPSt4}. Since $T$'s are their own inverse, $T_{12}$ can convert $W_1$ into $M_1$ as well. In fact, a transformation matrix $T_{ij}$ corresponds to $T$-dualities on $x_i$ and $x_j$ directions in $T^4$ and exchanges excitations in $W_{i,j} \leftrightarrow M_{i,j}$ classes. Similarly, $T_{1234}$ corresponds to $T$-dualities on all the four directions in $T^4$ and exchanges excitations in $W_{i} \leftrightarrow M_{i}$ classes $\forall i$ \cite{SeoPhD}. 
 
More explicitly, one can extend 16 by 16 matrices into 24 by 24 matrices, where each element denotes $\left(2p^1, w_1, 2p^2, w_2, \ldots, 2p^4, w_4; P_L \right)$. 
The action of $T_{ij}$ is seen better from 24 by 24 matrices $\mathbf{T}_{ij}$ which reads for example,
\be 
\mathbf{T}_{12}= \left( \begin{array}{cccc:c}
\sigma_1 & & & & \\
& \sigma_1 & & &  \\
  & & {{\mathsf{1\!\!1}}}_2 & & \\
    & & & {{\mathsf{1\!\!1}}}_2  & \\ \hdashline
        & & &    & T_{12} 
        \end{array}
\right), \qquad \mathbf{T}_{1234}= \left( \begin{array}{cccc:c}
\sigma_1 & & & & \\
& \sigma_1 & & &  \\
  & & \sigma_1  & & \\
    & & & \sigma_1   & \\ \hdashline
        & & &    & T_{1234} 
        \end{array}
\right).
\ee

Starting from a BPS object in $M_4$ class with $P_L= \left( 1, 1, 0^{14} \right)$, one has
\begin{eqnarray}
P_{L} &=&P_{L}\cdot T_{12}=P_{L}\cdot T_{13} =P_{L}\cdot T_{23} =\left( 1,1,0^{14}\right)  \\
P_{L}\cdot T_{1234} &=&P_{L}\cdot T_{14}=P_{L}\cdot T_{24}=P_{L}\cdot T_{34} \nonumber \\
&=&\left( \frac{1}{2},0,\frac{1}{2},0;
\frac{1}{2},0,-\frac{1}{2},0;\frac{1}{2},0,-\frac{1}{2},0;-\frac{1}{2},0,
-\frac{1}{2},0\right).
\end{eqnarray}%
The transformation by $T_{i4}$ or $T_{1234}$ turns $P_L$ of $M_4$ class into $P_L$ of $W_4$ class which has half-integer elements (with correct sign constraints) as seen in table \ref{table:hetBPSt4}.

Similarly, starting from a BPS object in $M_1$ class with $P_L= \left( 1,0^{7};1,0^{7}\right)$, one has
\begin{eqnarray}
P_{L} &=&P_{L}\cdot T_{23}=P_{L}\cdot T_{24}=P_{L}\cdot T_{34}=\left( 1,0^{7}; 1,0^{7}\right)
\\
P_{L}\cdot T_{1234} &=&P_{L}\cdot T_{12}=P_{L}\cdot T_{13}=P_{L}\cdot T_{14}   \nonumber  \\ &=& \left( \frac{1}{2}%
,\frac{1}{2},\frac{1}{2},-\frac{1}{2}; \frac{1}{2},-\frac{1}{2},-\frac{1}{2},%
-\frac{1}{2};0^{8}\right).
\end{eqnarray}%
The transformation by $T_{1j}$ or $T_{1234}$ turns $P_L$ of $M_1$ class into $P_L$ of $W_1$ class which has half-integer elements as seen in table \ref{table:hetBPSt4}.

\subsubsection{BPS and non-BPS molecules\label{BPSmole}}
As we can see the formulas for momenta given in \eqref{PLAw}, \eqref{t4mom}, \eqref{pVAw}, the momenta are all linear in $w, p, V$. Therefore when we form a 
BPS molecule by combining BPS atoms, momenta are additive. However mass and BPS-ness cannot be superposed. 
Here we will observe how the mass and BPS-ness change as we add multiple BPS atoms.

We will see that when we add multiple BPS states together, the masses can go down with Pythagorean type saving at best, 
\be P_{\rm bound} = \sum_i P_{i}, \quad m_{\rm bound} \ge \sqrt{ \sum_i m_{i}^2}, \ee
as long as $p, w$ of each individual states do not cancel out. 
  
Recall the masses of BPS atoms and the lightest non-BPS state 
\be m_{W_i} = {2R_i}, \quad   m_{M_i} =\frac{1}{2R_i},  \qquad m_{\emptyset} =\sqrt{8}. \ee
 As we will explain near \eqref{halfBPSM4}, adding two (quarter) BPS atoms of the same kind (also with the same sign choice for $p_R$ components) gives a half-BPS bound state with zero binding energy. The charge allows them to form a non-BPS bound state too, but energetically not allowed. The masses of bound states are given as: \be m_{M_i+M_i,\frac{1}{2}} = 2  m_{M_i} , \qquad m_{M_i+M_i,\emptyset} = \sqrt{ 4 m_{M_i}^2 + m_{\emptyset}^2 } , \ee
   where the numbers $\frac{1}{2}, \emptyset$ in the subscript denote amount of supersymmetry preserved.
   
Similarly we can add two BPS atoms of different kinds, where the excitations are along two different circles. 
The bound state can be a quarter-BPS molecule with Pythagorean energy saving, or a non-BPS molecule which are energetically allowed only in some moduli region. Their masses are: 
\bea m_{M_i+M_j, \frac{1}{4}} = \sqrt{m_{M_i}^2 +m_{M_j}^2 }, \quad m_{M_i+M_j, \emptyset} =  \sqrt{m_{M_i}^2 +m_{M_j}^2 +m_{\emptyset}^2 } \nonumber  ,\\
 m_{W_i+M_j, \frac{1}{4}} = \sqrt{m_{W_i}^2 +m_{M_j}^2   }, \quad  m_{W_i+M_j, \emptyset} =  \sqrt{m_{W_i}^2 +m_{M_j}^2 +m_{\emptyset}^2 }\nonumber ,\\
 m_{W_i+W_j, \frac{1}{4}}   = \sqrt{m_{W_i}^2 +m_{W_j}^2   }, \quad m_{W_i+W_j, \emptyset} = \sqrt{m_{W_i}^2 +m_{W_j}^2 +m_{\emptyset}^2 } . \label{twoatom} \eea
The first two lines of \eqref{twoatom} is explained near \eqref{M34BPS} and \eqref{M4W1BPS}. By taking $T$-dualities from the first two lines, we obtain the third line of \eqref{twoatom}. One can generalize this to having a mixture of up to 4 $W$ and $M$'s each with excitation along different circle of 4-torus, then we will have similar energy saving for the bound state
\be  m_{M_1+M_2+M_3+M_4, \frac{1}{4}} = \sqrt{m_{M_1}^2+m_{M_2}^2+m_{M_3}^2+m_{M_4}^2   }. \label{addfour} \ee
 Similar equality holds when we switch any of $M$ into $W$, or when we delete any of $M$'s.
 
 The next example is when we add two BPS atoms of different types, but where the excitations are along the same circle as explained near \eqref{MW1BPS}. The BPS bound state has zero binding energy, while the non-BPS bound state is energetically forbidden. Their masses are
\be m_{W_i+M_i, \frac{1}{4} } = m_{W_i} +m_{M_i}  \ge 2, \quad m_{W_i+M_i, \emptyset} =  \sqrt{ \left(m_{W_i} +m_{M_i} \right)^2 +m_{\emptyset}^2 } . \ee 

Similarly from \eqref{m11234q} we have below: 
\bea m_{W_i+M_i+M_j+M_k+M_l, \frac{1}{4} }   &=&\sqrt{ \left(m_{W_i} +m_{M_i} \right)^2 +m_{M_j}^2 +m_{M_k}^2 +m_{M_l}^2 }. \eea
(Again act $T$-dualities on above formula to obtain similar ones.)

As explained near \eqref{12BPS}, by adding 4 BPS atoms together as below, we can get a half-BPS bound state, as well as a quarter-BPS bound state. Their masses are given below:\bea m_{W_i+M_i+W_j+M_j, \frac{1}{2} } &= & \sqrt{ \left(m_{W_i} +m_{M_i} \right)^2+\left(m_{W_j} +m_{M_j} \right)^2} \ge \sqrt{8}, \nonumber \\ 
  m_{W_i+M_i+W_j+M_j , \frac{1}{4} } &=& \sqrt{ m_{W_i}^2 +m_{M_i}^2+ m_{W_j}^2 +m_{M_j}^2}\ge 2 \label{twoways}. \eea
Again the half-BPS state is more massive, paying by a heavier mass for $N_L=1$ as predicted near \eqref{HetMassT4}.
The equality holds at self-dual radii.

%

The rest of subsection gives the justification and examples of the above mass formulas.
    In mass formulas and charge tables below, $\emptyset$ means non-BPS combination. In each table, above the solid line, we list charge of BPS atoms to add up, and below the solid line, the charge for the bound state.
\paragraph{$M_4$ and $M_4$}
It is important to choose them so that $p_R$ does not cancel each other and add up to zero. For fixed $M_4$, there is a unique choice of $M_4$ to give half-BPS bound state, and the rest gives non-BPS. When adding states from the same class, we obtain either $P_{L,{\rm bound}}=0$ (only unique choice out of 32 cases) forming a half-BPS bound state, such as \be 
\renewcommand{\arraystretch}{1.5}
\begin{tabular}{|c|cccc:cccc:cccc:cccc|}\hline  $M_4$ & 
 $1$ & $1$ & $0$ & $0$ &  $0$ &  $0$ &  $0$ &  $0$ &  $0$ &  $0$ &  $0$ &  $0$ &  $0$ &  $0$ &  $0$ &  $0$ \\ \hdashline    $M_4$ & 
$-1$ & $-1$ & 0 & 0 &  $0$ &  $0$ &  $0$ &  $0$ &  $0$ &  $0$ &  $0$ &  $0$ &  $0$ &  $0$ &  $0$ &  $0$ \\ \hline $M_4+M_4, \frac{1}{2}$ & 
$0$    & $0$ & 0 & 0 &  $0$ &  $0$ &  $0$ &  $0$ &  $0$ &  $0$ &  $0$ &  $0$ &  $0$ &  $0$ &  $0$ &  $0$ \\ \hline 
\end{tabular}  \label{halfBPSM4}
\renewcommand{\arraystretch}{1}
, \ee
or a non-BPS state with $P_{L,{\rm bound}}^2=4$ such as
\bea &
\renewcommand{\arraystretch}{1.5}
\begin{tabular}{|c|cccc:cccc:cccc:cccc|}\hline $M_4$ & 
 $1$ & $1$ & $0$ & $0$ &  $0$ &  $0$ &  $0$ &  $0$ &  $0$ &  $0$ &  $0$ &  $0$ &  $0$ &  $0$ &  $0$ &  $0$ \\ \hdashline  $M_4$ & 
$0$ & $0$ & 1 & $-1$ &  $0$ &  $0$ &  $0$ &  $0$ &  $0$ &  $0$ &  $0$ &  $0$ &  $0$ &  $0$ &  $0$ &  $0$ \\ \hline   $M_4+M_4,{\emptyset}$ & 
$1$    & 1 & 1 & $-1$ &  $0$ &  $0$ &  $0$ &  $0$ &  $0$ &  $0$ &  $0$ &  $0$ &  $0$ &  $0$ &  $0$ &  $0$ \\ \hline 
\end{tabular}  
\renewcommand{\arraystretch}{1}   \nonumber \\
  {\rm or } &
\renewcommand{\arraystretch}{1.5}
\begin{tabular}{|c|cccc:cccc:cccc:cccc|}\hline $M_4$ & 
 $1$ & $1$ & $0$ & $0$ &  $0$ &  $0$ &  $0$ &  $0$ &  $0$ &  $0$ &  $0$ &  $0$ &  $0$ &  $0$ &  $0$ &  $0$ \\ \hdashline  $M_4$ & 
$1$ & $-1$ & 0 & 0 &  $0$ &  $0$ &  $0$ &  $0$ &  $0$ &  $0$ &  $0$ &  $0$ &  $0$ &  $0$ &  $0$ &  $0$ \\ \hline $M_4+M_4,{\emptyset}$ & 
$2$    & $0$ & 0 & 0 &  $0$ &  $0$ &  $0$ &  $0$ &  $0$ &  $0$ &  $0$ &  $0$ &  $0$ &  $0$ &  $0$ &  $0$ \\ \hline 
\end{tabular}  \label{nonBPSM42}
\renewcommand{\arraystretch}{1}.
\eea
Their masses are given by  
 \be m_{M_4+M_4,\frac{1}{2}} =\frac{1}{R_4}=2  m_{M_4}, \qquad m_{M_4+M_4,\emptyset} =\sqrt{\left(\frac{1}{R_4^2}+8\right)} =\sqrt{ 4 m_{M_4}^2 + m_{\emptyset}^2 } > 2 m_{M_4}. \ee

We conclude that by adding two quarter-BPS of same kind (and sign), we can also obtain half-BPS ($N_L=1$) with zero binding energy if $P_L$ cancel each other.
The charge allows them to form a non-BPS bound state too, but energetically not allowed since heavier than the constituents.

\paragraph{$M_3$ and $M_4$}
Both $P_{L,M_3}$ and $P_{L,M_4}$ have two nonvanishing elements. When adding them together, if they have no overlap in location of those elements just like
\be 
\renewcommand{\arraystretch}{1.5}
\begin{tabular}{|c|cccc:cccc:cccc:cccc|}\hline $M_4$ & 0
   &  $0$ & $0$ & $0$ &  1 & 1 &  $0$ &  $0$ &  $0$ &  $0$ &  $0$ &  $0$ &  $0$ &  $0$ &  $0$ &  $0$ \\ \hdashline  $M_3$ & 
$-1$  & 0 & $1$ & 0 &0 &  $0$ &  $0$ &  $0$ &  $0$ &  $0$ &  $0$ &  $0$ &  $0$ &  $0$ &  $0$ &  $0$ \\ \hline $M_3+M_4, \emptyset$ &
   $0$ & 1& 1 & 0 & 1 & 1 &  $0$ &  $0$ &  $0$ &  $0$ &  $0$ &  $0$ &  $0$ &  $0$ &  $0$ &  $0$ \\ \hline 
\end{tabular}  \label{M34nonBPS}, 
\renewcommand{\arraystretch}{1}
\ee
then $P_{L,{\rm bound}}^2=4$ and the bound state is non-BPS.
If they overlap and cancel each other along one direction as given below: 
\bea &
\renewcommand{\arraystretch}{1.5} 
\begin{tabular}{|c|cccc:cccc:cccc:cccc|}\hline $M_4$ & 
 $1$ & $1$ & $0$ & $0$ &  $0$ &  $0$ &  $0$ &  $0$ &  $0$ &  $0$ &  $0$ &  $0$ &  $0$ &  $0$ &  $0$ &  $0$ \\ \hdashline  $M_3$ & 
$0$ & $-1$ & 0 & $1$ &  $0$ &  $0$ &  $0$ &  $0$ &  $0$ &  $0$ &  $0$ &  $0$ &  $0$ &  $0$ &  $0$ &  $0$ \\ \hline $M_3+M_4, \frac{1}{4}$ & 
$1$    & $0$ & 0 & 1 &  $0$ &  $0$ &  $0$ &  $0$ &  $0$ &  $0$ &  $0$ &  $0$ &  $0$ &  $0$ &  $0$ &  $0$ \\ \hline 
\end{tabular} \label{M34BPS},
\renewcommand{\arraystretch}{1}
 \nonumber \\ {\rm or \ } &
\renewcommand{\arraystretch}{1.5}
\begin{tabular}{|c|cccc:cccc:cccc:cccc|}\hline $M_4$ & 
 $1$ & $1$ & $0$ & $0$ &  $0$ &  $0$ &  $0$ &  $0$ &  $0$ &  $0$ &  $0$ &  $0$ &  $0$ &  $0$ &  $0$ &  $0$ \\ \hdashline  $M_3$ & 
$-1$  & 0 & $1$ & 0 & 0 &  $0$ &  $0$ &  $0$ &  $0$ &  $0$ &  $0$ &  $0$ &  $0$ &  $0$ &  $0$ &  $0$ \\ \hline $M_3+M_4, \frac{1}{4}$ &
   $0$ & 1& 1 & 0 &  $0$ &  $0$ &  $0$ &  $0$ &  $0$ &  $0$ &  $0$ &  $0$ &  $0$ &  $0$ &  $0$ &  $0$ \\ \hline 
\end{tabular} ,
\renewcommand{\arraystretch}{1}
\eea
 then we obtain a quarter-BPS molecule. For fixed $P_{L,M_3}$, there are 4 choices of $P_{L,M_4}$ which satisfies quarter-BPS condition for the bound state: they come from 2 choices for direction to cancel $P_L$ 
 and 2 choices for the sign of the other element.  
  
  The masses of bound states are given as:
\bea m_{M_3+M_4, \frac{1}{4} }& =&\sqrt{ \frac{1}{4R_3^2}+ \frac{1}{4R_4^2}} = \sqrt{m_{M_3}^2 +m_{M_4}^2 }, \label{M34q} \nonumber \\
 m_{M_3+M_4, \emptyset}& =&\sqrt{ \frac{1}{4R_3^2}+ \frac{1}{4R_4^2} +8}= \sqrt{m_{M_3}^2 +m_{M_4}^2 +m_{\emptyset}^2 }. \label{mass34} \eea
In a certain region of moduli space (very small radii, e.g.), even the second choice \eqref{mass34} can be energetically allowed, if charges add up like \eqref{M34nonBPS}.

\paragraph{$M_4$ and $W_1$}

Again two quarter-BPS states add up to another quarter-BPS molecule as in
\be 
\renewcommand{\arraystretch}{1.5}
\begin{tabular}{|c|cccc:cccc:cccc:cccc|}\hline   $W_1$ &
   $-\frac{1}{2}$ & $\frac{1}{2}$ & $\frac{1}{2}$ & $\frac{1}{2}$ & $-%
\frac{1}{2}$ & $-\frac{1}{2}$ & $-\frac{1}{2}$ & $\frac{1}{2}$ &  $0$ &  $0$ &  $0$ & 0
&  $0$ &  $0$ &  $0$ &  $0$ \\ \hdashline      $M_4$ & 0
  &  $0$ &  $0$ &  $0$ & $ 1$ & 1 &  $0$ &  $0$ &  $0$ &  $0$ &  $0$ &  $0$ &  $0$ &  $0$ &  $0$ &  $0$ \\    \hline  $W_1+M_4,  \frac{1}{4}$ &
 $-\frac{1}{2}$ & $\frac{1}{2}$ & $\frac{1}{2}$ & $\frac{1}{2}$ & $
\frac{1}{2}$ & $ \frac{1}{2}$ & $-\frac{1}{2}$ & $\frac{1}{2}$ &  $0$ &  $0$ &  $0$ & 0
&  $0$ &  $0$ &  $0$ &  $0$ \\ \hline
\end{tabular} \label{M4W1BPS} ,
\renewcommand{\arraystretch}{1}\ee
  or a non-BPS state such as \be \renewcommand{\arraystretch}{1.5}
\begin{tabular}{|c|cccc:cccc:cccc:cccc|}\hline   $W_1$ & 
   $-\frac{1}{2}$ & $\frac{1}{2}$ & $\frac{1}{2}$ & $\frac{1}{2}$ & $-%
\frac{1}{2}$ & $-\frac{1}{2}$ & $-\frac{1}{2}$ & $\frac{1}{2}$ &  $0$ &  $0$ &  $0$ & 0
&  $0$ &  $0$ &  $0$ &  $0$ \\ \hdashline       $M_4$ & 0
  &  $0$ &  $0$ &  $0$ &  $0$ &  $0$ &  $0$ &  $0$ &    1 & 1 &  $0$ &  $0$ &  $0$ &  $0$ &  $0$ &  $0$ \\    \hline  $W_1+M_4, \emptyset$ &
 $-\frac{1}{2}$ & $\frac{1}{2}$ & $\frac{1}{2}$ & $\frac{1}{2}$ & $
-\frac{1}{2}$ & $ -\frac{1}{2}$ & $-\frac{1}{2}$ & $\frac{1}{2}$ &  1 & 1 &  $0$ & 0
&  $0$ &  $0$ &  $0$ &  $0$ \\ \hline
\end{tabular}  \label{M4W1nonBPS} 
\renewcommand{\arraystretch}{1}. \ee 
For fixed $P_{L, W_1}$, there are four locations $P_{L,M_4}$ can take, and the signs are fixed automatically, to form quarter-BPS bound state.
Masses are given by 
\bea m_{W_1+M_4,  \frac{1}{4}} &=&\sqrt{ {4R_1^2}+ \frac{1}{4R_4^2}} = \sqrt{m_{W_1}^2 +m_{M_4}^2   },\label{m14q} \nonumber \\
  m_{W_1+M_4, \emptyset} &=&\sqrt{  {4R_1^2}+ \frac{1}{4R_4^2} +8} = \sqrt{m_{W_1}^2 +m_{M_4}^2 +m_{\emptyset}^2 }.\eea

\paragraph{$M_1$ and $W_1$}
 The next example is when we add two BPS atoms of different types, but where the excitations are along the same circle as explained near \eqref{MW1BPS}. The BPS bound state has zero binding energy, while the non-BPS bound state is energetically forbidden everywhere in the moduli space. 
From level matching condition \eqref{HetMatchT4}, a quarter-BPS condition bound state forms if
$ \left( -  {p^{i}}  w_{i}   \right)=\frac{1}{2}$ and $P_{L,{\rm bound}}^2=3$, for example of the following form:
\be
 \renewcommand{\arraystretch}{1.5}
\begin{tabular}{|c|cccc:cccc:cccc:cccc|}\hline   $W_1$ &
   $-\frac{1}{2}$ & $\frac{1}{2}$ & $\frac{1}{2}$ & $\frac{1}{2}$ & $-%
\frac{1}{2}$ & $-\frac{1}{2}$ & $-\frac{1}{2}$ & $\frac{1}{2}$ &  $0$ &  $0$ &  $0$ & 0
&  $0$ &  $0$ &  $0$ &  $0$ \\ \hdashline   $M_1$ & 0
  &  $0$ &  $0$ &  $0$ & $1$ &  $0$ &  $0$ &  $0$ &  $0$ &  $0$ &  $0$ &  $0$ &1  &  $0$ &  $0$ &  $0$ \\   \hline $W_1+M_1, \frac{1}{4}$ &
   $-\frac{1}{2}$ & $\frac{1}{2}$ & $\frac{1}{2}$ & $\frac{1}{2}$ & $ 
\frac{1}{2}$ & $-\frac{1}{2}$ & $-\frac{1}{2}$ & $\frac{1}{2}$ &  $0$ &  $0$ &  $0$ & 0
& 1 &  $0$ &  $0$ &  $0$ \\ \hline
\end{tabular} \label{MW1BPS}, 
\renewcommand{\arraystretch}{1} \ee 
 while $P_{L,{\rm bound}}^2\ge 5$ such as \be
 \renewcommand{\arraystretch}{1.5}
\begin{tabular}{|c|cccc:cccc:cccc:cccc|}\hline   $W_1$ &
   $-\frac{1}{2}$ & $\frac{1}{2}$ & $\frac{1}{2}$ & $\frac{1}{2}$ & $-%
\frac{1}{2}$ & $-\frac{1}{2}$ & $-\frac{1}{2}$ & $\frac{1}{2}$ &  $0$ &  $0$ &  $0$ & 0
&  $0$ &  $0$ &  $0$ &  $0$ \\ \hdashline   $M_1$ & 0
  &  $0$ &  $0$ &  $0$ & $-1$ &  $0$ &  $0$ &  $0$ &  $0$ &  $0$ &  $0$ &  $0$ &1  &  $0$ &  $0$ &  $0$ \\   \hline $W_1+M_1, \emptyset$ &
   $-\frac{1}{2}$ & $\frac{1}{2}$ & $\frac{1}{2}$ & $\frac{1}{2}$ & $ 
-\frac{3}{2}$ & $-\frac{1}{2}$ & $-\frac{1}{2}$ & $\frac{1}{2}$ &  $0$ &  $0$ &  $0$ & 0
& 1 &  $0$ &  $0$ &  $0$ \\ \hline
\end{tabular} \label{MW1nonBPS}
\renewcommand{\arraystretch}{1} \ee 
gives non-BPS state. 
For fixed $P_{L}, p_L$ of $W_1$, $P_{L, M_1}$ has freedom to choose 8 locations for nonvanishing elements, and to choose sign for one of them. The signs in $p_{L, M_1}$ are already fixed because of the level matching condition.
 
    The masses are given as 
\bea m_{W_1+M_1, \frac{1}{4} } &=& { {2R_1 }+ \frac{1}{2R_1}} =m_{W_1} +m_{M_1} \ge 2, \nonumber \\
 m_{W_1+M_1, \emptyset}& =&  \sqrt{ \left(m_{W_1} +m_{M_1} \right)^2 +m_{\emptyset}^2 } > m_{W_1} +m_{M_1}. \eea 
 The BPS bound state has zero binding energy, and the non-BPS state is never allowed since it is heavier than the constituents.

Similarly a quarter-BPS bound state of $M_{1,4}$ and $W_1$ with $P_L$ such as
\eqref{M14W1BPS} BPS
\be 
\renewcommand{\arraystretch}{1.5}
\begin{tabular}{|c|cccc:cccc:cccc:cccc|}\hline    $W_1$ &
   $-\frac{1}{2}$ & $\frac{1}{2}$ & $\frac{1}{2}$ & $\frac{1}{2}$ & $-%
\frac{1}{2}$ & $-\frac{1}{2}$ & $-\frac{1}{2}$ & $\frac{1}{2}$ &  $0$ &  $0$ &  $0$ &0
&  $0$ &  $0$ &  $0$ &  $0$ \\ \hdashline    $M_4$ & 0
&  $0$ &  $0$ &  $0$ & $1$ & $ 1$ &  $0$ &  $0$ &  $0$ &  $0$ &  $0$ &  $0$ &  $0$ &  $0$ &  $0$ &  $0$ \\  \hdashline  $M_1$ & 
 1 &  $0$ &  $0$ &  $0$ &  $0$ &  $0$ &  $0$ &  $0$ & 1 &  $0$ &  $0$ &  $0$ &  $0$ &  $0$ &  $0$ &  $0$ \\ \hline  \hline $W_1+M_1+M_4, \frac{1}{4}$ &
   $ \frac{1}{2}$ & $\frac{1}{2}$ & $\frac{1}{2}$ & $\frac{1}{2}$ & $ 
\frac{1}{2}$ & $ \frac{1}{2}$ & $-\frac{1}{2}$ & $\frac{1}{2}$ & 1 &  $0$ &  $0$ & 0
&  $0$ &  $0$ &  $0$ &  $0$ \\ \hline
\end{tabular} \label{M14W1BPS}
\renewcommand{\arraystretch}{1}
 \ee
can form with mass 
\be m_{W_1+M_1+M_4, \frac{1}{4} } = \sqrt{ \left({ {2R_1 }+ \frac{1}{2R_1}}\right)^2 + \frac{1}{4R_4^2} } =\sqrt{ \left(m_{W_1} +m_{M_1} \right)^2 +m_{M_4}^2 }\label{m114q} . \ee 

By adding up $M_{1,2,3,4}$ and $W_1$, 
a quarter-BPS bound state can form if $P_{L,{\rm bound}}^2=3$ as
\be 
\renewcommand{\arraystretch}{1.5}
\begin{tabular}{|c|cccc:cccc:cccc:cccc|}\hline    $W_1$ &
   $-\frac{1}{2}$ & $\frac{1}{2}$ & $\frac{1}{2}$ & $\frac{1}{2}$ & $-%
\frac{1}{2}$ & $-\frac{1}{2}$ & $-\frac{1}{2}$ & $\frac{1}{2}$ &  $0$ &  $0$ &  $0$ & 0
&  $0$ &  $0$ &  $0$ &  $0$ \\ \hdashline    $M_1$ &
 1 &  $0$ &  $0$ &  $0$ &  $0$ &  $0$ &  $0$ &  $0$ & 1 &  $0$ &  $0$ &  $0$ &  $0$ &  $0$ &  $0$ &  $0$ \\ \hdashline    $M_2$ &0
   &  $0$ &  $0$ &  $0$ &  $0$ &  $0$ &  $0$ &  $0$ & $-1$ &  $0$ &  $0$ &  $0$ & 1 &  $0$ &  $0$ &  $0$ \\ \hdashline    $M_3$ &0
   &  $0$ &  $0$ &  $0$ &  $0$ &  $0$ &  $0$ &  $0$ &  $0$ &  $0$ &  $0$ &  $0$ & $-1$&  $0$ & 1 &  $0$ \\ \hdashline    $M_4$&0
   &  $0$ &  $0$ &  $0$ &  $0$ &  $0$ &  $0$ &  $0$ &  $0$ &  $0$ &  $0$ &  $0$ &  $0$ &  $0$ & $-1$ & 1 \\ \hline  \hline $W_1+\sum M_i, \frac{1}{4} $ &
   $ \frac{1}{2}$ & $\frac{1}{2}$ & $\frac{1}{2}$ & $\frac{1}{2}$ & $ -
\frac{1}{2}$ & $ -\frac{1}{2}$ & $-\frac{1}{2}$ & $\frac{1}{2}$ &  $0$ &  $0$ &  $0$ & 0
&  $0$ &  $0$ &  $0$ &  1  \\ \hline
\end{tabular}  
\renewcommand{\arraystretch}{1}
\label{M1234W1BPS} , \ee
with a mass
\bea m_{W_1+M_1+M_2+M_3+M_4, \frac{1}{4} } &=& \sqrt{ \left({ {2R_1 }+ \frac{1}{2R_1}}\right)^2 + \frac{1}{4R_1^2}  + \frac{1}{4R_2^2}  + \frac{1}{4R_3^2} + \frac{1}{4R_4^2} } \nonumber \\
 &=&\sqrt{ \left(m_{W_1} +m_{M_1} \right)^2 +m_{M_2}^2 +m_{M_3}^2 +m_{M_4}^2 }\label{m11234q}. \eea

\paragraph{A strange case of adding $M_{1,2}$ and $W_{1,2}$}
If the $P_L$ charges add up to be $P_{L,{\rm bound}}^2 =2$, such as 
\be  
\renewcommand{\arraystretch}{1.5}
\begin{tabular}{|c|cccc:cccc:cccc:cccc|}\hline    $W_1$ &
   $-\frac{1}{2}$ & $\frac{1}{2}$ & $\frac{1}{2}$ & $\frac{1}{2}$ & $-%
\frac{1}{2}$ & $-\frac{1}{2}$ & $-\frac{1}{2}$ & $\frac{1}{2}$ &  $0$ &  $0$ &  $0$ & 0
&  $0$ &  $0$ &  $0$ &  $0$ \\ \hdashline  $W_2$ & 
    $-\frac{1}{2}$ & $-\frac{1}{2}$ & $-\frac{1}{2}$ & $\frac{1%
}{2}$  &  $0$ &  $0$ &  $0$ &  $0$ & $-\frac{1}{2}$ & $\frac{1}{2}$ & $\frac{1}{2}$ &
$\frac{1}{2}$ &  $0$ &  $0$ &  $0$ & $0$ \\  \hdashline  $M_1$ & 
 $ 1$ &  $0$ &  $0$ &  $0$ &  $0$ &  $0$ &  $0$ &  $0$ & 1 &  $0$ &  $0$ &  $0$ &  $0$ &  $0$ &  $0$ &  $0$ \\ \hdashline  $M_2$ & 0
  &  $0$ &  $0$ & $-1$ &  $0$ &  $0$ &  $0$ & $-1$  &  $0$ &  $0$ &  $0$ &  $0$ &  $0$ &  $0$ &  $0$ &  $0$ \\ \hline   
  $\frac{1}{2}, \frac{1}{4} $ &
 0 & 0 & 0 & 0 & $  
-\frac{1}{2}$ & $-\frac{1}{2}$ &$-\frac{1}{2}$ & $-\frac{1}{2}$ & $\frac{1}{%
2}$ & $\frac{1}{2}$ & $\frac{1}{2}$ & $\frac{1}{2}$ &  $0$ &  $0$ & $0$ &  $0$ \\ \hline
\end{tabular}%
\renewcommand{\arraystretch}{1}
\label{12BPS} ,
\ee
then we can have both half and quarter-BPS bound states: if $ \left( -  {p^{i}}  w_{i}   \right)=0$ we have a quarter-BPS. If $ \left( -  {p^{i}} w_{i}   \right)=1$ we have half-BPS. 
 
 If $P_{L,{\rm bound}}^2=4$ just like
\be  
\renewcommand{\arraystretch}{1.5}
\begin{tabular}{|c|cccc:cccc:cccc:cccc|}\hline 
 $W_1$ &  $-\frac{1}{2}$ & $\frac{1}{2}$ & $\frac{1}{2}$ & $\frac{1}{2}$ & $-\frac{1}{2}$ & $-\frac{1}{2}$ & $-\frac{1}{2}$ & $\frac{1}{2}$ &  $0$ &  $0$ &  $0$ &  $0$ &  $0$ &  $0$ &  $0$ &  $0$ \\ \hdashline 
 $W_2$ &    $-\frac{1}{2}$ & $-\frac{1}{2}$ & $-\frac{1}{2}$ & $\frac{1}{2}$  &  $0$ &  $0$ &  $0$ &  $0$ & $-\frac{1}{2}$ & $\frac{1}{2}$ & $\frac{1}{2}$ & $\frac{1}{2}$ &  $0$ &  $0$ &  $0$ & $0$ \\  \hdashline 
$M_1$ & $ 1$ &  $0$ &  $0$ &  $0$ &  $0$ &  $0$ &  $0$ &  $0$ & 1 &  $0$ &  $0$ &  $0$ &  $0$ &  $0$ &  $0$ &  $0$ \\ \hdashline 
 $M_2$&  $0$ &  $0$ & 1 &  $0$ &  $0$ &  $0$ & 1 &  $0$ &  $0$ &  $0$ &  $0$ &  $0$ &  $0$ &  $0$ &  $0$ &  $0$ \\ \hline  $\frac{1}{4}$
 & 0 & 0 & 1 &1 & $ -\frac{1}{2}$ & $-\frac{1}{2}$ &$ \frac{1}{2}$ & $ \frac{1}{2}$ & $\frac{1}{2}$ & $\frac{1}{2}$ & $\frac{1}{2}$ & $\frac{1}{2}$ &  $0$ &  $0$ & $0$ &  $0$ \\ \hline
\end{tabular}%
\renewcommand{\arraystretch}{1}
\label{12BPS2}, \ee
only quarter-BPS is possible, provided $ \left( -  {p^{i}}  w_{i}   \right)=1$. If $P_{L,{\rm bound}}^2 \ge 6$ just like
\be  
\renewcommand{\arraystretch}{1.5}
\begin{tabular}{|c|cccc:cccc:cccc:cccc|}\hline   
  $W_1$ &  $-\frac{1}{2}$ & $\frac{1}{2}$ & $\frac{1}{2}$ & $\frac{1}{2}$ & $-%
\frac{1}{2}$ & $-\frac{1}{2}$ & $-\frac{1}{2}$ & $\frac{1}{2}$ &  $0$ &  $0$ &  $0$ &0
&  $0$ &  $0$ &  $0$ &  $0$ \\ \hdashline $W_2$ & 
    $-\frac{1}{2}$ & $-\frac{1}{2}$ & $-\frac{1}{2}$ & $\frac{1%
}{2}$  &  $0$ &  $0$ &  $0$ &  $0$ & $-\frac{1}{2}$ & $\frac{1}{2}$ & $\frac{1}{2}$ &
$\frac{1}{2}$ &  $0$ &  $0$ &  $0$ & $0$ \\  \hdashline  $M_1$ &
 $ 1$ &  $0$ &  $0$ &  $0$ &  $0$ &  $0$ &  $0$ &  $0$ & 1 &  $0$ &  $0$ &  $0$ &  $0$ &  $0$ &  $0$ &  $0$ \\ \hdashline $M_2$ & 0
  &  $0$ &  $0$ &  1 &  $0$ &  $0$ &  $0$ & $-1$  &  $0$ &  $0$ &  $0$ &  $0$ &  $0$ &  $0$ &  $0$ &  $0$ \\ \hline  $\emptyset$
  & 0 & 0 & 0 & 2 & $  
-\frac{1}{2}$ & $-\frac{1}{2}$ &$-\frac{1}{2}$ & $-\frac{1}{2}$ & $\frac{1}{%
2}$ & $\frac{1}{2}$ & $\frac{1}{2}$ & $\frac{1}{2}$ &  $0$ &  $0$ & $0$ &  $0$ \\ \hline
\end{tabular}%
\renewcommand{\arraystretch}{1}
\label{12nonBPS}, \ee
the bound state is non-BPS. 

The mass of  
half-BPS case 
\bea m_{W_1+M_1+W_2+M_2, \frac{1}{2} }&=& \sqrt{ \left(m_{W_1} +m_{M_1} \right)^2+\left(m_{W_2} +m_{M_2} \right)^2} \nonumber \\
& = & \sqrt{ \left({ {2R_1 }+ \frac{1}{2R_1}}\right)^2 +  \left({ {2R_2 }+ \frac{1}{2R_2}}\right)^2 }  \ge \sqrt{8} \label{m12h} \eea
is a bit bigger than that of the
quarter-BPS case 
\bea m_{W_1+M_1+W_2+M_2 , \frac{1}{4} } &=&\sqrt{ m_{W_1}^2 +m_{M_1}^2+ m_{W_2}^2 +m_{M_2}^2}\nonumber \\
&=& \sqrt{  {4R_1^2 }+ \frac{1}{4R_1^2}  + {4R_2^2 }+ \frac{1}{4R_2^2}  }  \ge  2 \label{m12q} \eea
 because of extra excitation on left-mover ($N_L=1$) as anticipated from \eqref{HetMassT4}.

The most important lesson here is that the energy saving is at best of Pythagorean type. There is no further dramatic energy reduction. Also it is interesting that we can have both quarter and half-BPS bound states. Now we nailed down the properties on the BPS side (both atoms and molecules), we are ready to test stability of non-BPS state.

\subsection{Stability of non-BPS states \label{stbnon}}
Here we will study stability of non-BPS states against all possible decay channels allowed by charge conservation and mass non-creation. As promised earlier, various selection rules about conserved charge gets reduced to the problem of even and odd integers. Each entry can be integer or half-integer only. Therefore when a non-BPS state with half-integer charge
decays its products must include a state with half-integers. It may seem like a weak constraint, but $T$-duality matrices multiply rules, and they are already powerful enough to give a large region of exact stability of a non-BPS state ruling out any other decays. 
 \subsubsection{Selection rule from charge conservation on $\Gamma^{16}$ \label{sec:HeteroticOrganizer}}
\begin{figure}[!h]
\centerline{\includegraphics[width=1.3  in]{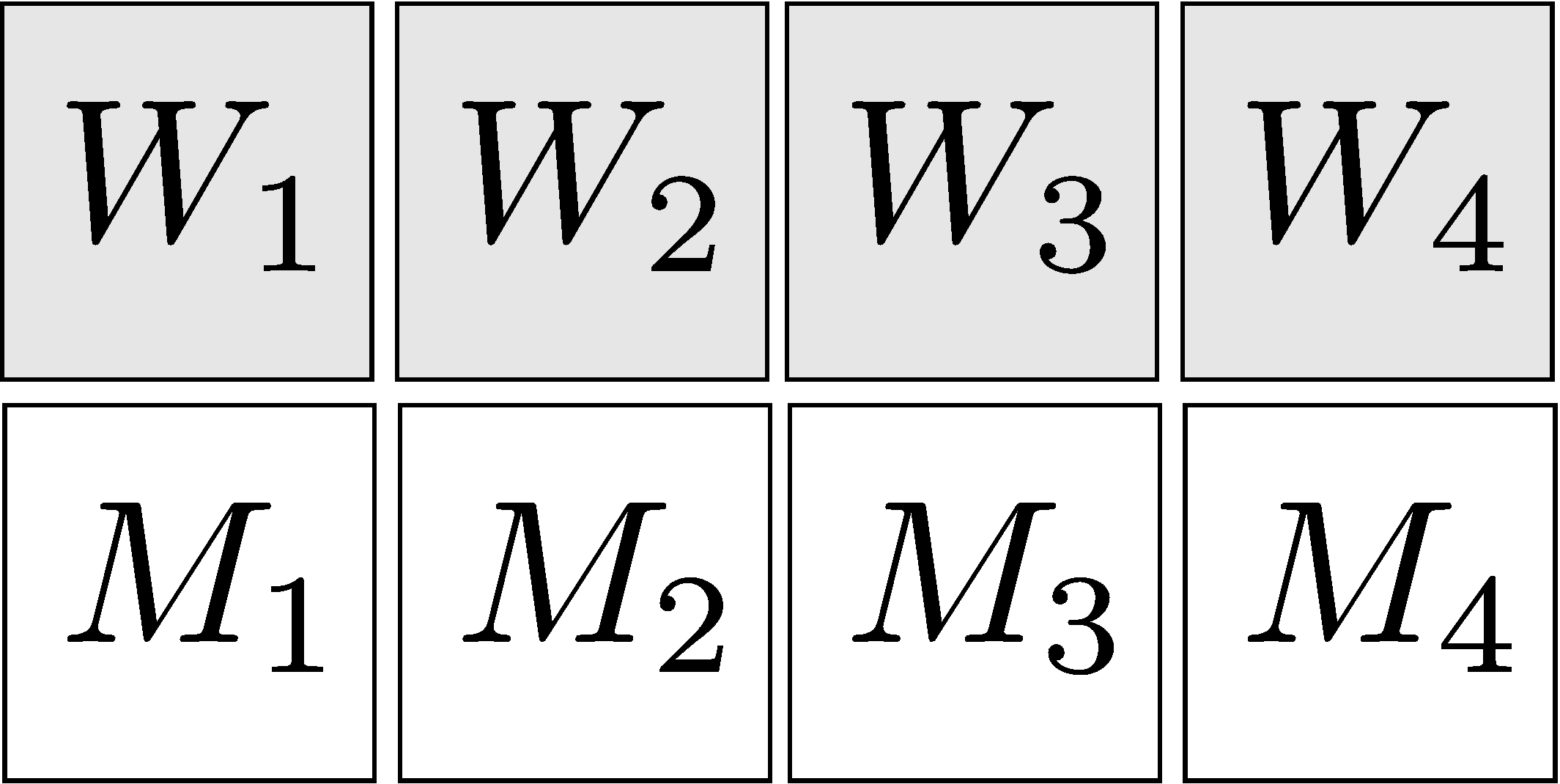}}
\caption{For a non-BPS state with $p_R=p_L=0$ and $P_L \in \left( {\mathbb{Z} } +\frac{1}{2} \right)^{16}$, possible BPS-decay channels must contain one of the \colorbox{lightshade}{shaded} objects, $W_i$'s here.}
\label{fig:8boxW}
\end{figure}

Take a non-BPS state of $p_R=0, P_L \in \left( {\mathbb{Z} } +\frac{1}{2} \right)^{16}$, then decay products must contain some state which carries half-integers in some of these 16 entries of $P_L$. The BPS-states in $M_i$'s have only integer entries in $P_L$, while BPS-states in $W_i$'s have eight half-integer entries in $P_L$. Therefore, the decay channel of the non-BPS state with $P_L \in \left( {\mathbb{Z} } +\frac{1}{2} \right)^{16}$ must contain some of $W_i$'s as in {\bf figure} \ref{fig:8boxW}. 
 
 Each BPS atom must occur in pairs for two reasons:
 \begin{itemize}
 \item Charge in $p_R$ needs to be cancelled.
 \item If a BPS atom gives half-integers in 8 (of 16) locations, another BPS atom of the same class must be added to add half-integers in the rest of 8 locations.
 \end{itemize}
Consider a non-BPS state with  
\begin{equation}
p_L=p_R=0, \quad P_{L}  = \left( \frac{1}{2},\frac{1}{2},\frac{1}{2},-\frac{1}{2%
}; \left(  \frac{1}{2},-\frac{1}{2},-\frac{1}{2},-\frac{1}{2}  \right)^2 ;-\frac{1}{2},-\frac{1}{2},-\frac{1}{2},\frac{1}{2}%
\right)  \label{D3IIa}
\end{equation}%
and mass $m_h = \sqrt{8({\frac{1}{2}}P_{L}^{2}-1)%
} =\sqrt{8}$, where the superscript after the $( \ )$ denotes a set of elements being repeated.
Its BPS-decay products must contain states with $W_i$ excitations. If it contains a state with $W_i$ excitation, it has to contain another state with $W_i$ with opposite $p_R$ charge, because at the end of the day the non-BPS state (lightest) has no $p_R$ charge. Care should be taken so that the $P_L$ charges do not cancel each other. In other words, each $W_i$ state gives half-integers at 8 locations of $P_L$, but we need half-integers at 16 places, so they better be at different places. $P_{L, W_i}$ vectors should be orthogonal to each other.

 The lightest possible collection of BPS-decay products are a pair of $W_i$ objects as discussed \cite{GaberdielLecture}, with total mass $2\times \sqrt{8({\frac{1}{2}}P_{R}^{2})} =4|P_{R}| =4{R_{hi}}$. No other lighter decays are allowed because of the form of the $P_L \in \left( {\mathbb{Z} } +\frac{1}{2} \right)^{16}$. On top of those, the BPS decay products may involve $M_i$ excitations as long as they all cancel each other at the end in $p_R$, but this kind of tampering does not decrease the energy.
The stability region for this against all the possible decays allowed by charge conservation is therefore a corner of moduli space with
\begin{equation}
\sqrt{8} < 4{R_{hi}}, \qquad \forall i. \label{alllarge}
 \end{equation}

  \begin{figure} [!h]
\centerline{\includegraphics[width=1.3  in]{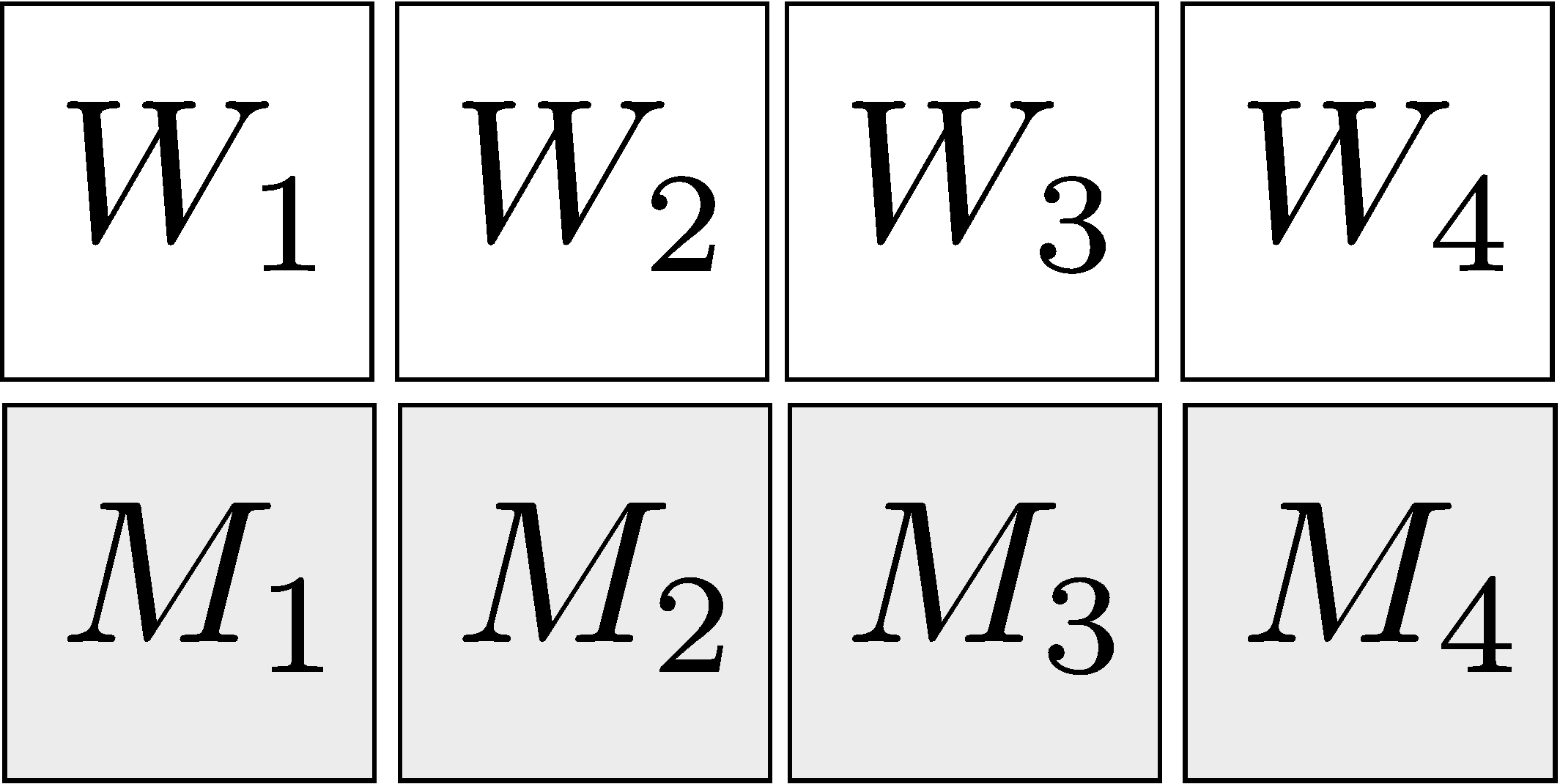}}
\caption{For a non-BPS state with $p_R=p_L=0$ and $P_L \cdot  T_{1234} \in \left( {\mathbb{Z} } +\frac{1}{2} \right)^{16}$, possible BPS-decay channels must contain one of the \colorbox{lightshade}{shaded} objects, $M_i$'s here.}
\label{fig:8boxM}
\end{figure}
 Similarly, one may ask whether there are non-BPS objects whose decay products must contain some of $M_i$'s instead. This answer is yes, due to symmetry as expected from $T$-dualities.  
  A non-BPS state with $p_R=p_L=0, P_L \cdot  T_{1234}  \in \left( {\mathbb{Z} } +\frac{1}{2} \right)^{16}$ can decay only into sets of BPS states that contain some of $M_i$'s, as shown in {\bf figure} \ref{fig:8boxM}.
 For example, a non-BPS state with  
 \begin{eqnarray}
p_R&=&p_L=0, \qquad P_{L} =\left( 2,0^{15}\right) ,  \nonumber \\
P_{L}\cdot T_{1234} &=&\left( \frac{1}{2},\frac{1}{2},\frac{1}{2},-\frac{1}{2%
}; \left(  \frac{1}{2},-\frac{1}{2},-\frac{1}{2},-\frac{1}{2}  \right)^2 ;-\frac{1}{2},-\frac{1}{2},-\frac{1}{2},\frac{1}{2}%
\right) \label{D1IIA}
\end{eqnarray}%
can decay into any of $M_i$'s as in \cite{BG,GaberdielLecture}, and we have shown that {\emph{any}} collection of BPS decay products {\it must contain} a pair of any of $M_i$'s.  The mass of non-BPS state, before decay, is $\sqrt{8 ({\frac{1}{2}} P_L^2 - 1)}  = \sqrt{8}$.
The mass on the BPS side, after decay, is $2 \times \sqrt{8({\frac{1}{2}} p_R^2)}  = 4 |p_R| =\frac{1}{R_{hi}}$
where $i \in \{1, 2, 3, 4 \}$.
The stability region for this object against every possible BPS decay is
\begin{equation}
\sqrt{8} < \frac{1}{R_{hi}}, \qquad  \forall i \in \{1, 2, 3, 4 \} \label{allsmall}
\end{equation}
which holds in another corner of moduli space.

 \begin{figure}[!h]
\centerline{\includegraphics[width= 1.3 in]{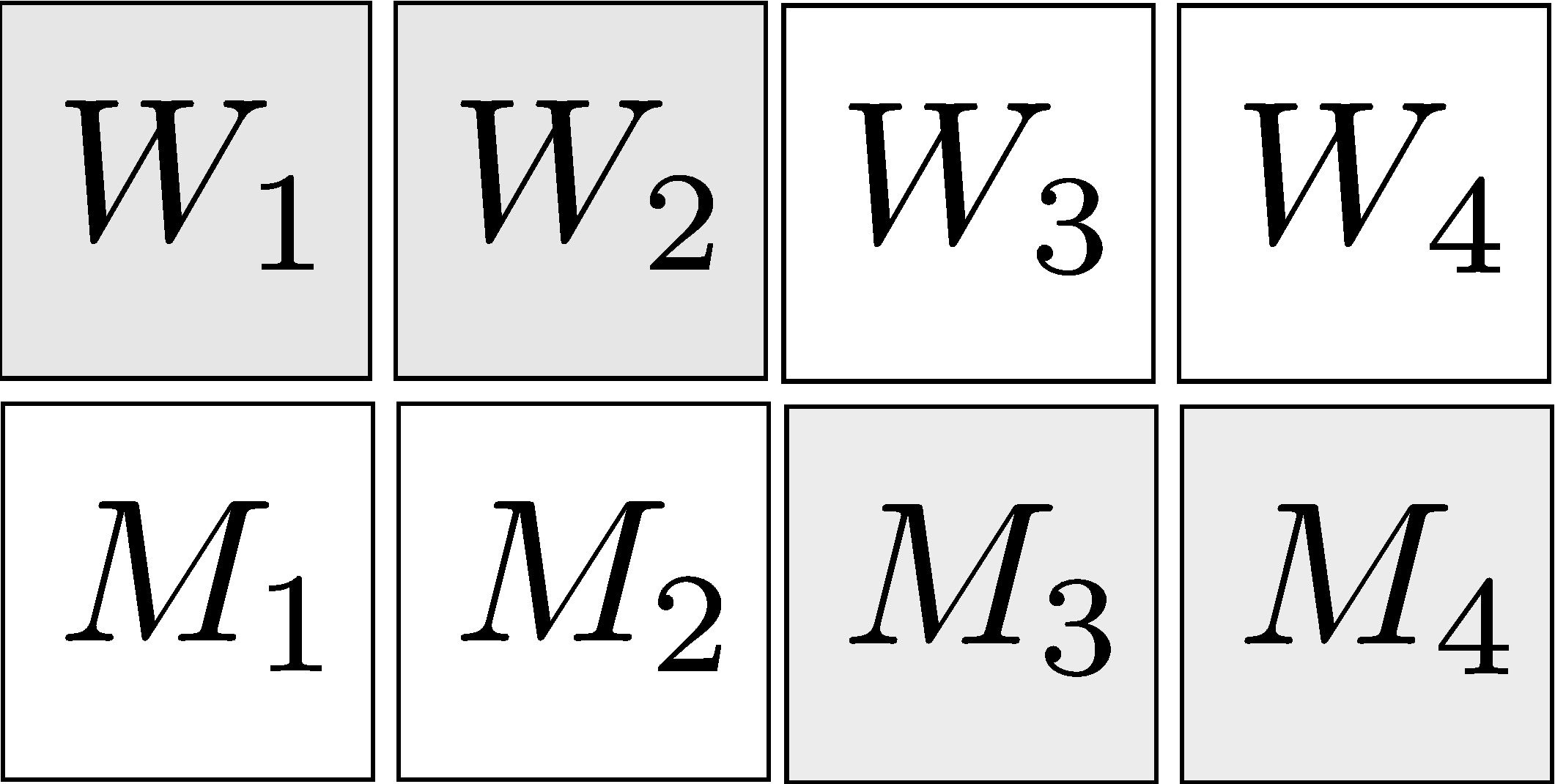}}
\caption{For a non-BPS state with $p_R=p_L=0$ and $P_L  \cdot T_{34} \in \left( {\mathbb{Z} } +\frac{1}{2} \right)^{16}$, possible BPS-decay channels must contain one of the \colorbox{lightshade}{shaded} objects, $W_1, W_2, M_3$, and $M_4$ here.}
\label{fig:8box34}
\end{figure} Decay products of a non-BPS object with 
\begin{eqnarray} 
p_R &=& p_L=0, \qquad P_{L} =\left( 1,1,1,-1,0^{12}\right)  , \qquad  P_{L}\cdot T_{12} =\left( 2,0^{15}\right), \nonumber \\
P_{L}\cdot T_{34} &=&\left( \frac{1}{2},\frac{1}{2},\frac{1}{2},-\frac{1}{2%
}; \left(  \frac{1}{2},-\frac{1}{2},-\frac{1}{2},-\frac{1}{2}  \right)^2 ;-\frac{1}{2},-\frac{1}{2},-\frac{1}{2},\frac{1}{2}%
\right) \label{D1in2a}
\end{eqnarray}
must contain some of $W_1, W_2, M_3, M_4$ as in {\bf figure} \ref{fig:8box34}.
Similarly, BPS decay products of a non-BPS state with $p_R=p_L=0$ and $P_L \cdot T_{ij}=\left( (\pm {\frac{1}{2}})^{16} \right)$ must contain some of $M_i, M_j, W_k, W_l$ where $\{i,j,k,l\}=\{1,2,3,4\}$.
For later convenience, we will divide the above into two cases, singling out the index $4$.
The first case is a non-BPS object with  
\bea p_R =p_L=0, \qquad P_L \cdot T_{k4}
= ( 0^a, \pm 2 , 0^b),  \qquad P_L \cdot T_{ij}=(  (\pm {\frac{1}{2}
})^{16}). \label{d1} \eea
 The lightest possible BPS decay products must contain some of $W_i, W_j, M_k, M_4$, and the non-BPS stability condition is
 \begin{equation}
 \sqrt{8} < \frac{1}{R_{hk}} , 4{R_{hi}},4{R_{hj}}, \frac{1}{R_{h4}} , \qquad \{i,j,k\}=\{1,2,3\}. \label{slls}
 \end{equation}
The second case is a non-BPS object with 
\bea p_R =p_L=0, \qquad P_L \cdot T_{ij}
= (  0^a, \pm 2 , 0^b) , \qquad P_L \cdot T_{k4} =(  (\pm {\frac{1}{2}
})^{16}), \label{d3} \eea
  the lightest possible BPS decay products must contain some of $M_i, M_j, W_k, W_4$, and the non-BPS stability condition is
 \begin{equation}
 \sqrt{8} < 4{R_{hj}}, \frac{1}{R_{hk}} , \frac{1}{R_{hi}}, 4{R_{h4}} , \qquad \{i,j,k\}=\{1,2,3\}. \label{lssl}
 \end{equation}

\begin{figure}
\centerline{\includegraphics[width=.35\textwidth]{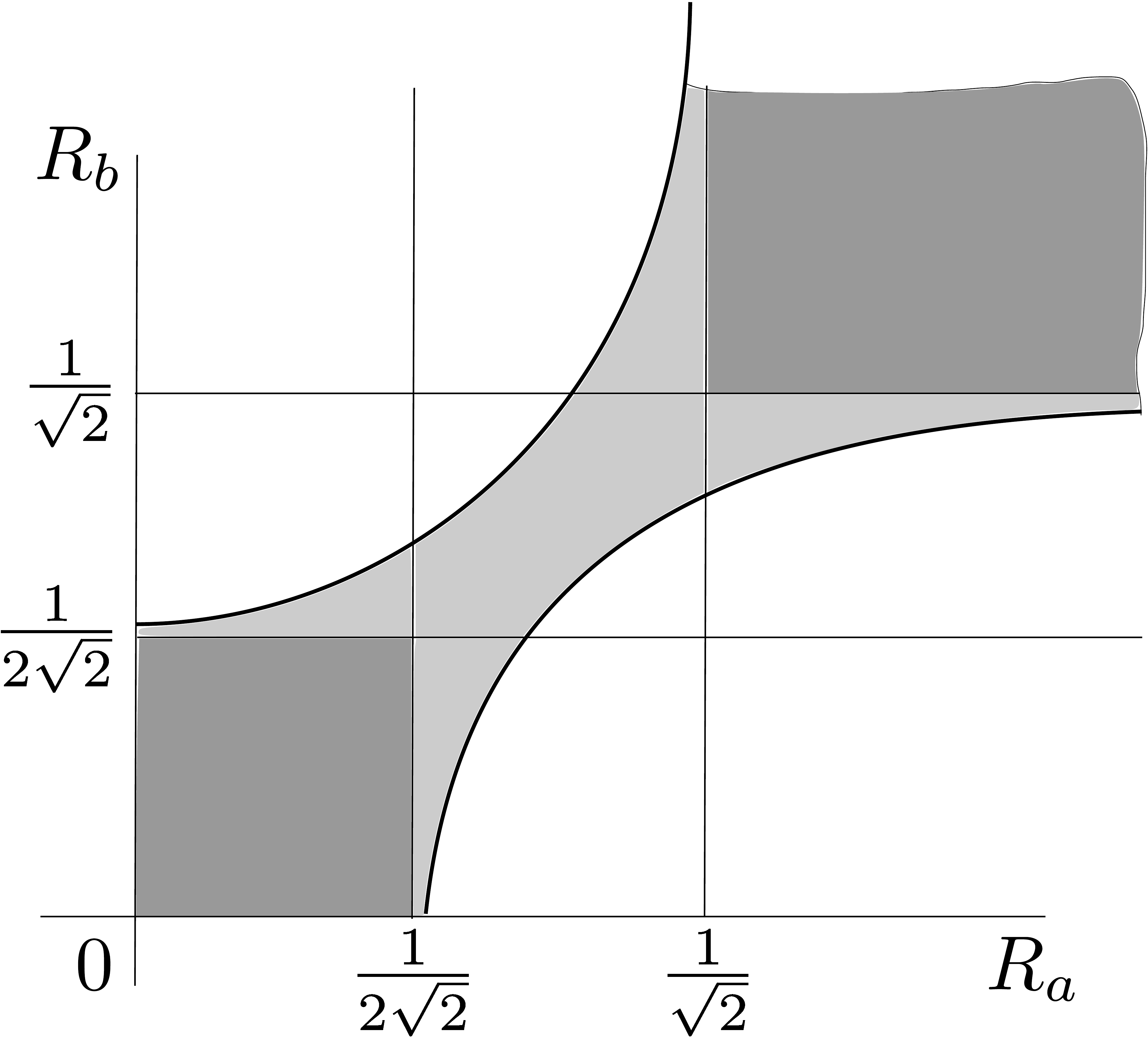}}
\caption{A non-BPS state of charge $P_L = \left( \left(   \frac{1}{2} \right)^{16} \right)$ is stable in all the shaded regions. Drawn is a 2d slice varying two radii $R_a$ and $R_b$ of the $T^4$, fixing both of the other two radii $R_c$ and $R_d$ very large or small. As we choose less extreme values for $R_c$ and $R_d$, the \colorbox{lightshade}{light shade} will get {\it larger}.
  An extra non-BPS state of \eqref{heavyNon}
 becomes exactly stable in the \colorbox{darkshade}{dark shades}.}
\label{fig:2Dphase}
\end{figure}

The results from \eqref{alllarge}, \eqref{allsmall}, \eqref{slls}, and \eqref{lssl} can be summarized as follows: A stable non-BPS object exists in every other corner of moduli space, where even number of radii are small and the rest even number of radii are large. In the dark shades in {\bf figure} \ref{fig:2Dphase}, one kind  of non-BPS states we considered becomes exactly stable. 

\subsubsection{Stability region of a robust non-BPS state in heterotic string theory \label{HetRegion}}

In the previous section, we showed that the form of $P_L$ can restrict possible BPS-decay modes. To maximize this effect, we now study a non-BPS object with $P_L  \cdot T_{ } \in \left( {\mathbb{Z} } +\frac{1}{2} \right)^{16}$ for all the eight $T$'s. For example, a non-BPS state with \bea p_R=p_L=0, \qquad P_{L}  = \left( \left( {\frac{1}{2}}\right) ^{16}\right) \label{heavyNon} \eea satisfies  $P_L  \cdot T_{ } \in \left(\pm \frac{1}{2} \right)^{16}$ for all the eight $T$'s, and its decay products must have overlap with all of the these eight groups
\begin{itemize}
\item $\{W_i|\forall i \}, \{M_i| \forall i\}$ 
\item and $ \{M_i, M_j, W_k, W_l\}$ with $\{i,j,k,l\} =\{1,2,3,4\}$
\end{itemize} 
 \begin{figure}[h]
\centerline{\includegraphics[width=\textwidth]{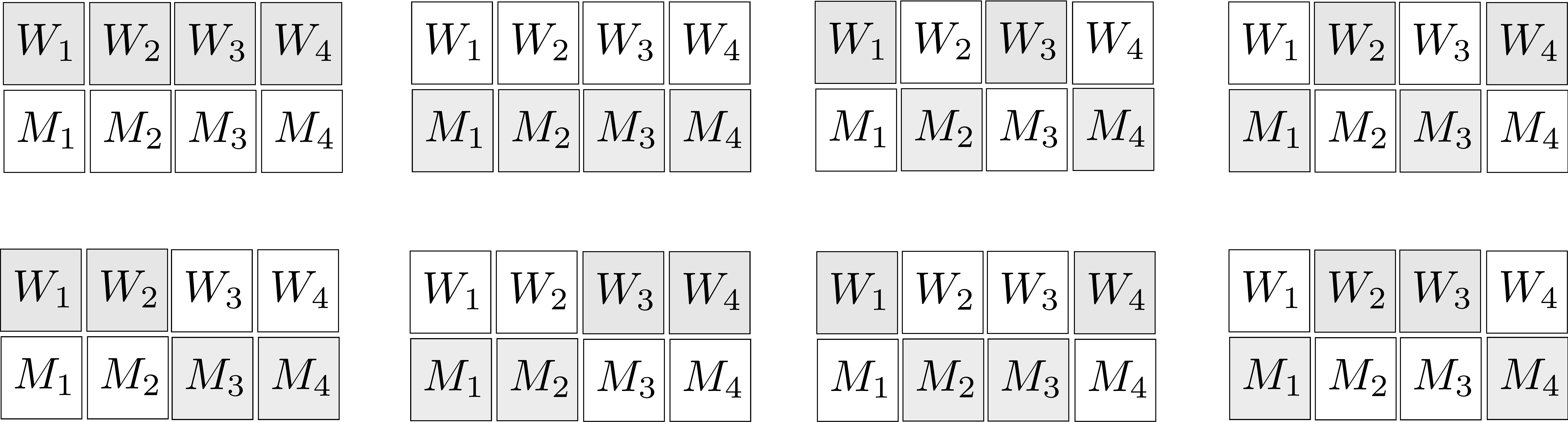}} 
\caption{A non-BPS object with $p_R=p_L=0$ and $P_L  \cdot T_{ } \in \left( {\mathbb{Z} } +\frac{1}{2} \right)^{16}$ for all eight $T$'s can decay only into sets of BPS states that have overlap with all of these eight groups.}
\label{fig:8box8necessary}
\end{figure}
as depicted in {\bf figure} \ref{fig:8box8necessary}. (This type of non-BPS state was also considered in \cite{BG}, and here we will exhaust all possible decay channels to delimitate the exact stability region continuing along methods of \cite{SeoPhD}.)

This severe restriction on decay channel comes from a ${{\mathbb{Z} }_2}^8$ parity, where each which ${{\mathbb{Z} }_2}$  determines whether $P_L  \cdot T_{ } \in   {\mathbb{Z} }^{16}$ or $P_L  \cdot T_{ } \in \left( {\mathbb{Z} } +\frac{1}{2} \right)^{16}$. This is reminiscent of ${{\mathbb{Z} }_2}$ symmetry of a $SO(32)$ spinor representation of heterotic string theory in 10d,  which does not decay due to conserved charge \cite{SpinorNotDecay}, as discussed in subsection \ref{unComp}.

\begin{figure}[h]
\centerline{\includegraphics[width=\textwidth]{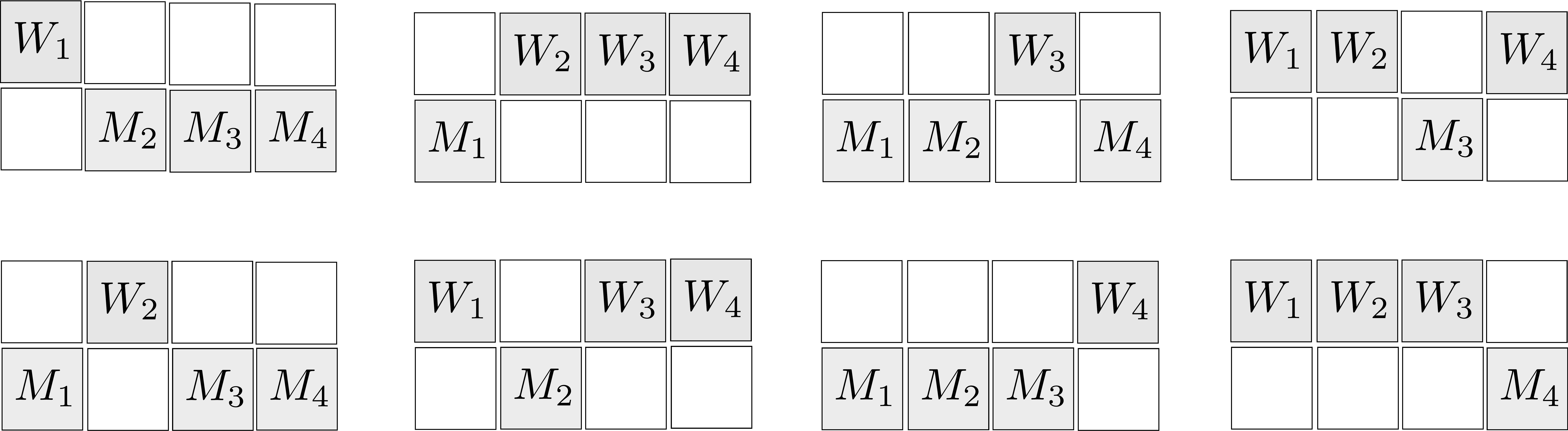}} 
\caption{Charge allows a non-BPS state with $p_R=p_L=0$ and $P_{L} =\left( \left( {\frac{1}{2}}\right) ^{16}\right)$ to decay into $ 2(W_{i}+M_{j}+M_{k}+M_{l})$ or $2(W_{i}+W_{j}+W_{k}+M_{l})$ with $\left\{ i,j,k,l \right\}  =\left\{ 1, 2, 3, 4 \right\} $.}
\label{fig:8box8cheese}
\end{figure}

The lightest collections of BPS decay products are in following two kinds:
\begin{itemize}
\item $2(M_{a}+W_{a})$ as in {\bf figure} \ref{fig:8box4heavy}.
   \begin{figure}[h]
\centerline{\includegraphics[width=\textwidth]{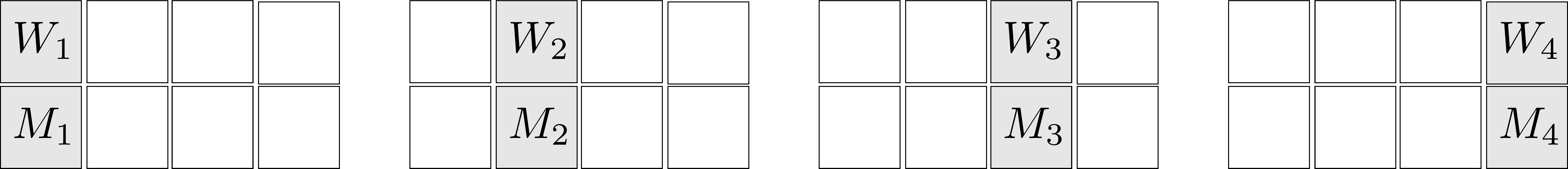}}
\caption{Charge conservation allows a non-BPS state with $p_R=p_L=0$ and $P_{L} =\left( \left( {\frac{1}{2}}\right) ^{16}\right)$ to decay into $M_i$ and $W_i$ pairs with $i \in \{1,2,3,4 \}$, but energy prohibits those decays.}
\label{fig:8box4heavy}
\end{figure}
\item  as in {\bf figure} \ref{fig:8box8cheese}, \be \emptyset \rightarrow 2(W_{i}+M_{j}+M_{k}+M_{l}),  \quad {\rm or} \quad 2(W_{i}+W_{j}+W_{k}+M_{l}) \label{decayofzero} \ee with $\left\{ i,j,k,l \right\}  =\left\{ 1, 2, 3, 4 \right\} $.
\end{itemize}

The mass on BPS side of the first decay channel
$
\frac{1}{R_{i}}+4R_{i} \geq 4>\sqrt{8}
$
is always heavier than the original non-BPS state, so this decay is excluded by energy. Even if we try decaying into two BPS molecules $2(M_{i}+W_{i}+M_{j}+W_{j})$, their mass is still $\geq 4$, as seen in second line of \eqref{twoways}.
 The mass of the second set of decay products is (forming BPS molecules of \eqref{addfour} type to save energy)
\begin{eqnarray}
\left(16R_{i}^2+\frac{1}{R_{j}^2}+\frac{1}{R_{k}^2}+\frac{1}{R_{l}^2} \right)^{\frac{1}{2}} \quad {\rm and} \quad  
\left(16R_{i}^2+16R_{j}^2+16R_{k}^2+\frac{1}{R_{l}^2} \right)^{\frac{1}{2}}.
\end{eqnarray}%
Therefore, the non-BPS object of \eqref{heavyNon} is exactly stable against {\emph{any}} decay (Since it is the lightest possible non-BPS state the decay product cannot include a non-BPS state, therefore ruling out BPS decay channel is enough.) {\emph{if and only if}} both of following hold:
\begin{eqnarray}
 16R_{i}^2+\frac{1}{R_{j}^2}+\frac{1}{R_{k}^2}+\frac{1}{R_{l}^2}  >   8 \label{spinor1} \qquad  {\rm and} \qquad
16R_{i}^2+16R_{j}^2+16R_{k}^2+\frac{1}{R_{l}^2}   >   8 . 
\end{eqnarray}%

  \begin{figure}[h]
\centerline{\includegraphics[width=.350\textwidth]{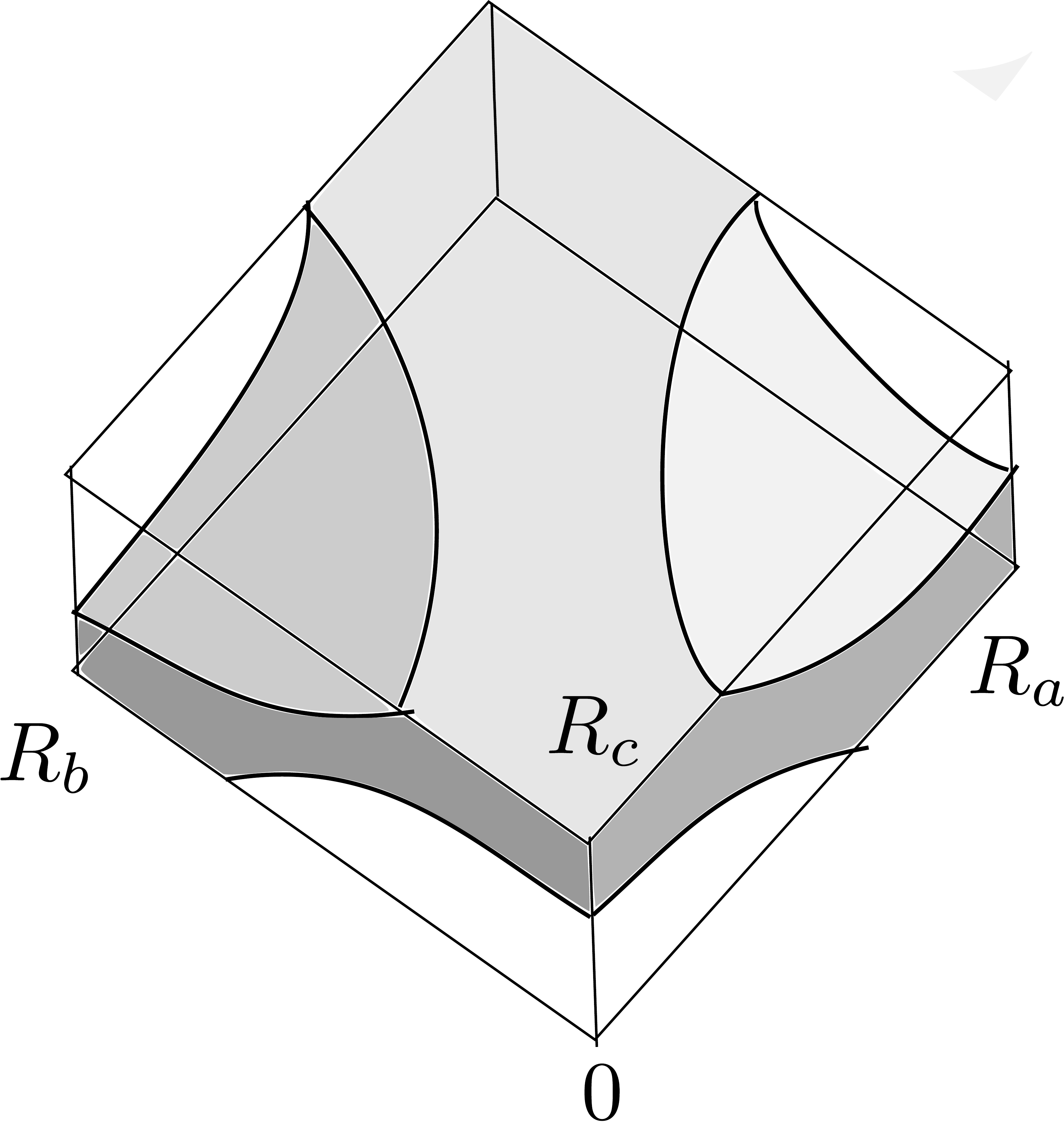}}
\caption{A cheese diagram: a 3d slice of stability region of a non-BPS state of charge $P_L = \left( \left(  \frac{1}{2} \right)^{16} \right)$. }
\label{fig:3Dphase}
\end{figure}
 If we consider the moduli space of $T^4$ as a 4d cube, then among 16 corners, this non-BPS state of \eqref{heavyNon} will be exactly stable in alternate corners and in the connecting region between them. The stability region looks like 4d cheese in a shape of a cube with every other corner (where odd number of radii are large and odd number of radii are small) eaten.  See {\bf figures} \ref{fig:2Dphase} and \ref{fig:3Dphase} for the 2d and 3d projection of stability region of this non-BPS state against decay into energetically competing BPS sides $2(W_{a}+M_{b}+M_{c}+M_{d})$ or $2(W_{a}+W_{b}+W_{c}+M_{d})$.  
 
 {\bf Figure} \ref{fig:2Dphase} is a 2d slice varying two radii $R_a$ and $R_b$ of the $T^4$, fixing both of the other two radii $R_c$ and $R_d$ very large or small - namely $R_c, R_d > \frac{1}{\sqrt{2}}$ or $R_c, R_d <  \frac{1}{2\sqrt{2}}$. If $R_c > \frac{1}{\sqrt{2}}$ and $R_d <  \frac{1}{2\sqrt{2}}$, then it will roughly appear as shown, with one of the axis now denoting $\frac{1}{4R}$ instead of $R$. As we choose less extreme values for fixed radii $R_c$ and $R_d$, the {light shade} will get {\it larger}.
  {\bf Figure} \ref{fig:3Dphase}
is a 3d slice of phase structure varying three radii $R_a$, $R_b$, $R_c$ of the $T^4$, fixing $R_d$ large, $R_d > \frac{1}{\sqrt{2}}$.
  If they instead chose $R_d <  \frac{1}{2\sqrt{2}}$ to be small, then it will roughly look similar to this, with one of the axis denoting $\frac{1}{4R}$ instead of $R$. As we choose less extreme values for $R_d$, the stability region will get {\it thicker}.
 Each of uneaten 8 corners (where even number of radii are large and even number of radii are small) corresponds to where we have one kind of exactly stable non-BPS heterotic string states, chosen among  \eqref{alllarge}, \eqref{allsmall}, \eqref{slls}, and \eqref{lssl}.

 Non-BPS states made of D-branes in ${{\mathbb{Z} }_2} \times {{\mathbb{Z} }_2}$ orientifolds with torsion are studied in \cite{ShiuTorsionRegion}. The stability region is computed using boundary state formalism and demanding the tachyons to be massless \cite{ShiuTorsionRegion}. The non-BPS stability region delineated by \eqref{spinor1}  has a similar shape to those computed in \cite{ShiuTorsionRegion,QuirozStefanskiTorsion}.   
In type IIA string theory, BPS D2-branes wrapped on 2-cycles of a Calabi-Yau 3-fold are studied, and a similar looking phase diagram appeared by considering decays between non-BPS $\widehat{\rm D1}$-brane and non-BPS $\widehat{\rm D3}$-brane \cite{MajumderSenIIAmoduliCY3}. (The symbol $\widehat{\phantom{D}}$ over $\widehat{\rm D}$-brane denotes that a D-brane has wrong dimensions and is a non-BPS object. D-even-branes (D-odd-branes) are BPS (non-BPS) objects in type IIA string theory.)

Before we end this subsection, let us demonstrate that the charge indeed allows the decay of the non-BPS state of \eqref{heavyNon} can indeed decay into the channel of \eqref{decayofzero}, such as 
\be
\renewcommand{\arraystretch}{1.5}
 \begin{tabular}{|c|cccc:cccc:cccc:cccc|}\hline $W_1$ &
   $-\frac{1}{2}$ & $\frac{1}{2}$ & $\frac{1}{2}$ & $\frac{1}{2}$ & $-%
\frac{1}{2}$ & $-\frac{1}{2}$ & $-\frac{1}{2}$ & $\frac{1}{2}$ &  $0$ &  $0$ &  $0$ &
&  $0$ &  $0$ &  $0$ &  $0$ \\ \hdashline  $W_1$ &
 0 &  $0$ &  $0$ &  $0$ &  $0$ &  $0$ &  $0$ &  $0$ & $-\frac{1}{2}$ & $\frac{1}{2}$ & $\frac{1}{2}$ &
$\frac{1}{2}$ & $-\frac{1}{2}$ & $-\frac{1}{2}$ & $-\frac{1}{2}$ & $\frac{1%
}{2}$ \\ \hline $M_2$ &
 $-1$ &  $0$ &  $0$ &  $0$ & $-1$ &  $0$ &  $0$ &  $0$ &  $0$ &  $0$ &  $0$ &  $0$ &  $0$ &  $0$ &  $0$ &  $0$ \\ \hdashline  $M_2$ &
0  &  $0$ &  $0$ &  $0$ &  $0$ &  $0$ &  $0$ &  $0$ & $-1$ &  $0$ &  $0$ &  $0$ & $-1$ &  $0$ &  $0$ &  $0$ \\ \hline $M_3$ &
  0&  $0$ &  $0$ &  $0$ & $-1$ &  $0$ & $-1$ &  $0$ &  $0$ &  $0$ &  $0$ &  $0$ &  $0$ &  $0$ &  $0$ &  $0$ \\ \hdashline $M_3$ &
 0 &  $0$ &  $0$ &  $0$ &  $0$ &  $0$ &  $0$ &  $0$ &  $0$ &  $0$ &  $0$ &  $0$ & $-1$ &  $0$ & $-1$ &  $0$ \\ \hline $M_4$ &
 0 &  $0$ &  $0$ &  $0$ & $1$ & $-1$ &  $0$ &  $0$ &  $0$ &  $0$ &  $0$ &  $0$ &  $0$ &  $0$ &  $0$ &  $0$ \\  \hdashline $M_4$ &
 0 &  $0$ &  $0$ &  $0$ &  $0$ &  $0$ &  $0$ &  $0$ &  $0$ &  $0$ &  $0$ &  $0$ & $1$ & $-1$ &  $0$ &  $0$ \\ \hline  \hline $\emptyset$ &
 $\frac{1}{2}$ & $\frac{1}{2}$ & $\frac{1}{2}$ & $\frac{1}{2}$ & $  
\frac{1}{2}$ & $\frac{1}{2}$ & $\frac{1}{2}$ & $\frac{1}{2}$ & $\frac{1}{%
2}$ & $\frac{1}{2}$ & $\frac{1}{2}$ & $\frac{1}{2}$ & $\frac{1}{2}$ & $ 
\frac{1}{2}$ & $\frac{1}{2}$ & $\frac{1}{2}$  \\ \hline
\end{tabular}.
\renewcommand{\arraystretch}{1}
\ee

 \subsection{Compactification on smaller dimensional tori \label{smaller}}
Now we consider heterotic string compactified on $T^d$ with $d \le 3$.
As explained near \eqref{nowp}, we will have massless BPS states with nonvanishing charge, which was absent for $d=4$ case above.
 With less number of circles and Wilson lines, we will see slightly different charge rules, but still we will have a non-BPS state with a large exact stability region. 
\subsubsection{Stability region of a non-BPS state in heterotic string theory on $T^3$ \label{t3}}


Still using the notation of \eqref{Enotation} for directions in $\Gamma^{16}$, we use three Wilson lines out of four given in \eqref{FixWilsonBi}. Without loss of generality, we can choose first 3 as below:
\be 
A^{1} =  \frac{1}{2} \sum_{b,c,d=0}^{1} E_{1bcd}, \quad
A^{2}= \frac{1}{2} \sum_{a,c,d=0}^{1}E_{a1cd} , \quad
A^{3} = \frac{1}{2} \sum_{a,b,d=0}^{1}E_{ab1d}  \label{FixWilsonBiT3},
\ee
which breaks the gauge group from $SO(32)$ into $SO(2^{5-3})^{2^3}=SO(4)^{8}$.
Most other equations we can inherit from $T^4$ case, only modifying the range of dummy variables.

 In {\bf table} \ref{table:hetBPSt3}, we list BPS atoms in heterotic string theory on $T^3$.
First two columns correspond to $p_L$ and $P_L$ of heterotic string states. A symbol $W_i$ denotes a set of these BPS objects with $w_i=1$ and $w_j=p=0$. Similarly, a set $M_i$ consists of the BPS excitation modes with minimal physical momentum $p^i =\frac{1}{2}$ in one of $T^3$ directions, with no other excitations $p^j = w =0$. For each capitalized index $A,B,C,D$ and $D^\prime$ , we have two choices (0 or 1), and $\pm$ and $e_i$ also gives two choices of sign. Therefore 64 choices for each row (4 indices and two signs), except for the last one, which has 32 choices  (three indices and two signs). 

   \begin{table}[!h]
\begin{center}
\renewcommand{\arraystretch}{1.5}
\begin{tabular}{|c|c||c|}
\hline
$p_{L}$ & $P_{L}$ & symbol   \\ \hline
$ \left( R_{h1},0,0  \right)  $ &  $\frac{e_1}{2}   \left( \sum_{c,d=0}^1 E_{A1cd}-2E_{A1CD}\right)+\frac{e_0}{2}    \left( \sum_{c,d=0}^1 E_{A0cd}-2E_{A0\bar{C}D^\prime}\right)$ & $W_{1}$   \\ \hdashline
$\left(0,R_{h2},0  \right)$ & $\frac{e_1}{2}  \left( \sum_{c,d=0}^1 E_{1Bcd}-2E_{1BCD}\right)+\frac{e_0}{2}   \left( \sum_{c,d=0}^1 E_{0Bcd}-2E_{0B\bar{C}D^\prime}\right)$  & $W_{2}$  \\ \hdashline
$\left(0,0,R_{h3}  \right)$ & $\frac{e_1}{2}  \left( \sum_{a,d=0}^1 E_{a1Cd}-2E_{A1CD}\right)+\frac{e_0}{2}  \left( \sum_{a,d=0}^1 E_{a0Cd}-2E_{\bar{A}0CD^\prime}\right)$ & $W_{3}$    \\ \hline
$\left(\frac{1}{4R_{h1}},0,0  \right)$ & $e_1 E_{1BCD} +e_0  E_{0BCD^\prime} $& $M_{1}$  \\ \hdashline
$\left(0,\frac{1}{4R_{h2}},0  \right)$ & $ e_1 E_{A1CD}+e_0  E_{A0CD^\prime}  $ &$M_2$\\ \hdashline
$\left(0,0,\frac{1}{4R_{h3}} \right)$ & $ e_1 E_{AB1D} +e_0 E_{AB0D^\prime} $  & $M_{3}$    \\ \hline 
$0$  &  $ e_1 E_{ABC1} + e_0 E_{ABC0}$ & massless BPS \\ \hline \end{tabular}%
\renewcommand{\arraystretch}{1}
\caption{BPS atoms in heterotic string theory on $T^3$} \label{table:hetBPSt3}
\end{center}
\end{table}

 Most importantly we have massless BPS states which were absent for $T^4$ compactification. A massless state without $p$ or $w$ excitations needs to satisfy  \eqref{nowp} (which is copied here for convenience)   \be 2 P_L  A^i =  \left( e_a \delta_{a_i 1}  + e_b \delta_{b_i 1} \right)   \equiv     \delta_{a_i 1}  + \delta_{b_i 1}  \qquad  ({\rm mod}\ 2)  \ee  for any $i\in \{1,2,3 \}$, while $\vec{a} \ne \vec{b}$. That happens if and only if $a_i=b_i$ for $i\in \{1,2,3 \}$ and $a_4 \ne b_4$. These massless BPS states are listed in the last row of the {\bf table} \ref{table:hetBPSt3}, and the form of $P_{L,0}$ 
 looks the same as $P_{L,M_4}$ of $M_4$ in heterotic string on $T^4$ shown in {\bf table} \ref{table:hetBPSt4}, where the subscript $0$ for $P_L$ means zero mass, which is different from $\emptyset$ for zero supersymmetry.

Just as in {\bf table} \ref{table:hetBPSt4}, we will verify that each row
satisfies \eqref{pQuanta}.
Starting from $P_{L,M_1}$, it is immediately seen that 
\be 2 P_{L,M_1} A^1 \equiv 1 , \quad 2 P_{L,M_1} A^{2,3} \equiv 0 . \ee
For $P_{L,W_1}$, we get three following relations with each of three Wilson lines,
 \bea 2 P_{L,W_1}A^1 &=&\delta_{A1} \left( {e_1}  + {e_0}  \right)  \equiv 0, \nonumber \\
2 P_{L,W_1}A^2 &=&\frac{e_1}{2}   \left( \sum_{c,d=0}^1 1-2 \right) = e_1 \equiv 1, \nonumber \\
2 P_{L,W_1} A^3 &=&\frac{e_1}{2}   \left( 2-2 \delta_{C1} \right)+\frac{e_0}{2}    1 \left( 2-2 \delta_{\bar{C}1}\right).
 \eea

We inherit all $T_{ij}$ and $U_{ij}$ matrices from $T^4$ case of \eqref{T12def}, as long as `4' does not appear in the index.
Again, $T_{ij}$ performs $T$-dualities along $x_i$ and $x_j$ directions in $T^3$ and exchanges excitations in $W_{i,j} \leftrightarrow M_{i,j}$ classes.   
   Massless BPS states transform among themselves under $T$. However, if we used the forbidden $T^{i4}$, then it will take a form with half-integer elements (like $W_4$ of $T^4$ case) at the {\emph{wrong}} location, which is no longer allowed due to lack of the fourth Wilson line $A^4$.
    Appendix \ref{alter} discusses alternative choice of $T_{ij}$'s. Among all tori considered in the paper, $T^3$ case is most restricted due to lack of the fourth Wilson line.

\begin{figure}[h]
\centerline{\includegraphics[width= \textwidth]{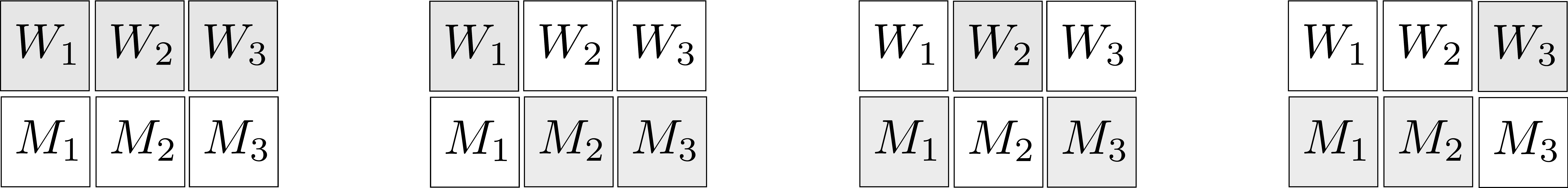}} 
\caption{A non-BPS object with $p_R=p_L=0$ and $P_L  \cdot M_{ } \in \left( {\mathbb{Z} } +\frac{1}{2} \right)^{16}$ for all eight $T$'s can decay only into sets of BPS states that have overlap with all of these eight groups.}
\label{fig:6box4necessary}
\end{figure}

\paragraph{Stability region of the lightest non-BPS state in heterotic string theory}
 A non-BPS state with \be p_R=p_L=0, P_{L}  = \left( \left( {\frac{1}{2}}\right) ^{16}\right) \ee satisfies  $P_L  \cdot T_{ } \in \left(\pm \frac{1}{2} \right)^{16}$ for all the eight $T$'s, and its decay products must have overlap with all of the these four groups $\{W_1, W_2, W_3\}$ and $\{M_i, M_j, W_k \}$ with $\{i,j,k\} =\{1,2,3\}$ as depicted in {\bf figure} \ref{fig:6box4necessary}.

\begin{figure}[h]
\centerline{\includegraphics[width=\textwidth]{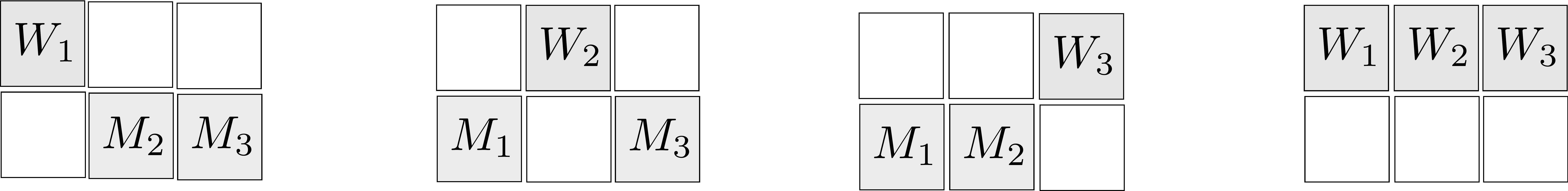}} 
\caption{Charge allows a non-BPS state with $p_R=p_L=0$ and $P_{L} =\left( \left( {\frac{1}{2}}\right) ^{16}\right)$ to decay into $2(W_{a}+M_{b}+M_{c})$ with $\left\{ a, b, c  \right\}  =\left\{ 1, 2, 3  \right\} $ or $2(W_{1}+W_{2}+W_{3})$.}
\label{fig:6box4cheese}
\end{figure}

  The lightest collections of BPS decay products are in following two kinds:
\begin{itemize}
\item $2(M_{a}+W_{a})$ as in {\bf figure} \ref{fig:6box3heavy}.
 \begin{figure}[h]
\centerline{\includegraphics[width=.7 \textwidth]{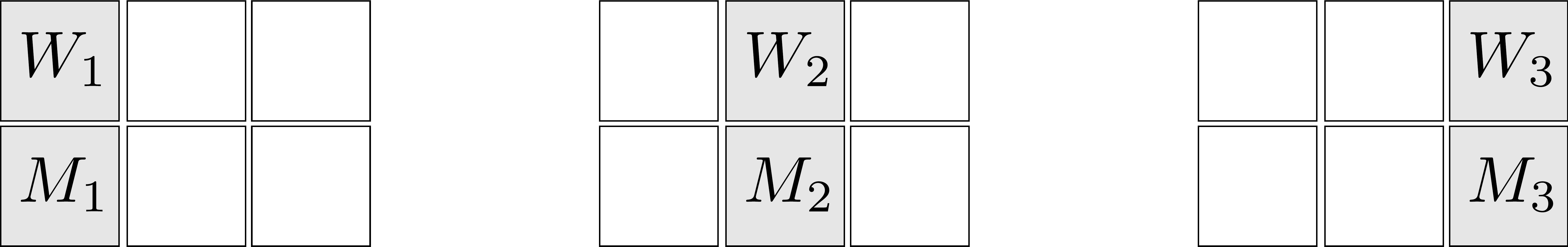}}
\caption{Charge allows a non-BPS state with $p_R=p_L=0$ and $P_{L} =\left( \left( {\frac{1}{2}}\right) ^{16}\right)$ to decay into $M_a$ and $W_a$ pairs with $a \in \{1,2,3 \}$, but energy prohibits those decays.}
\label{fig:6box3heavy}
\end{figure}
\item $2(W_{a}+M_{b}+M_{c})$ or $2(W_{a}+W_{b}+W_{c})$ with $\left\{ a, b, c \right\}  =\left\{ 1, 2, 3 \right\} $ as in {\bf figure}
\ref{fig:6box4cheese}.
\end{itemize}

Mass on BPS side of the first kind
\begin{equation}
\frac{1}{R_{a}}+4R_{a} \geq 4>\sqrt{8}
\end{equation}
is always heavier than the original non-BPS state, so this decay is excluded by energy.

Masses of the second type of decay products sum up to (forming two BPS molecules)
\begin{eqnarray}
\left(16R_{a}^2+\frac{1}{R_{b}^2}+\frac{1}{R_{c}^2} \right)^{\frac{1}{2}}  \qquad  {\rm and} \qquad
\left(16R_{a}^2+16R_{b}^2+16R_{c}^2  \right)^{\frac{1}{2}}
\end{eqnarray}%
respectively .
Therefore, this non-BPS object is exactly stable against decay into BPS states {if and only if} both of following hold:
\begin{eqnarray}
 16R_{a}^2+\frac{1}{R_{b}^2}+\frac{1}{R_{c}^2}    >   8    \qquad  {\rm and} \qquad
 16R_{a}^2+16R_{b}^2+16R_{c}^2   >   8 . 
\end{eqnarray}

The first type of decay products can take following charges for example:
\be 
\renewcommand{\arraystretch}{1.5}
 \begin{tabular}{|c|cccc:cccc:cccc:cccc|}\hline  $W_1$ &
   $ \frac{1}{2}$ & $\frac{1}{2}$ & $\frac{1}{2}$ & $-\frac{1}{2}$ & $ %
\frac{1}{2}$ & $-\frac{1}{2}$ & $ \frac{1}{2}$ & $\frac{1}{2}$ &  $0$ &  $0$ &  $0$ &0
&  $0$ &  $0$ &  $0$ &  $0$ \\ \hdashline  $W_1$ & 0
  &  $0$ &  $0$ &  $0$ &  $0$ &  $0$ &  $0$ &  $0$ & $ \frac{1}{2}$ & $ \frac{1}{2}$ & $-\frac{1}{2}$ &
$\frac{1}{2}$ & $- \frac{1}{2}$ & $ \frac{1}{2}$ & $ \frac{1}{2}$ & $ \frac{1%
}{2}$ \\ \hline $M_1$ & 0
   &  $0$ &  $0$ &  $0$ &  $0$ & $1$ &  $0$ &  $0$ &  $0$ &  $0$ &  $0$ &  $0$ &  $1$ &  $0$ &  $0$ &  $0$ \\  \hdashline  $M_1$ & 0
  &  $0$ &  $0$ &  $1$ &  $0$ &  $0$ &  $0$ &  $0$ &  $0$ &  $0$ & $1$ &  $0$ &  $0$ &  $0$ &  $0$ &  $0$ \\ \hline    \hline $\emptyset$ &
 $\frac{1}{2}$ & $\frac{1}{2}$ & $\frac{1}{2}$ & $\frac{1}{2}$ & $  
\frac{1}{2}$ & $\frac{1}{2}$ & $\frac{1}{2}$ & $\frac{1}{2}$ & $\frac{1}{%
2}$ & $\frac{1}{2}$ & $\frac{1}{2}$ & $\frac{1}{2}$ & $\frac{1}{2}$ & $ 
\frac{1}{2}$ & $\frac{1}{2}$ & $\frac{1}{2}$  \\ \hline
\end{tabular}%
\renewcommand{\arraystretch}{1}
\label{} \ee  

The second type of decay products can take following charges for example:
\be 
\renewcommand{\arraystretch}{1.5}
 \begin{tabular}{|c|cccc:cccc:cccc:cccc|}\hline $W_1$ &
   $ \frac{1}{2}$ & $\frac{1}{2}$ & $\frac{1}{2}$ & $-\frac{1}{2}$ & $ %
\frac{1}{2}$ & $-\frac{1}{2}$ & $ \frac{1}{2}$ & $\frac{1}{2}$ &  $0$ &  $0$ &  $0$ &
0&  $0$ &  $0$ &  $0$ &  $0$ \\ \hdashline   $W_1$ &
 0 &  $0$ &  $0$ &  $0$ &  $0$ &  $0$ &  $0$ &  $0$ & $ \frac{1}{2}$ & $-\frac{1}{2}$ & $\frac{1}{2}$ &
$\frac{1}{2}$ & $ \frac{1}{2}$ & $ \frac{1}{2}$ & $ \frac{1}{2}$ & $-\frac{1%
}{2}$ \\ \hline   $M_2$ &
 $-1$ &  $0$ &  $0$ &  $0$ &  $0$ & $1$ &  $0$ &  $0$ &  $0$ &  $0$ &  $0$ &  $0$ &  $0$ &  $0$ &  $0$ &  $0$ \\  \hdashline  $M_2$ &
 0 &  $0$ &  $0$ &  $0$ &  $0$ &  $0$ &  $0$ &  $0$ &  $0$ &  $0$ & $1$ &  $0$ &  $0$ &  $0$ &  $0$ & $1$  \\ \hline  $M_3$ &
 $1$ &  $0$ &  $0$ & $1$ &  $0$ &  $0$ &  $0$ &  $0$ &  $0$ &  $0$ &  $0$ &  $0$ &  $0$ &  $0$ &  $0$ &  $0$ \\ \hdashline  $M_3$ &
 0  &  $0$ &  $0$ &  $0$ &  $0$ &  $0$ &  $0$ &  $0$ &  $0$ & $1$  & $-1$ &  $0$ &  $0$ &  $0$ &  $0$ &  $0$ \\ \hline  \hline $\emptyset$ &
 $\frac{1}{2}$ & $\frac{1}{2}$ & $\frac{1}{2}$ & $\frac{1}{2}$ & $  
\frac{1}{2}$ & $\frac{1}{2}$ & $\frac{1}{2}$ & $\frac{1}{2}$ & $\frac{1}{%
2}$ & $\frac{1}{2}$ & $\frac{1}{2}$ & $\frac{1}{2}$ & $\frac{1}{2}$ & $ 
\frac{1}{2}$ & $\frac{1}{2}$ & $\frac{1}{2}$  \\ \hline
\end{tabular}%
\renewcommand{\arraystretch}{1}
\label{} 
\ee

A non-BPS state with $P_L=(2,0^{15})$ (which had a corner of stability region in $T^4$ case) decays into massless BPS states, listed in the last row of the {\bf table} \ref{table:hetBPSt3}, everywhere in the moduli space.

\subsubsection{Stability region of a non-BPS state in heterotic string theory on $T^2$ \label{t2}}
 
Still using the notation of \eqref{Enotation}, we use three Wilson lines out of four given in \eqref{FixWilsonBi}. Without loss of generality, we can choose first 2.
\be 
A^{1} =  \frac{1}{2} \sum_{b,c,d=0}^{1} E_{1bcd}, \quad
A^{2}= \frac{1}{2} \sum_{a,c,d=0}^{1}E_{a1cd}    \label{FixWilsonBiT2},
\ee
  with gauge group to be 
 $SO(2^{5-2})^{2^2}=SO(8)^{4}$.
\paragraph{BPS atoms }

 \begin{table}[!h]
\begin{center}
\renewcommand{\arraystretch}{1.5}
\begin{tabular}{|c|c||c|}
\hline
$p_{L}$ & $P_{L}$ & symbol   \\ \hline
$ \left( R_{h1},0   \right)  $ &  $ \frac{e_1}{2}    \left( \sum_{c,d=0}^1 E_{A1cd}-2E_{A1CD}\right)+\frac{e_0}{2}   \left( \sum_{c,d=0}^1 E_{A0cd}-2E_{A0C^\prime D^\prime}\right)$ & $W_{1}$   \\ \hdashline
$\left(0,R_{h2}   \right)$ & $ \frac{e_1}{2}   \left( \sum_{c,d=0}^1 E_{1Bcd}-2E_{1BCD}\right)+\frac{e_0}{2}  \left( \sum_{c,d=0}^1 E_{0Bcd}-2E_{0BC^\prime D^\prime}\right)$  & $W_{2}$  \\ \hline
$\left(\frac{1}{4R_{h1}},0   \right)$ & $e_1 E_{1BCD} +e_0 E_{0BC^\prime D^\prime} $& $M_{1}$  \\ \hdashline
$\left(0,\frac{1}{4R_{h2}}   \right)$ & $ e_1 E_{A1CD} +e_0  E_{A0C^\prime D^\prime}  $ &$M_2$  \\ \hline 
$0$  &  $e_1 E_{ABCD} + e_0 E_{ABC^\prime D^\prime}$, $(c^\prime d^\prime)\ne (c,d)$ & massless BPS \\ \hline \end{tabular}%
\renewcommand{\arraystretch}{1}
\caption{BPS atoms in heterotic string theory on $T^2$} \label{table:hetBPSt2}
\end{center}
\end{table}

In {\bf table} \ref{table:hetBPSt2}, we list BPS atoms. A symbol $W_i$ denotes a set of these BPS objects with $w_i=1$ and $w_j=p=0$. Similarly, a set $M_i$ consists of the BPS excitation modes with minimal physical momentum $p^i =\frac{1}{2}$ in one of $T^2$ directions, with no other excitations. First four rows each have freedom to choose five indices and 2 signs ($2^7$), and the last one has choice to choose $A, B$ and two signs ($2^4$) multiplied by choosing $C, D$ and $C^\prime, D^\prime$ ($4\times 3$). Therefore there are total $3\times 2^6$
choices. Massless BPS states have same $P_L$ as linear combination of $M_3$ and $M_4$ of heterotic string on $T^4$.
  We again take $T_{12}$ from \eqref{T12def}, which exchanges excitations in $W_{1,2} \rightarrow M_{1,2}$ classes, while leaving massless BPS states transform among themselves.
    
\begin{figure}[h!]
\centerline{\includegraphics[width=.4\textwidth]{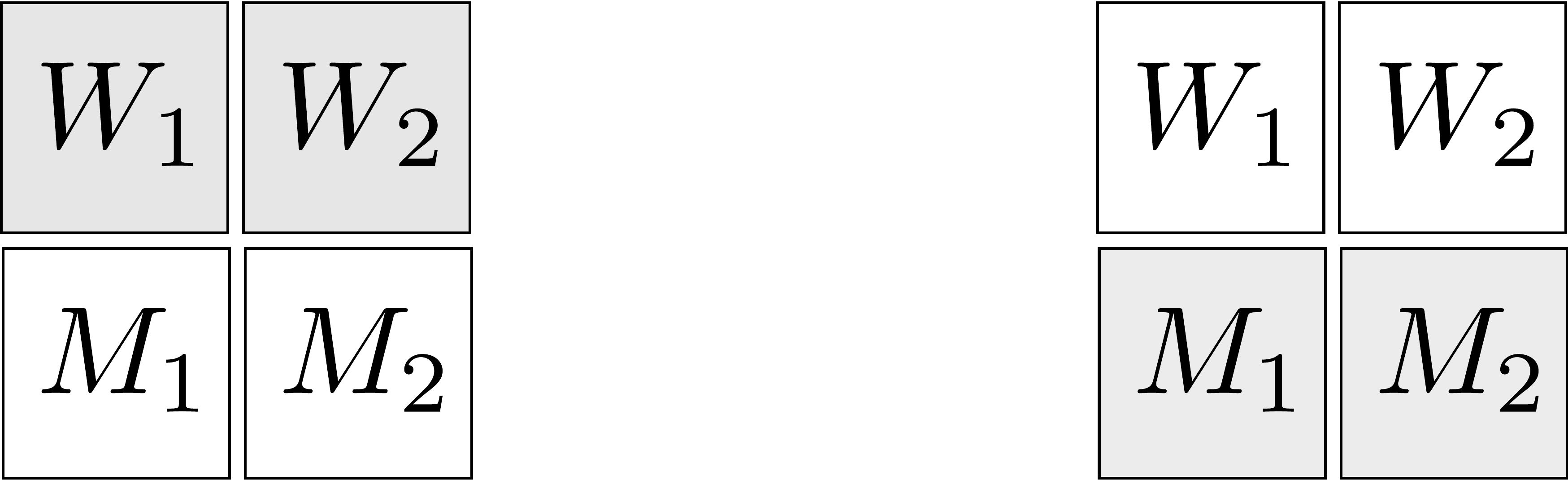}} 
\caption{A non-BPS object with $p_R=p_L=0$ and $P_L  \cdot M_{ } \in \left( {\mathbb{Z} } +\frac{1}{2} \right)^{16}$ for all eight $T$'s can decay only into sets of BPS states that have overlap with all of these eight groups.}
\label{fig:4box2necessary}
\end{figure}

\paragraph{Stability of non-BPS states}
 For example, a non-BPS state with $p_R=p_L=0$  and $ P_{L}  = \left( \left( {\frac{1}{2}}\right) ^{16}\right) $
 satisfies  $P_L  \cdot T_{ } \in \left(\pm \frac{1}{2} \right)^{16}$ for all the eight $T$'s, and its decay products must have overlap with all of the these two groups $\{W_1, W_2\},   \{M_1, M_2 \}$ as depicted in {\bf figure} \ref{fig:4box2necessary}.

  The lightest collections of BPS decay products are in following two kinds:
\begin{itemize}
\item $2(M_{a}+W_{a})$ as in {\bf figure} \ref{fig:4box2heavy}.
 \begin{figure}[h]
\centerline{\includegraphics[width=.4\textwidth]{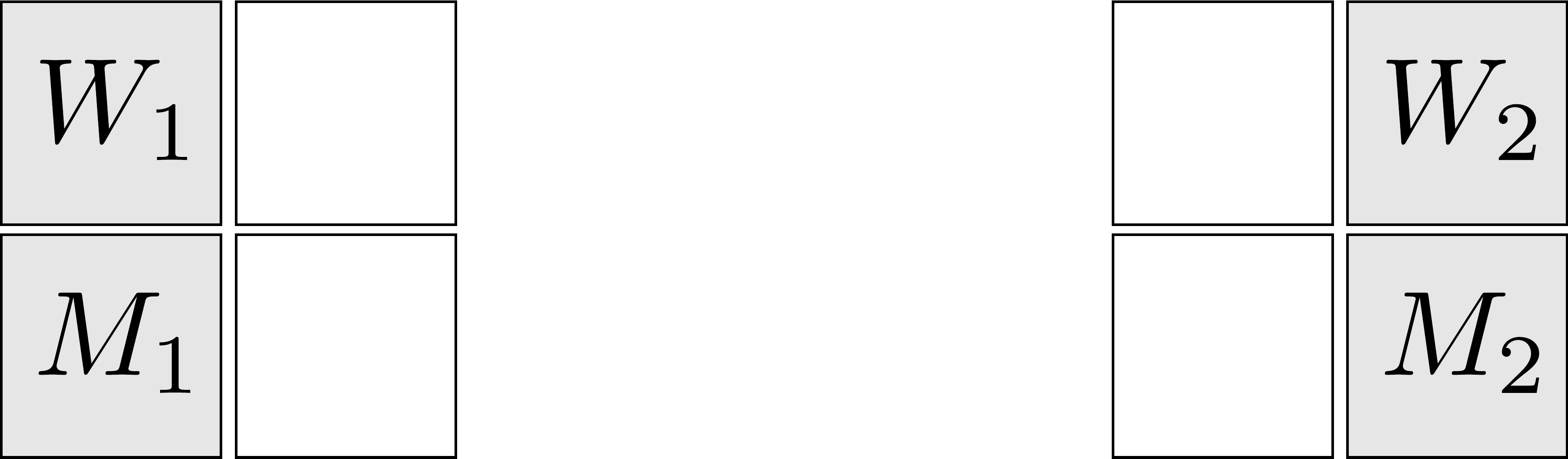}}
\caption{Charge allows a non-BPS state with $p_R=p_L=0$ and $P_{L} =\left( \left( {\frac{1}{2}}\right) ^{16}\right)$ to decay into $M_a$ and $W_a$ pairs with $a \in \{1,2  \}$, but energy prohibits those decays.}
\label{fig:4box2heavy}
\end{figure}
\item $2(W_{a}+M_{b} )$  with $\left\{ a, b \right\}  =\left\{ 1, 2 \right\} $ as in {\bf figure}
\ref{fig:4box2cheese}.
 \begin{figure}[h]
\centerline{\includegraphics[width=.4\textwidth]{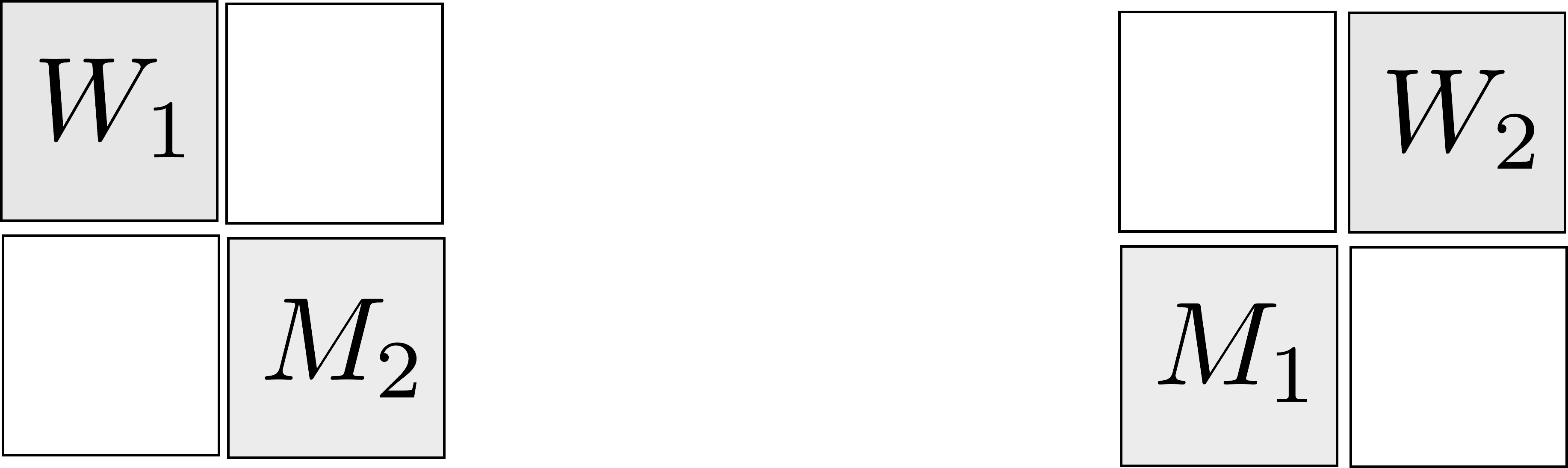}} 
\caption{Charge allows a non-BPS state with $p_R=p_L=0$ and $P_{L} =\left( \left( {\frac{1}{2}}\right) ^{16}\right)$ to decay into $2(W_{a}+M_{b} )$  with $\left\{ a, b  \right\}  =\left\{ 1, 2  \right\} $.}
\label{fig:4box2cheese}
\end{figure}
\end{itemize}

Mass on BPS side of the first kind
$
\frac{1}{R_{a}}+4R_{a} \geq 4>\sqrt{8}
$
is always heavier than the original non-BPS state, so this decay is excluded by energy.
 Masses of the second type of decay products sum up to
$
\left(16R_{a}^2+\frac{1}{R_{b}^2}  \right)^{\frac{1}{2}}  .
$
Therefore, this non-BPS object is exactly stable against decay into BPS states {if and only if} both of following hold:
\begin{eqnarray}
 16R_{1}^2+\frac{1}{R_{2}^2}    >   8   \qquad  {\rm and} \qquad
  16R_{2}^2+\frac{1}{R_{1}^2}    >   8 .  
\end{eqnarray}%

The first type of decay products may carry charges as below for example:
\be 
\renewcommand{\arraystretch}{1.5}
 \begin{tabular}{|c|cccc:cccc:cccc:cccc|}\hline $W_1$ &
   $ \frac{1}{2}$ & $\frac{1}{2}$ & $\frac{1}{2}$ & $-\frac{1}{2}$ & $ 
\frac{1}{2}$ & $-\frac{1}{2}$ & $ \frac{1}{2}$ & $\frac{1}{2}$ &  $0$ &  $0$ &  $0$ &0
&  $0$ &  $0$ &  $0$ &  $0$ \\ \hdashline $W_1$ & 0
  &  $0$ &  $0$ &  $0$ &  $0$ &  $0$ &  $0$ &  $0$ & $ \frac{1}{2}$ & $\frac{1}{2}$ & $-\frac{1}{2}$ &
$\frac{1}{2}$ & $ \frac{1}{2}$ & $-\frac{1}{2}$ & $ \frac{1}{2}$ & $\frac{1%
}{2}$ \\ \hline $M_1$ & 0
  &  $0$ &  $0$ &  $ 1$ &  $0$ &  $0$ &  $0$ &  $0$ &  $0$ &  $0$ & $1$ &  $0$ &  $0$ &  $0$ &  $0$ &  $0$ \\  \hdashline $M_1$ & 0
  &  $0$ &  $0$ &  $0$ &  $0$ & $ 1$ &  $0$ &  $0$ &  $0$ &  $0$ &  $0$ &  $0$ &  $0$ & $ 1$ &  $0$ &  $0$ \\ \hline  \hline $\emptyset$ &
 $\frac{1}{2}$ & $\frac{1}{2}$ & $\frac{1}{2}$ & $\frac{1}{2}$ & $  
\frac{1}{2}$ & $\frac{1}{2}$ & $\frac{1}{2}$ & $\frac{1}{2}$ & $\frac{1}{%
2}$ & $\frac{1}{2}$ & $\frac{1}{2}$ & $\frac{1}{2}$ & $\frac{1}{2}$ & $ 
\frac{1}{2}$ & $\frac{1}{2}$ & $\frac{1}{2}$  \\ \hline
\end{tabular}%
\renewcommand{\arraystretch}{1},
\label{ } \ee 
while the second type of decay products may carry charges as below for example:
\be 
\renewcommand{\arraystretch}{1.5}
 \begin{tabular}{|c|cccc:cccc:cccc:cccc|}\hline $W_1$ &
   $ \frac{1}{2}$ & $\frac{1}{2}$ & $\frac{1}{2}$ & $-\frac{1}{2}$ & $ 
\frac{1}{2}$ & $-\frac{1}{2}$ & $ \frac{1}{2}$ & $\frac{1}{2}$ &  $0$ &  $0$ &  $0$ &0
&  $0$ &  $0$ &  $0$ &  $0$ \\ \hdashline $W_1$ & 0
  &  $0$ &  $0$ &  $0$ &  $0$ &  $0$ &  $0$ &  $0$ & $ \frac{1}{2}$ & $\frac{1}{2}$ & $-\frac{1}{2}$ &
$\frac{1}{2}$ & $ \frac{1}{2}$ & $-\frac{1}{2}$ & $ \frac{1}{2}$ & $\frac{1%
}{2}$ \\ \hline  $M_2$ & 0
  &  $0$ &  $0$ &  $ 1$ &  $0$ & $ 1$ &  $0$ &  $0$ &  $0$ &  $0$ &  $0$ &  $0$ &  $0$ &  $0$ &  $0$ &  $0$ \\  \hdashline $M_2$ & 0
  &  $0$ &  $0$ &  $0$ &  $0$ &  $0$ &  $0$ &  $0$ &  $0$ &  $0$ &   $1$ &  $0$ &  $0$ & $ 1$ &  $0$ &  $0$   \\ \hline  \hline $\emptyset$ &
 $\frac{1}{2}$ & $\frac{1}{2}$ & $\frac{1}{2}$ & $\frac{1}{2}$ & $  
\frac{1}{2}$ & $\frac{1}{2}$ & $\frac{1}{2}$ & $\frac{1}{2}$ & $\frac{1}{%
2}$ & $\frac{1}{2}$ & $\frac{1}{2}$ & $\frac{1}{2}$ & $\frac{1}{2}$ & $ 
\frac{1}{2}$ & $\frac{1}{2}$ & $\frac{1}{2}$  \\ \hline
\end{tabular}%
\renewcommand{\arraystretch}{1}
\label{ } \ee  

\subsubsection{Stability region of a non-BPS state in heterotic string theory on $T^1$\label{t1}}
Heterotic string theory on a circle was studied in \cite{GinspargHetTorus}.
Also \cite{Bergman:1997py} explained its light BPS states and dualities. 
Still using the notation of \eqref{Enotation}, we use three Wilson lines out of four given in \eqref{FixWilsonBi}. Without loss of generality, we can choose first one
$
A^{1} =  \frac{1}{2} \sum_{b,c,d=0}^{1} E_{1bcd} .$
with a gauge group to be
 $SO(2^{5-1})^{2^1}=SO(16)^{2}$.
 
\paragraph{BPS atoms }
   
 \begin{table}[!h]
\begin{center}
\renewcommand{\arraystretch}{1.5}
\begin{tabular}{|c|c||c|}
\hline
$p_{L}$ & $P_{L}$ & symbol   \\ \hline
$  R_{h1}    $ &  $\left( \left( \pm \frac{1}{2} \right)^8, 0^8 \right) $, or $\left(0^8, \left( \pm \frac{1}{2} \right)^8 \right) $ where (-) appears even times  & $W_{1}$   \\   \hline
$ \frac{1}{4R_{h1}}    $ & $e_1 E_{1BCD} +e_0 E_{0B^\prime C^\prime D^\prime} $& $M_{1}$  \\   \hline 
$0$ &  $ e_1 E_{abcd} +e_0 E_{AB^\prime C^\prime D^\prime}$, $(B^\prime, C^\prime D^\prime)\ne (B,C,D)$ & massless BPS \\ \hline \end{tabular}%
\renewcommand{\arraystretch}{1}
\caption{BPS atoms in heterotic string theory on $T^1$} \label{table:hetBPSt1}
\end{center}
\end{table}

In {\bf table} \eqref{table:hetBPSt1} BPS atoms are listed. A symbol $W_1$ denotes a set of these BPS objects with $w_1=1$ and $p_1=0$. Similarly, a set $M_1$ consists of the BPS excitation modes with minimal physical momentum $p^1 =\frac{1}{2}$ with no other excitation. First line has $2^7$ choices to choose signs for 7 of them, multiplied by extra factor of 2 for choosing which form to take. The second has 2 signs and 6 index, again giving $2^8$. The last one has $A$ to choose, and 8 times 7 to choose $(B^\prime, C^\prime D^\prime)$ and $(B,C,D)$. These massless BPS states has same $P_L$ as a linear combination of $M_2$, $M_3$ and $M_4$ of heterotic string on $T^4$. 
        
\paragraph{Stability of non-BPS states} 

A non-BPS state with $w=p=0, P_L = \left( \left( \frac{1}{2}\right)^{16} \right)$ is a lightest non-BPS state, whose 
 lightest collection of decay products is $2( W_{1})$ with 
mass $4R_{1}$. Therefore 
$
  R_{1}^2  > \frac{1}{2}
$
is the stability region of the non-BPS state. 
 
\section{Type IIA string theory on K3 from the heterotic string perspective\label{HetIIAduality}}

This section reviews heterotic string theory on $T^4$ and type IIA string theory on an
orbifold limit of a K3 surface and discusses the duality between them.

\subsection{A duality chain between heterotic theory and type IIA string theory \label{dualitychain}}

Type IIA string theory on an orbifold limit of a K3 surface is dual to heterotic string theory on $T^4$ \cite{WittenSDualityDimensions,AspinwallK3}, through the following chain of dualities \cite{BG}
\begin{equation}
\frac{\mbox{het}}{T^{4}}\;\;\overset{S}{\longrightarrow }\frac{\mbox{I}\;}{%
T^{4}}\;\overset{T^{4}}{\longrightarrow }\frac{\mbox{IIB}\;}{T^{4}/{\mathsf{%
Z\!\!Z}}_{2}^{\prime }}\;\overset{S}{\longrightarrow }\frac{\mbox{IIB}\;}{%
T^{4}/{\mathsf{Z\!\!Z}}_{2}^{\prime \prime }}\;\overset{T}{\longrightarrow }%
\frac{\mbox{IIA}\;}{T^{4}/{\mathsf{Z\!\!Z}}_{2}}\;\, . \label{chain}
\end{equation}
Here the $\mathbb{Z}_2$ actions are given as
\begin{equation}
{\mathsf{Z\!\!Z}}_{2}^{\prime }  =  (1,\Omega \mathcal{I}_{4}) ,\qquad
{\mathsf{Z\!\!Z}}_{2}^{\prime \prime}  =  (1,(-1)^{F_{L}}\mathcal{I}_{4}) ,\qquad
{\mathsf{Z\!\!Z}}_{2}  =  (1,\mathcal{I}_{4}),
\end{equation}
where the operator $\Omega $
reverses world-sheet parity and $F_{L}$ is the left-moving part of the
spacetime fermion number.  The operator
$\mathcal{I}_{4}$ implements reflection in all 4 compact directions $x_i$'s of a 4-torus \begin{equation}
\mathcal{I}_{4} :(x_1,x_2,x_3,x_4) \rightarrow (-x_1,-x_2,-x_3,-x_4). \end{equation} Since orbifolding in each direction gives 2 fixed points, the action of $\mathcal{I}_{4}$ on a 4-torus gives $2^4=16$ fixed points on an orbifold limit of a K3 surface.

This chain employs S-duality between type I
and heterotic string theories, self-S-duality of type IIB string theory, $T$-duality between
type I and IIB string theories along all four $x_i$ directions of $T^4$, and $T$-duality between type IIB and IIA string theories along $x_4$ direction.

Assuming a diagonal metric
tensor for $T^4$,
the coupling constant $g_h$ of heterotic string theory and the radii $R_{hi}$'s of $T^4$ are
written in terms of the coupling constant $g_A$ of type IIA string theory and the moduli $R_{Ai}$'s of an orbifold limit of a K3 surface as \cite{BG}
\begin{equation}
g_{h}=\frac{V_{A}}{8g_{A}R_{A4}},
\qquad
R_{hj}={\frac{1}{2}}\frac{\sqrt{V_{A}}}{R_{Aj}R_{A4}},
\qquad
R_{h4}=\frac{\sqrt{V_{A}}}{2}, \label{radii}
\end{equation}
with \begin{equation}
V_{A}= R_{A1}R_{A2}R_{A3}R_{A4}, \qquad V_{h}=
R_{h1}R_{h2}R_{h3}R_{h4}.\end{equation}
Radii along $x_j$ ($j \in\{ 1,2,3 \}$) directions and $x_4$ direction have different formula in \eqref{radii} due to an extra $T$-duality along $x_4$ direction between type IIA and IIB string theories in the duality chain of \eqref{chain}.
The masses of BPS states in type IIA and heterotic string theories are related to each other by \cite{BG}
\begin{equation}
m_{h}=\frac{ \sqrt{ V_{h}} }{g_{h}}m_{A}. \label{hetIIAmassBPS}
\end{equation}%

Consider compactification of type IIA string theory on an orbifold limit of a K3 surface, $T^4/{{\mathsf{Z\!\!Z} }_2}$. D-even-branes are BPS states in type IIA string theory, and their masses may be expressed as the tension $\frac{1}{g_A}$ times the volume of D-brane
\begin{equation}
m_{{\rm BPS}, A} = \frac{ {\rm volume}}{g_A}.
\end{equation}

A bulk D0-brane has a unit volume in 0-dimension and has the mass of $\frac{1}{g_A}$. Fractional D0-branes sit
at the 16 fixed points of an orbifold limit of a K3 surface. After blowing up each fixed point into a 2-sphere, we can wrap D2-branes over these 2-cycles. The fractional D0-brane can be
thought of as a D2-brane wrapping a vanishing 2-cycle, which comes from a resolution of an orbifold singularity.
 Fractional D0-branes have one-half unit volume in 0-dimension because of the ${\mathsf{Z\!\!Z}}_{2}$ orbifolding in the K3 surface. Their mass
is  $\frac{1}{2g_A}$, which is half of that of a bulk D0-brane.
 A D4-brane wrapping the whole K3 surface has mass $\frac{V_A}{2g_A}$. D2-branes wrapping the torus $T^2$ in $x_j$ and $x_k$ directions have mass $\frac{%
R_{Aj} R_{Ak}}{2g_A}$.

By matching the charges and the masses, \cite{BG} lists mappings between BPS states in heterotic and type IIA string theories. Heterotic states in $W_4$ and $M_4$ are dual to D4-branes and fractional D0-branes. Heterotic states in $W_j$ and $M_j$ are dual to D2-branes over 2-cycles over $x_k, x_l$ (denoted as ${\rm D2}_{kl}$) and $x_j, x_4$ directions (denoted as ${\rm D2}_{j4}$), with $\{j,k,l\} =\{1,2,3\}$.

$T_{ij}$ and $U_{ij}$ matrices can still be interpreted similarly in type IIA string theory. $U_{ij}$ matrices swap the roles of $x_i$ and $x_j$ directions of K3, and $T_{ij}$ matrices perform $T$-dualities along $x_i$ and $x_j$ directions.

\subsection{BPS bound states in type IIA \label{BPSmole2a}}
Here we will review the BPS bound states in type IIA picture which appeared in subsection \ref{BPSmole} in heterotic picture. 

First, we added two quarter-BPS atoms of same kind to obtain a half-BPS molecule in \eqref{halfBPSM4}. In type IIA, it translates as adding two fractional D0-branes at a fixed point to obtain a bulk D0-brane. In both dual theories, the binding energy is zero. Applying the same logic, one could add two D2 or D4-branes of same kind to obtain another BPS object of same dimensionality with zero binding energy. 

In 
\eqref{M34BPS}, we saw examples where two quarter-BPS states add up to another quarter-BPS molecule, where the BPS molecule saves energy just a little bit by the Pythagorean theorem.  
In type IIA, they correspond to a D0 brane at a fixed point and a D2 brane spanning $x^{3,4}$ directions with negative binding energy. By one $T$-duality, this can be interpreted as two orthogonal D1-branes, which form a bound state of a single D1-brane with shorter length than that of two original D1-branes. We saw a similar example in \eqref{M4W1BPS}, where two quarter-BPS states add up to another quarter-BPS molecule. They are a 
D0 brane at a fixed point and a ${\rm D2_{23}}$ (a D2 brane spanning $x^{2,3}$ directions), and the same logic applies for constructing a BPS molecule which is D2 brane with less energy. For both cases, the mass of the bound state is expected from the first lines of \eqref{M34q} and \eqref{m14q} to take following form: 
  \be m_{\rm D0_f+D2} =  \sqrt{m_{\rm D0_f}^2 +m_{\rm D2}^2   } . \ee
  The subscript $f$ denotes fractional D0 brane, as opposed to a bulk D0 brane, however we will drop this subscript since only fractional D0 branes appear in the equations. 
  
A different kind of binding happens for \eqref{MW1BPS}, where two quarter-BPS states add up to form a quarter-BPS state, but with zero binding energy.
In IIA language, they are a D2-brane along $x^{2,3}$ directions and another along $x^{1,4}$ directions. By 2 $T$-dualities, the system is equivalent to a D4 and D0, and the bound state has zero binding energy.

A related example appears in \eqref{M14W1BPS}, which correspond to 
a D0 at a fixed point, D2-branes along $x^{1,4}$ and $x^{2,3}$ directions. The negative binding energy is due to the interaction between D0 and D2, but not because of the interaction between two D2-branes. From \eqref{m114q}, we expect the mass to be \be m_{\rm D0+D2_{14}+D2_{23} } =  \sqrt{ \left(m_{\rm D2_{14}} +m_{\rm D2_{23}} \right)^2 +m_{\rm D0}^2 }\label{ } .\ee

Again the case in \eqref{M1234W1BPS} corresponds to having a 
D0 at a fixed point and D2-branes along $x^{2,3}, x^{1,4} ,x^{2,4} ,x^{3,4}$ directions, with a small negative binding energy. From \eqref{m11234q}, the mass becomes 
 \be  m_{\rm D0+D2_{23} +D2_{14}+D2_{24}+D2_{34}} = \sqrt{ \left(m_{\rm D2_{23}} +m_{\rm D2_{14}} \right)^2 +m_{\rm D0}^2 +m_{\rm D2_{24}}^2 +m_{\rm D2_{34}}^2 }. \label{ } \ee 

The examples of \eqref{12BPS} and \eqref{12BPS2} are translated as binding 
D2-branes along $x^{2,3}, x^{1,3}, x^{1,4}, x^{2,4}$ directions with {\emph {varying}} binding energy. From \eqref{m12h} and \eqref{m12q}, the masses are given as
 \bea m_{{\rm D2_{14}+D2_{23}+D2_{13}+D2_{24} }, \frac{1}{2} } &=&  \sqrt{  \left( m_{\rm D2_{14}} +m_{\rm D2_{23}}\right)^2+\left(m_{\rm D2_{13}} +m_{\rm D2_{24}}\right)^2}, \nonumber \\
 m_{{\rm D2_{14}+D2_{23}+D2_{13}+D2_{24}} , \frac{1}{4}}   &=&\sqrt{  m_{\rm D2_{14}}^2+m_{\rm D2_{23}}^2+m_{\rm D2_{13}}^2+m_{\rm D2_{24}}^2} ,  \eea
where we kept the amount of supersymmetry preserved on the heterotic side, to distinguish. By $T$-dualities, this can be translated into 
\bea m_{{\rm D0_{f}+D2_{12}+D2_{34}+D4_{1234} }, \frac{1}{2} } &=&  \sqrt{  \left( m_{\rm D0_{f}} +m_{\rm D4_{1234}}\right)^2+\left(m_{\rm D2_{12}} +m_{\rm D2_{34}}\right)^2}, \nonumber \\
 m_{{\rm D0_{f}+D2_{12}+D2_{34}+D4_{1234}} , \frac{1}{4}}   &=&\sqrt{  m_{\rm D0_{f}}^2+m_{\rm D4_{1234}}^2+m_{\rm D2_{12}}^2+m_{\rm D2_{34}}^2} ,  \eea
where we denoted by $f$ to emphasize that these are fractional D0 branes (as opposed to bulk D0 branes).
Other integer subscript denotes directions D branes are spanning.

Also from \eqref{addfour}, we have 
  \be  m_{{\rm D2_{14}+D2_{24}+D2_{34}+D0_{f}} , \frac{1}{4}} = \sqrt{m_{ {D2_{14}}}^2+m_{\rm D2_{24}}^2+m_{\rm D2_{34}}^2+m_{\rm D0_{f}}^2   }. \label{ } \ee
 
\subsection{A stable non-BPS bound state \label{stab2anon}}
 The non-BPS state of \eqref{D3IIa} on heterotic string side corresponds to a non-BPS $\widehat{\rm D3}$-brane in type IIA string theory stretched along $x_1, x_2, x_3$, as discussed in \cite{GaberdielLecture}. The BPS decay channels on the heterotic side are $W$'s: under string-string duality they correspond to a pair of D4-$\overline{{\rm D4}}$, or a pair of D2-$\overline{{\rm D2}}$ spanning $i$ and $j$ directions with $ i,j \in \{1,2,3 \} $. Similarly the non-BPS state of \eqref{D1IIA} is interpreted as a non-BPS $\widehat{\rm D1}$-brane stretched along $x_4$ direction \cite{BG,GaberdielLecture}. This is also justified from the fact that the possible BPS decay products allowed by charge conservation on the heterotic side ($M$'s) are mapped under string-string duality into a pair of wrapped D0-$\overline{{\rm D0}}$, or a
pair of D2-$\overline{{\rm D2}}$ spanning $x_i$ and $x_4$ directions with $i\in \{1,2,3 \}$. The stability region in type IIA is worked out in \cite{BG} and as explained in \cite{GaberdielLecture}, which qualitatively matches with that in heterotic string theory. 

The previous paragraph is summarized in \eqref{d13dual}: 
  \bea \emptyset & { \rm of  } & \eqref{D3IIa}  
  \left\{ 
 \begin{array}{lcl} 
 \rightarrow 2(W_{i})   \Leftrightarrow   {{\rm D4}}{\rm -}\overline{{\rm D4}}_{1234}, {{\rm D2}}{\rm -}\overline{{\rm D2}}_{12, 23, 13}
 \end{array} 
  \right>   \sim   \widehat{\rm D3}_{123} \nonumber \\ 
  \emptyset  & {\rm of  } & \eqref{D1IIA}  
  \left\{ 
 \begin{array}{lcl} 
 \rightarrow 2(M_{i})   \Leftrightarrow   {{\rm D0}}{\rm -}\overline{{\rm D0}}, {{\rm D2}}{\rm -}\overline{{\rm D2}}_{14,24,34}
 \end{array} 
  \right>   \sim   \widehat{\rm D1}_{4} \label{d13dual}
 \eea 
 Starting from the non-BPS state on the heterotic side, $\rightarrow$ denotes the BPS decay products on the heterotic side, and after $\Leftrightarrow$ BPS objects on the type IIA side are given from string-string duality of subsection \ref{dualitychain}. Finally after $\sim$ is the expected non-BPS state which is expected to form by forming type IIA BPS objects listed between $\Leftrightarrow$ and $\sim$.
 
Similarly, the non-BPS state of \eqref{D1in2a} on the heterotic side is expected to map to $\widehat{\rm D1}_3$-brane of type IIA, as below
   \be \emptyset  {\ \rm of \ } \eqref{D1in2a}
  \left\{ 
 \begin{array}{lcl} 
 \rightarrow 2(W_1+ W_2+ M_3+ M_4) & \Leftrightarrow & {{\rm D0}}{\rm -}\overline{{\rm D0}}, {{\rm D2}}{\rm -}\overline{{\rm D2}}_{13,23,34}
 \end{array} 
  \right> \sim \widehat{\rm D1}_{3}.
 \ee 
 Similar results hold for the non-BPS objects of \eqref{d1} and \eqref{d3}:
     \bea \emptyset & { \rm of  } 
    & \eqref{d1}
  \left\{ 
 \begin{array}{lcl} 
 \rightarrow 2(W_i+W_j+ M_k+M_{4}) & \Leftrightarrow & {{\rm D0}}{\rm -}\overline{{\rm D0}}, {{\rm D2}}{\rm -}\overline{{\rm D2}}_{jk,ik,k4}
 \end{array} 
  \right> \sim \widehat{\rm D1}_{k}, \nonumber \\
  \emptyset & { \rm of  } & \eqref{d3}
  \left\{ 
 \begin{array}{lcl} 
 \rightarrow 2(M_i+M_j+W_k+ W_4) & \Leftrightarrow & {{\rm D4}}{\rm -}\overline{{\rm D4}}_{ijk4}, {{\rm D2}}{\rm -}\overline{{\rm D2}}_{i4,j4,ij}
 \end{array} 
  \right> \sim \widehat{\rm D3}_{ij4}.
 \eea   
Lastly, for the stubborn non-BPS state of \eqref{heavyNon}, the following relation holds:
 \be \emptyset  {\ \rm of \ } \eqref{heavyNon}
 \left\{ 
 \begin{array}{lcl} 
   \rightarrow 2(M_{a}+W_{a}) & \Leftrightarrow &   {{\rm D2}}{\rm -}\overline{{\rm D2}}_{ij,kl} \\
  \rightarrow 2(M_{4}+W_{4}) & \Leftrightarrow & {{\rm D4}}{\rm -}\overline{{\rm D4}}, {{\rm D0}}{\rm -}\overline{{\rm D0}} \\
 \rightarrow 2(M_{1}+M_{2}+M_{3}+W_{4}) & \Leftrightarrow & {{\rm D4}}{\rm -}\overline{{\rm D4}}, {{\rm D2}}{\rm -}\overline{{\rm D2}}_{14,24,34}\\ 
 \rightarrow 2(W_{1}+W_{2}+W_{3}+M_{4}) & \Leftrightarrow & {{\rm D0}}{\rm -}\overline{{\rm D0}}, {{\rm D2}}{\rm -}\overline{{\rm D2}}_{23,13,12}\\ 
 \rightarrow 2(W_{i}+M_{j}+M_{k}+M_{4}) & \Leftrightarrow & {{\rm D0}}{\rm -}\overline{{\rm D0}}, {{\rm D2}}{\rm -}\overline{{\rm D2}}_{jk,j4,k4}\\ 
 \rightarrow 2(M_{i}+W_{j}+W_{k}+W_{4}) & \Leftrightarrow & {{\rm D4}}{\rm -}\overline{{\rm D4}}, {{\rm D2}}{\rm -}\overline{{\rm D2}}_{i4,ik,ij}
 \end{array} 
 \right> \sim \begin{array}{l}
 \widehat{\rm D3}_{ijk}+\widehat{\rm D1}_{l}?\\
  {{\rm NS5}}{\rm -}\overline{{\rm NS5}}?
  \end{array}
 \ee 
The non-BPS state of heterotic string theory given in \eqref{heavyNon} 
may correspond to a bound state of a non-BPS $\widehat{\rm D1}$-brane and $\widehat{\rm D3}$-brane or $\widehat{\rm D3}$ with nontrivial magnetic flux \cite{GaberdielLecture}. Or it may be understood as a bound state of an NS5-brane and an anti-NS5-brane \cite{Bergman:1999kz}. The former is likely from the observation of the decay channels, but a conclusive test is yet to come. 
 
\section{Conclusion and open problems\label{Conclusion}}

We studied BPS and non-BPS states in heterotic string theory
compactified on $T^d$ with $0\le d \le 4$, and found a non-BPS state with a large exact stability region. 
We classified BPS atoms with elementary excitations, and wrote down the explicit expression for their charges. 
We organized conservation of the charges using a set
of $16 \times 16$ transformation matrices, which performs even numbers of $T$-dualities. Among other things, the charge conservation gets reduced to various parity problems.
We constructed a non-BPS state which is rather robust against decays into BPS states. We identified its huge
stability region in moduli space, proving that no other
decays into BPS states are possible. We looked at a particular non-BPS state, and gave its stability region against all possible decays. We do not know conclusively whether there can be another non-BPS state whose stability region is larger. 

Our non-BPS state does not require fine tuning of moduli for its stability, and it will be interesting to use this in a
string-inspired model building. The study of non-BPS objects and
their stability may also provide non-trivial tests of weak-strong
duality between heterotic string theory compactified on $T^4$ and type
IIA string theory on an orbifold limit of a K3 surface. 
 
We provided a detailed study of heterotic string on $T^4$ listing all possible quantum numbers. Based on charge and masses of BPS objects, one can map BPS objects of heterotic string theory to that of type IIA side on K3. It will be nice to build a detailed state-by-state map, which can guide how the detailed information on the heterotic side transforms into that on the type IIA side (such as brane location and flux etc). Then we can better understand the results we listed for half- and quarter- BPS and non-BPS molecules for type IIA side, which came from the heterotic computation. 

Our studies suggest that heterotic string theory may prove to be a fertile path to study non-BPS states: both BPS and non-BPS masses can be computed exactly and the charge conservation rules are explicit. It will be interesting to learn about K-theory and tachyons in heterotic string theory. When a non-BPS state decays into a set of lighter BPS states, where does the extra energy go? Level matching and mass formulas in heterotic sides give useful information for objects before and after decays, but it does not tell us much about what happens {\it during} the process. We want to have the dynamic picture as well as static pictures. We can also do a similar study in ${\cal N}=1, d=4$, instead of ${\cal N}=2, d=6$, by compactifying on a different manifold.
 
  \paragraph{Suggested recipe for studying non-BPS stability of heterotic string theory on other manifolds} 
  \begin{enumerate}
  \item Find proper Wilson lines. Based on the form of Wilson lines, find a nice notation for directions in $\Gamma^{16}$.
  \item Write down rules for $P_L, p_{L,R}$ which satisfies an equation like \eqref{pQuanta}.
  \item Classify BPS building blocks, and examine the lightest non-BPS states. 
  \item Search for automorphism (such as $T$-duality) within $\Gamma^{16}$, to make the most out of the symmetry in the system. (Can we write down something similar to $T$-duality transformation matrices for heterotic string theory on K3 or CY3?)
  \item Examine whether charge conservation can reduce to a simpler problem, such parity of integer etc.
  \item Starting from light non-BPS state, restrict decay channels as much as possible.
  \item Compare masses and delineate the stability region. BPS masses are given by $m = 2|p_R|$, while the lightest non-BPS mass is given by $m= \sqrt{ 8 ( N_R -C_R)} =\sqrt{8}$. Therefore BPS masses depend on moduli while non-BPS mass is constant.
  \end{enumerate}
    
Exactly stable non-BPS states have been studied in various string theories, but mostly those involving D-branes (such as IIA, IIB, I, etc) rather than heterotic string theory. 
A non-BPS D0-brane is stable in type I string theory \cite{SenD0I,SenD0Iinteractions,WittenDK}, which is realized as a stable non-BPS pair of D1-$\overline{{\rm D1}}$ in type IIB string theory \cite{SenIspinorIIAD}.  This stability holds in a particular region of moduli space. At the boundary of the stability region, tachyons become massless, the force between non-BPS objects vanishes, and there is exact degeneracy in the Bose-Fermi spectrum \cite{SenProofSpacetimeSusyTachyon,BoseFermiNoForce}. Non-BPS states and their stability against certain decay channels have been discussed for type IIA string theory compactified over a K3 surface \cite{SenK3orbifoldD} and a Calabi-Yau three-fold \cite{StefanskiCY3}.  
 There are also stable non-BPS brane-antibrane constructions in type IIB string theory using D4-branes and $\overline{{\rm D4}}$-branes hung between NS5-branes \cite{MukhiTong,MukhiDDbarRepulsion}. Also see \cite{Tatar:2000jm} for making D5- anti D5 stable by varying B field and magnetic fluxes, and massless tachyons.  Can we make similar progress on heterotic side as well?
Following much study on brane-antibrane system, non-BPS brane, and tachyons, K theory was developed which gives a new way to keep track of conserved charges \cite{WittenDK,Witten:2000cn,Horava:1998jy}. Heterotic string theory has explicit rules about conserved charge and BPS-ness. Can there be any extra secrets like K-theory, on the heterotic side?
 
 Although heterotic string theory is connected to other theories by duality, when it comes to non-BPS states, each theory needs to be studied on their own. Duality may not be a sufficient test of stability of non-BPS objects. For example, a non-BPS D0-brane is unstable in type IIB string theory, but stable in type I string theory \cite{SenD0I,SenD0Iinteractions,WittenDK} because the orientifold action projects out tachyon modes in type I \cite{WittenDK}. Also string-string duality being a strong-weak duality, each theory has a different validity regime. It is important to study both sides.
   
One ambitious plan would be to study wall-crossing phenomena between BPS and non-BPS states. We also have a real-codimension-one locus - boundary of non-BPS marginal stability region - in the moduli space, but strictly speaking in a subspace of moduli space. If we considered the full moduli space, taking into account off-diagonal part of the metric and B-field, then we could claim to have a wall of marginal crossing between BPS and non-BPS states.

\acknowledgments
 
This work evolved from a final project ``Stable non-BPS states in string theory'' with Ilya Nikokoshev for a physics course ``Topics in String Theory'' at Harvard University in 2006 Spring offered by Prof Cumrun Vafa. I thank Ilya for the fun collaboration and Cumrun for the exciting class. Ilya and I learned a lot from discussions with Megha Padi, Kirill Saraikin, Shing-Tung Yau, and especially Lubos Motl. 

Further explorations of 2010 spring lead to preliminary results published in \cite{SeoPhD} and 
it was pleasant and helpful to discuss with 
Lara Anderson, Chris Beem, Hengyu Chen, Keshav Dasgupta, Ron Donagi, Matthias Gaberdiel, Josh Guffin, John Harnad,
  Ian-Woo Kim, Sungjay Lee, Alex Maloney, Hirosi Ooguri, Chang-Soon Park, Ashoke Sen, Jaewon Song, and Cumrun Vafa.   
The author received helpful suggestions from Ron Donagi, Mboyo Esole, Momin Malik, Chang-Soon Park, Jon Tyson, and especially Matthias Gaberdiel on the earlier manuscript and from Keshav Dasgupta from the recent manuscript.
I also thank Minou and Minhye Lee for encouragement and support during the final stage of this work.

This research was supported in part by NSF grants PHY-0244821 and DMS-0244464, by the Korea Foundation for Advanced Studies, and by NSERC grants.

\appendix
\section{Further properties of $U_{ij}$ and $T_{ij}$ matrices \label{0eigen}}

 \subsection{Alternative choices for $T_{ij}$'s \label{alter}}
In defining $T_{ij}$'s we used \eqref{T12def} and \eqref{LRdef}, but it is not the unique choice, we could modify \eqref{T12def} and still enjoy all the nice properties, which are 
\begin{itemize}
\item $T_{ij}$ transforms $P_L$ charges as if there were $T$-dualities performed on $x^i$ and $x^j$ directions.
\item and $T_{ij}$'s to square to an identity.
\end{itemize}
 Recall that we used $L$ and $R$ defined in \eqref{LRdef} in order to write down $T_{12}$ in \eqref{T12def}, which are copied here for convenience below:
 \begin{equation}
L  =  {\frac{1}{2}} \left(
\begin{array}{cccc}
+1 & +1 & +1 & -1 \\
+1 & +1 & -1 & +1 \\
+1 & -1 & +1 & +1 \\
-1 & +1 & +1 & +1%
\end{array}%
\right), \qquad R = {\frac{1}{2}} \left(
\begin{array}{cccc}
+1 & -1 & -1 & -1 \\
-1 & +1 & -1 & -1 \\
-1 & -1 & +1 & -1 \\
-1 & -1 & -1 & +1%
\end{array}%
\right).  
\end{equation}
We can use their slightly modified versions $L^\prime, R^\prime$ defined as below:
\begin{equation}
L^\prime  = L {u_{34}} = {\frac{1}{2}} \left(
\begin{array}{cccc}
+1 & +1 & +1 & -1 \\
+1 & -1 & +1 & +1 \\
+1 & +1 & -1 & +1 \\
-1 & +1 & +1 & +1%
\end{array}%
\right), \qquad R^\prime = R  {u_{34}} = {\frac{1}{2}} \left(
\begin{array}{cccc}
+1 & -1 & -1 & -1 \\
-1 & -1 & +1 & -1 \\
-1 & +1 & -1 & -1 \\
-1 & -1 & -1 & +1%
\end{array}%
\right). \label{LRdefMod} 
\end{equation}

The key points of $L, R, L^\prime, R^\prime$ are
\begin{itemize}
\item that they square to the identity matrix
\item and that each row and each column contains odd number of negative elements
\item and that each row of $R$ can be obtained from each row of $L$ at the same height by reverting the order of elements, like a mirror image (and reverting the overall sign, optionally).
\end{itemize} 
 Alternative choices of $T_{12}$ are 
\begin{equation}
T_{AB}   =   \left(
\begin{array}{cccc}
A & 0 & 0 & 0 \\
0 & 0 & B & 0 \\
0 & B & 0 & 0 \\
0 & 0 & 0 & A%
\end{array} \right),  
\end{equation}
where $(A,B)=\pm (L,\pm R), \pm (R,\pm L),  \pm (R^\prime ,\pm L^\prime), \pm (L^\prime ,\pm R^\prime)$ where none of $\pm$'s are correlated. 
These work perfectly well for the cases of heterotic string theories compactfied on $T^1, T^2, T^4$. However if compactified on $T^3$, the lack of the fourth Wilson line forces one to distinguish $E_{abc0}$ from $E_{abc1}$ and only unprimed ones work. If we discard the primed matrices and stay with $(A,B)=\pm (L,\pm R), \pm (R,\pm L)$, then $T_{AB}$ has this property: for a fixed $a, b, c  \in \{0,1 \}$, $abc0_{(2)}$'th row and $abc1_{(2)}$'th has difference only between $a^\prime b^\prime c^\prime 0_{(2)}$'th row and $a^\prime b^\prime c^\prime 1_{(2)}$'th element for a fixed $a^\prime, b^\prime, c^\prime \in \{0,1 \}$.

\subsection{Eigenvectors and eigenvalues of $U_{ij}$ and $T_{ij}$ matrices  \label{eigen}}
Here we will list eigenvectors and eigenvalues of various matrices which appear in the paper. 
First, recall the definition of ${u_{34}}$ from \eqref{sigma4def},\begin{equation} {u_{34}} = 
\left(
\begin{array}{cccc}
1 & 0 & 0 & 0 \\
0 & 0 & 1 & 0 \\
0 & 1 & 0 & 0 \\
0 & 0 & 0 & 1%
\end{array}%
\right),  \end{equation}
 which is copied here for reading convenience.
Eigenvectors of ${u_{34}}$ are 
\begin{itemize}
\item $
\left\{
\begin{array}{r}
  \left( 1,0,0,0\right) \\
\frac{1}{\sqrt{2}}\left( 0,1,1,0\right) \\
\left( 0,0,0,1\right) 
\end{array}
\right.$ 
with an eigenvalue $1$ and \item
$\frac{1}{\sqrt{2}}\left( 0,-1,1,0\right) $ with an eigenvalue $ -1$.
\end{itemize}
From this information, eigenvectors and eigenvalues of $U_{34}$ of \eqref{Udef} are obtained similarly: its eigenvectors are 
\begin{itemize}
\item $
\left\{
\begin{array}{r}
\left( 0^{4k};1,0,0,0;0^{12-4k}\right)  \\ 
\frac{1}{\sqrt{2}}\left( 0^{4k};0,1,1,0;0^{12-4k}\right)    \\
\left( 0^{4k};0,0,0,1;0^{12-4k}\right)   
\end{array} \right.$ with an eigenvalue $1$, and
\item $\frac{1}{\sqrt{2}}\left( 0^{4k};0,-1,1,0;0^{12-4k}\right)$ with an eigenvalue $-1$, where $k \in \{0,1,2,3 \}$. 
\end{itemize}
Using the notation of \eqref{Enotation}, this can be rewritten as 
\begin{itemize}
\item $ E_{ab11}, \frac{1}{\sqrt{2}}\left(   E_{ab10}+E_{ab01} \right), E_{ab00}$ with an eigenvalue $1$, and
\item $ \frac{1}{\sqrt{2}}\left( -  E_{ab10}+E_{ab01} \right)$  with an eigenvalue $-1$, where $a, b=0, 1$. 
\end{itemize}
 Eigenvectors of other $U_{ij}$'s can be written out similarly. For example, those of $U_{13}$ are   
 \begin{itemize}
\item $ E_{1b1d}, \frac{1}{\sqrt{2}}\left(   E_{0b1d}+E_{1b0d} \right), E_{0b0d}$ with an eigenvalue $1$, and
\item $ \frac{1}{\sqrt{2}}\left( -  E_{0b1d}+E_{1b0d} \right)$  with an eigenvalue $-1$, where $ b, d=0, 1$. 
\end{itemize}

Omitting the procedure, we list eigenvectors of $T_{12}$ below: 
\begin{itemize}
\item $
\left\{
\begin{array}{l}
  \left( 1,-1,-1,1;0^{12}\right) \\
\left( 0^{12};1,-1,-1,1\right) \\
\left( 0^{4};1,1,1,-1;0,0,0,2;0^{4}\right) \\
\left( 0^{4};1,1,-1,1;0,0,2,0;0^{4}\right) \\
\left( 0^{4};1,-1,1,1;0,2,0,0;0^{4}\right) \\
\left( 0^{4};-1,1,1,1;2,0,0,0;0^{4}\right) 
\end{array}
\right.$ with an eigenvalue $-1$, and 
\item $
\left\{
\begin{array}{l}
\left( 1,1,0,0;0^{12}\right)  \\
\left( 1,0,1,0;0^{12}\right) \\
\left( 1,0,0,-1;0^{12}\right) \\
\left( 0^{12};1,1,0,0\right) \\
\left( 0^{12};1,0,1,0\right) \\
\left( 0^{12};1,0,0,-1\right) \\
\left( 0^{4};1,1,1,-1;0,0,0,-2;0^{4}\right) \\
\left( 0^{4};1,1,-1,1;0,0,-2,0;0^{4}\right)  \\
\left( 0^{4};1,-1,1,1;0,-2,0,0;0^{4}\right)  \\
\left( 0^{4};-1,1,1,1;-2,0,0,0;0^{4}\right) 
\end{array}
\right.$ with an eigenvalue $1$,
\end{itemize}
and eigenvectors of $T_{1234}$ are
\begin{itemize}
\item $
\left\{
\begin{array}{l}
 \left( -1,1,1,1;1,1,1,-3;0^{7},2\right) \\
\left( 1,1,-1,1;-1,1,1,-1;0^{6},2,0\right) \\
\left( 1,-1,1,1;-1,1,1,-1;0^{5},2,0^{2}\right) \\
\left( 1,1,1,-1;-3,1,1,1;0^{4},2,0^{3}\right) \\
\left( 1,-1,-1,1;1,1,1,-1;0^{3},2,0^{4}\right) \\
\left( 1,1,-3,1;1,-1,1,1;0^{2},2,0^{5}\right) \\
\left( 1,-3,1,1;1,1,-1,1;0^{1},2,0^{6}\right) \\
\left( 1,-1,-1,1;-1,1,1,1;2,0^{7}\right) 
\end{array}
\right.$ with an eigenvalue $-1$, and 
\item 
$
\left\{
\begin{array}{l}
\left( -1,1,1,1;1,1,1,1;0^{7},-2\right) \\
\left( 1,1,-1,1;-1,1,-3,-1;0^{6},-2,0\right) \\
\left( 1,-1,1,1;-1,-3,1,-1;0^{5},-2,0^{2}\right) \\
\left( 1,1,1,-1;1,1,1,1;0^{4},-2,0^{3}\right) \\
\left( 1,-1,-1,-3;1,1,1,-1;0^{3},-2,0^{4}\right) \\
\left( 1,1,1,1;1,-1,1,1;0^{2},-2,0^{5}\right) \\
\left( 1,1,1,1;1,1,-1,1;0^{1},-2,0^{6}\right) \\
\left( -3,-1,-1,1;-1,1,1,1;-2,0^{7}\right) 
\end{array}
\right.$ with an eigenvalue $1$. 
\end{itemize}
 It is straightforward to write them in the notation of \eqref{Enotation} and to write them down for all other $T_{ij}$'s with minor modifications.
 
\bibliographystyle{JHEP}
\bibliography{HetNonBPS}

 \end{document}